\documentclass[fleqn,usenatbib]{mnras}

\usepackage{newtxtext,newtxmath}
\usepackage[T1]{fontenc}

\usepackage{amsmath}
\usepackage{bm}
\usepackage{xcolor}
\usepackage{graphicx}
\usepackage{geometry}
\usepackage{mathtools}
\usepackage{tikz}

\newcommand{\pa}{\partial}
\newcommand{\mb}{\boldsymbol}

\newcommand{\bgeq}{\begin{equation}}
\newcommand{\edeq}{\end{equation}}
\newcommand{\RNum}[1]
{\uppercase\expandafter{\romannumeral #1\relax}}

\title[DRWI in Turbulent Dust-Trapping Rings]{The Dusty Rossby Wave Instability (DRWI): Linear Analysis and Simulations of Turbulent Dust-Trapping Rings in Protoplanetary Discs}

\author[H. Liu and X.-N. Bai]{
Hanpu Liu$^{1,2}$\thanks{liuhanpu@stu.pku.edu.cn} and
Xue-Ning Bai$^{3,4}$\thanks{xbai@tsinghua.edu.cn}
\\
$^{1}$Kavli Institute for Astronomy and Astrophysics, Peking University, Beijing 100871, People's Republic of China\\
$^{2}$Department of Astronomy, Peking University, Beijing 100871, People's Republic of China\\
$^{3}$Institute for Advanced Study, Tsinghua University, Beijing 100084, China\\
$^{4}$Department of Astronomy, Tsinghua University, Beijing 100084, China
}

\date{Accepted 2023 August 24. Received 2023 August 24; in original form 2023 July 14}

\pubyear{2023}

\begin{document}
\label{firstpage}
\pagerange{\pageref{firstpage}--\pageref{lastpage}}
\maketitle

\begin{abstract}
%Planetesimals may preferentially form in the ubiquitous dust rings of protoplanetary discs, where a radial pressure maximum is believed to enhance dust concentration. 
Recent numerical simulations have revealed that dust clumping and planetesimal formation likely proceed in ring-like disc substructures, where dust gets trapped in weakly turbulent pressure maxima.
%Numerous protoplanetary discs have shown enhanced dust concentration in one or more rings, which may serve as sites for planetesimal formation. 
The streaming instability has difficulty operating in such rings with external turbulence and no pressure gradient. To explore potential paths to planetesimal formation in this context,
%we introduce a new local shearing sheet model of the dust-trapping ring and analyse its linear stability.
we analyse the stability of turbulent dust-trapping ring under the shearing sheet framework.
We self-consistently establish the pressure maximum and the dust ring in equilibrium, the former via a balance of external forcing versus viscosity and the latter via dust drift versus turbulent diffusion. We find two types of $\gtrsim H$-scale instabilities ($H$ being the pressure scale height), which we term the dusty Rossby wave instability (DRWI). Type~\RNum{1} is generalised from the standard RWI, which is stationary at the pressure maximum and dominates in relatively sharp pressure bumps. Type~\RNum{2} is a newly identified travelling mode that requires the presence of dust. It can operate in relatively mild bumps, including many that are stable to the standard RWI, and its growth rate is largely determined by the equilibrium gas and dust density gradients. %Type~\RNum{1} is generalised from the standard RWI while Type~\RNum{2} is first identified. Type~\RNum{1} is stationary at the pressure maximum and dominates in relatively sharp pressure bumps. Type~\RNum{2} operates in relatively mild bumps, \xb{with} its maximum growth rate largely determined by the equilibrium gas and dust density gradients.
%The misalignment between perturbed pressure and total density gradients acts as an effective baroclinity that regulates the vortensity budget; based on this, we interpret the mechanism of the Type~\RNum{2} DRWI as two travelling Rossby waves coupled with a ``dust wave'' (baroclinity-modulated vortensity patterns), the three interacting closely to achieve concurrent propagation and growth. 
We further conduct two-fluid simulations that verify the two types of the DRWI.
While Type~\RNum{1} leads strong to dust concentration into a large gas vortex similar to the standard RWI, the dust ring is preserved in Type~\RNum{2}, and meanwhile exhibiting additional clumping within the ring.
%The two types of DRWI are verified in simulations, and their 
%non-linear evolution leads to compact dust vortices that likely trigger dust clumping.
The DRWI suggests a promising path towards formation of planetesimals/planetary embryos and azimuthally asymmetric dust structure from turbulent dust-trapping rings.
%(currently 234 words)
\end{abstract}


%%%%%%%%%%%%%%%%%%%%%%%%%%%%%%%%%%%%%%%%%%%%%%%%%%

%%%%%%%%%%%%%%%%% BODY OF PAPER %%%%%%%%%%%%%%%%%%

\section{Introduction}
%Tentative outline: instabilities and planetesimal formation

%RWI literature

%Dust bump (theory + observation)

%This work

It has recently been established that ring-like substructures are ubiquitous among extended protoplanetary discs (PPDs), as revealed by ALMA (\citealt{ALMA15}; \citealt{Andrews18}; for a review, see \citealt{Andrews20}). While the formation mechanisms of such ring-like substructures are debated \citep[see, e.g.,][for a review]{Bae22}, they are believed to reflect dust trapping in turbulent pressure bumps \citep[e.g.,][]{Dullemond18,Rosotti20}.
Such dust-trapping sites not only retain the dust by preventing or slowing down radial drift \citep[e.g.,][]{Pinilla2012}, but also allow dust density to build up, and it has been speculated to be preferred sites for planetesimal formation \citep[e.g.,][]{Pinilla2017,Dullemond18}.

%Ring-like substructures are ubiquitous among protoplanetary discs (PPDs): %since the discovery of multiple rings in the PPD HL Tau \citep{ALMA15}, 
%high spatial resolution images observed by the Atacama Large Millimeter/submillimeter Array (ALMA) have shown an increasing occurrence and variety of rings (e.g., \citealt{Andrews16}, \citealt{Isella16}, \citealt{Loomis17}; for an overview, see \citealt{Bae22}), which are essentially concentrated dust particles usually believed as a result of dust drifting towards pressure maxima. Such dust-trapping sites offer a natural solution to the persistent problem that planetesimal formation is likely outpaced by the fast radial drift and loss of dust particles. 
%Indeed, theories have emerged that consider, for example, the streaming instability \citep[SI;][]{Youdin05} in rings that trigger the growth of super-kilometer-sized solids \citep{Carrera21,Carrera22}.

Conventionally, planetesimal formation is believed to be triggered by the streaming instability \citep[SI;][]{Youdin05} between gas and marginally or weakly coupled dust
%The SI is characterised by rapid radial concentration of dust 
as a result of reciprocal dust-gas aerodynamic drag. 
%Given a pressure gradient with an adequate 
The source of free energy behind the SI arises from the background radial pressure gradient, which induces relative drift between gas and dust. Once the dust abundance (vertically-integrated dust-to-gas mass ratio)
%dust-gas surface density ratio 
exceeds a certain threshold (depending on dust size, typically
%usually 
$\gtrsim0.02$, \citealt{Bai10}; but see \citealt{Li21}), the SI is found in simulations to lead to efficient dust clumping, with clumps dense enough to form planetesimals directly by gravitational collapse \citep[e.g.,][]{Johansen09,Carrera15,Yang17}.
%and widely accepted in the literature. 
However, if turbulent dust-trapping ring-like substructures are common as found in observations, the 
%mechanism 
streaming instability paradigm for planetesimal formation in such dust rings
faces two challenges.
%, though. 
First, most existing simulations did not include external turbulence, but studies have found that modest turbulence of viscous parameter $\alpha\sim10^{-3}$ suffices to impede the development of the SI \citep{Chen20,Umurhan20} and SI-induced clumping \citep{Umurhan20}. 
%Also, in the context of disc rings, 
Second,
the SI does not operate %occur
at the pressure maxima where most of the dust is concentrated \citep[but see][]{Auffinger18,Hsu2022}, although SI-induced dust clumping remains efficient in low-pressure-gradient regions near pressure bumps \citep[without external turbulence]{Carrera21}.
%although the instability might set in before dust is fully trapped by a newly formed pressure bump \citep{Carrera21}.

More realistic models of dust rings should incorporate turbulence as well as a certain driving mechanism that leads to ring formation, and recent simulations 
%in realistic ring settings 
along this line suggested instabilities beyond the SI paradigm. For instance, \cite{Huang20} found ``meso-scale'' instability triggered by dust feedback in a pressure bump where the disc transitions between low and high viscosity mimicking the dead zone outer boundary, which leads to the the formation of dust clumps. Similar instability has also been found at planetary gap edges \citep{Surville20,Yang20}.
In the presence of turbulence due to the vertical shear instability (VSI), \cite{Lehmann2022} found
%dust density enhancement at an initially seeded pressure bump with frequent formation of vortices, which is thought to relate to the VSI-induced dust corrugation motion.
strongly enhanced dust-trapping into VSI-induced vortices when there is an initial pressure bump. Moreover,
\cite{Xu22a,Xu22b} conducted %simulated 
hybrid particle-gas non-ideal magnetohydrodynamic (MHD) simulations in outer PPDs and found efficient dust clumping
%and formation of fine-scale filaments 
in the presence of a ring-like pressure maximum, which was formed
%first 
by zonal flows 
%and later by 
or external forcing. The dust rings are also observed to split into finer-scale filaments.
%Facilitated 
With dust clumping found in environments unfavorable to the SI, these results imply %mechanisms yet to be identified. 
additional mechanisms could be responsible to trigger instabilities that potentially lead to dust clumping.
%The filaments in \cite{Xu22b} are also potentially linked to observed low-contrast asymmetric substructures in PPDs % The latter method had the advantage of self-consistency in realising a balance between radial trapping and turbulent diffusion of dust. They found

A closely related physical process in the context of the dust ring is the Rossby wave instability \citep[RWI;][]{Lovelace99}. Planet-induced radial pressure variations are found to give rise to the RWI \citep{deVal-Borro2007,Lyra2009,Lin14,Bae2016,Cimerman2023}, which requires the presence of a local vortensity (also known as potential vorticity) minimum 
%which essentially requires a deep local minimum of vortensity to operate
\citep{Li2000}. \cite{Ono16} conducted detailed linear parametric studies in a global 2D barotropic disc, showing the necessary and sufficient condition for the onset and a physical interpretation of the RWI. The linear behaviour of this instability in the presence of turbulence and dust, however, has not been rigorously explored.%is not fully understood yet.

% Many extended PPDs are also known to have asymmetric substructures\citep[e.g.,][]{vanderMarel13,Perez14,Hashimoto21}. Among the possible explanations is the hydrodynamic instabilities that create dust-trapping vortices. The Rossby wave instability \citep[RWI;][]{Lovelace99} is one such mechanism, which operates in the presence of of particular interest in the context of a pressure bump. The RWI triggers vortex formation given a local minimum in the radial profile of the vortensity, which a pressure maximum naturally provides for. While \cite{Ono16,Ono18} performed extensive two-dimensional linear stability analyses and simulations on the conditions of the RWI in a gaseous system, the role of this mechanism in the presence of dust-gas interaction deserves further study.

%Observations: ring-like substructures are ubiquitous. They are expected to be pressure bumps trapping dust, and it is speculated that such dust trapping rings further lead to planetesimal formation.

%SI (refer to Ziyan's intro): widely accepted in the literature, but it does not work in pressure bump (maxima), nor in the presence of strong turbulence.

%Simulations found "meso-scale" instability (Huang+2020), Ziyan's work find dust clumping in turbulent dust trapping rings, (with some description, e.g., filaments) which may imply there is some other mechanism that triggers dust clumping.

In this work, we analyse the stability of turbulent dust-trapping rings.
%Compared to \cite{Ono16}, 
Our analysis generalises the work of \cite{Ono16} in a local shearing-sheet setting,
and consider several additional physical ingredients for a self-consistent and realistic dusty ring model. Specifically, similar to \cite{Xu22b}, we simultaneously introduce external forcing and gas viscosity (to mimick turbulence), the balance of which sustains a pressure bump that models an axisymmetric ring.
%Hydrodynamic effects involving the dust are also included, 
We include an additional dust fluid, particularly incorporating
%in particular 
the two-way drag between dust and gas, and the new formulation of dust concentration diffusion that properly ensures momentum conservation and Galilean invariance. Although limited in a local shearing sheet, our analysis 
%offers a possible explanation to the new instability in \cite{Xu22b} and a generalisation of \cite{Ono16}.
represents a major first step towards a comprehensive understanding of dust-trapping rings.

%A closely related problem is the RWI.... Our analysis presnets several additional physical ingredients (forcing, viscosity, dust diffusion).

This work is organised as follows. In Section~\ref{sec theory}, we formulate and assemble the physical ingredients of our analysis. We obtain equilibrium solutions of the dust-trapping rings in Section~\ref{sec equilibrium}, which serves as the basis of the linear perturbation developed in Section~\ref{sec formulation of perturbation}. In Section~\ref{sec results}, we show the two types of Rossby wave-like instabilities emerging from our linear analysis, describing their phenomenology, parametric dependence and important physical ingredients. We name them the "dusty Rossby wave instability" (DRWI). %We provide a physical interpretation of the new instability in Section~\ref{sec mechanism}. 
The DRWI is then numerically tested in Section~\ref{sec simulation}, in which we also briefly explore its nonlinear evolution. We summarise our findings and discuss implications, caveats and future work in Section~\ref{sec conclusions}. %conclude in Section~\ref{sec conclusions}.

\section{Theory}
\label{sec theory}

\subsection{Formation of a Pressure Bump from Forcing}
\label{subsec pressure bump}
We take a shearing-sheet formulation, which is constructed by following fluid
motion around a reference radius $R_0$ from the central object, and writing down equations in the corotating frame with respect to this radius in Cartesian coordinates. By doing so, it ignores curvature and only applies to regions around $R_0\pm\Delta L$ with
$\Delta L\ll R$. This is applicable for thin discs whose pressure scale height $H\ll R$,
and has been widely used for local models of accretion discs. In this radially narrow region, we also ignore the disc background pressure gradient, assuming that the local bump forms a pressure maximum on which our local sheet is centered. %The widest, mildest bump in our parameter study later would overcome a power-law background pressure $\propto R^{-11/4}$ to form a pressure maximum as long as $H/R\leq0.04$. 
This assumption also implies no net dust radial flux through the bump region (an ``isolated'' bump), suitable if another dust trap resides outside the bump in question. We will return to these assumptions in Section~\ref{subsec discussion}.

We customarily choose the $x$ axis along the radial, $y$ axis along the azimuthal, and $z$ axis along the vertical directions. In particular, at the reference radius $R_0$, we set $x=0$, where the angular velocity is denoted by $\Omega_0$. 
For simplicity, we assume an isothermal equation of state with isothermal sound speed $c_s$. The pressure is given by $P=\rho_gc_s^2$, and the pressure scale height $H=c_s/\Omega_0$. The fluid equations including viscosity now read
\begin{gather}
\frac{\pa\rho_g}{\pa t}+\nabla\cdot(\rho_g{\mb v}_g)=0\ ,\\
\frac{\pa{\mb v}_g}{\pa t}+({\mb v}_g\cdot\nabla){\mb v}_g=-\frac{\nabla P}{\rho_g}
+[2{\mb v}_g\times{\mb\Omega_0}+3\Omega_0^2 x{\mb e}_x]+\nu\nabla^2{\mb v}_g+f_0(x){\mb e}_y\ ,
%P=\rho_g c_s^2\ ,
\end{gather}
where we have also included a forcing term $f_0(x)$, to be discussed later. We use the subscript $``_g"$ to denote gas quantities, to be distinguished later from dust and combined one-fluid quantities.
Note that the form of viscosity adopted here differs from the standard Naiver-Stokes
viscosity, which captures the essence of viscosity without complicating the analysis.
Here we consider a 2D system and ignore the vertical dimension (i.e., being vertically-integrated), and $\rho_g$ essentially represents a surface density.

We consider a unit system such that time is normalised to $\Omega_0^{-1}$ and velocity is normalised to $c_s$. Then, the natural units for length is $H$. We simply choose $\Omega_0=c_s=H=1$. The standard $\alpha-$prescription for viscosity takes the form $\nu=\alpha c_sH$, where $\alpha$ is taken to be a constant and for protoplanetary discs, it is expected that $\alpha\sim10^{-4}$ to $10^{-3}$ \citep[see, e.g.,][for a review]{Lesur2022}, but there is also evidence for stronger turbulence in some systems \citep[e.g.,][]{Flaherty2020}.

We further subtract background Keplerian shear from the velocity, or ${\mb v}={\mb v}'-(3/2)\Omega_0{\mb e}_y$ \citep[known as orbital advection, or the FARGO algorithm,][]{Masset2000,Stone2010}. The equations then become
%To simplify further, we write the velocity as ${\mb v}={\mb v}_0+{\mb v}'$, where
%${\mb v}_0=-(3/2)\Omega_0x$ is the background Keplerian velocity in the shearing sheet.
%By subtracting background shear (known as the FARGO algorithm, ref?), the equations become
\begin{gather}
\frac{\pa\rho_g}{\pa t}+\nabla\cdot(\rho_g{\mb v}'_g)-\frac{3}{2}\Omega_0x\frac{\pa\rho_g}{\pa y} = 0\ ,\label{eq gas den}\\
\frac{\pa{\mb v}'_g}{\pa t}+({\mb v}'_g\cdot\nabla){\mb v}'_g-\frac{3}{2}\Omega_0x\frac{\pa{\mb v}'_g}{\pa y}
= -\frac{\nabla P}{\rho_g}+\nu\nabla^2{\mb v}'_g + \nonumber \\
\quad[2\Omega_0 v'_{gy}{\mb e}_x-\frac{1}{2}\Omega_0 v'_{gx}{\mb e}_y]+f_0(x){\mb e}_y\ .\label{eq gas mom old}
\end{gather}
In equilibrium without forcing, i.e., $f_0(x)=0$ for all $x$, we simply have $\rho_g=$ const., ${\mb v}'_g=0$. 

Now consider adding forcing %\textbf{to form a pressure bump. Note that we ignore the presence of a background pressure gradient, assuming our local box is centered on the pressure maximum. The pressure bump is imposed by exerting}
by imposing 
a positive torque at $x<0$ and a negative
torque at $x>0$, which is achieved by applying a force $f_0(x)$ in the $\hat{y}$ direction, being an odd function about $x=0$. This would modify the equilibrium state to create a pressure bump in the center of the box.
In reality, it mimics the effect of zonal flows or the presence of a planet, both of which will drive a density bump in the disc. The new equilibrium state is the background state that we shall consider for linear stability analysis, and for this state we include a subscript $``_0"$. Clearly, there is no radial flow, thus $v'_{x0}=0$. The solution for $\rho_g$ and $v'_y$ is determined by the forcing profile according to
\begin{gather}
    c_s^2\frac{\pa}{\pa x}\ln\rho_{g0}=2\Omega_0v'_{0y}\ , \\
    f_0(x)+\nu\frac{\pa^2}{\pa x^2}v'_{0y}=0\ . \label{eq f0 v0y}
\end{gather}

It is straightforward to see that the relation between forcing and the resulting density
profile is given by
\begin{equation}
f_0(x)=-\frac{\nu c_s^2}{2\Omega_0}\frac{\pa^3}{\pa x^3}\ln\rho_{g0}\ . \label{eq forcing definition}
\end{equation}
Assuming pressure varies on scales of $H$, an order-of-magnitude estimate shows that
the forcing term is $f_0\sim\alpha c_s\Omega_0$, while the pressure gradient term is on the
order of $\nabla P/\rho_g\sim c_s\Omega_0$. Therefore, for $\alpha\ll1$ which is expected to
apply in protoplanetary discs, very modest forcing can drive substantial pressure variation.
In practice, we consider a Gaussian bump, given by
\bgeq
\rho_{g0}=\rho_b[1+A\exp(-x^2/2\Delta w^2)]\ , \label{eq pressure bump profile}
\edeq
where $\rho_b$ is the background density. From here one can determine the forcing profile.
However, after taking logarithm, evaluation of the third order derivative results in substantial complication.
Instead, we may consider the limit where $A$ is relatively small ($A\lesssim1$), and instead assert
\bgeq
\rho_{g0}=\rho_b\exp[A\exp(-x^2/2\Delta w^2)]\ ,
\edeq
and this will yield
\bgeq
f_0(x)=\frac{A}{2}\alpha c_s\Omega_0\bigg(\frac{H}{\Delta w}\bigg)^3
\bigg(\frac{x^3}{\Delta w^3}-\frac{3x}{\Delta w}\bigg)\exp(-x^2/2\Delta w^2)\ .
\edeq
With this setup, the only parameters are $A$ and $\Delta w$ for the bump profile, and $\alpha$ for viscosity.

\subsection{Dust Diffusion and Concentration}
\label{subsec dust diffusion and concentration}
Dust is considered as a pressureless fluid, subjecting to gas drag and turbulent diffusion.
The gas drag is characterised by a stopping time $t_s$, which depends on dust size. This
is usually non-dimentionalised by defining a Stokes number $St\equiv\Omega_0t_s$.
The equations of dust fluid motion read
\begin{gather}
\frac{\pa\rho_d}{\pa t}+\nabla\cdot(\rho_d{\mb v}'_d)-\frac{3}{2}\Omega_0x\frac{\pa\rho_d}{\pa y} = 0\ , \label{eq dust den}\\
%\nabla\cdot\bigg[\rho_gD\nabla\bigg(\frac{\rho_d}{\rho_g}\bigg)\bigg]\ ,\\
\frac{\pa{\mb v}'_d}{\pa t}+({\mb v}'_d\cdot\nabla){\mb v}'_d-\frac{3}{2}\Omega_0x\frac{\pa{\mb v}'_d}{\pa y} =
\frac{1}{\rho_d}\nabla\cdot(\rho_d{\bm v_{\rm dif}}{\bm v_{\rm dif}}) + \nonumber\\
\quad2\Omega_0 v'_{dy}{\mb e}_x-\frac{1}{2}\Omega_0 v'_{dx}{\mb e}_y  - \frac{{\mb v}'_d-{\mb v}'_g}{t_s}\ , \label{eq dust mom}
\end{gather}
the dust diffusion velocity $\bm v_{\rm dif}$ defined by
\begin{equation}
    \bm v_{\rm dif} = -\frac{\rho_g}{\rho_d}D\nabla\left(\frac{\rho_d}{\rho_g}\right) = \frac{D}{f_d}\nabla\ln f_g\ , \label{eq vdif}
\end{equation}
where $D$ denotes the dust diffusion coefficient, and
\begin{equation}
    f_g=1-f_d\equiv\rho_g/(\rho_g + \rho_d)
\end{equation} is the gas mass fraction.
Note that what is being diffused is dust concentration, rather than dust density, so that diffusion drives the dust to achieve constant dust-to-gas density ratio. 

%Different from usual treatments (e.g., what paper to cite?), we represent dust concentration diffusion in the momentum equation while keeping the dust density conserved. In essence, our dust velocity term includes $\bm v_{\rm dif}$ in itself. The implied physical change is that now the velocity difference ${\mb v}'_d-{\mb v}'_g$, which determines the dust-gas interaction, is assumed to involve the dust concentration diffusion velocity.
Different from usual treatments, we represent dust concentration diffusion in the momentum equation while keeping the dust density conserved. This formulation is motivated by \cite{Tominaga2019,Huang22}, who pointed out the inconsistencies in the conventional treatment of adding the concentration diffusion term in the continuity equation that violates momentum conservation and Galilean invariance. Our treatment closely follows that of \cite{Huang22}, exemplified in their Equation (A1), but with one difference in that our dust velocity term includes $\bm v_{\rm dif}$ in itself, i.e., our ${\mb v}_d$ now represents their ${\mb v}_d+{\mb v}_{\rm dif}$ on the left hand side. Note that the drag term is proportional to ${\mb v}'_d-{\mb v}'_g$ which involves the dust concentration diffusion velocity.

The concentration diffusion coefficient $D$ is generally closely related to the turbulent gas kinematic viscosity (on the same order), at least for tightly coupled particles. While our formulation leaves flexibility for the specific expression of $D$, for the rest of the paper, we assume that the coefficient is simply proportional to the gas mass fraction (mimicking the reduction of turbulence strength in the presence of strong dust mass loading, cf., \citealt{Xu22b}), or 
\begin{equation}
    D(f_g) = \nu f_g = \alpha c_sHf_g\ . \label{eq D def}
\end{equation}
We also experimented an alternative expression of $D=\nu (\rm const.)$, which yields qualitatively the same results for the equilibrium solutions and linear perturbation behaviours.

As the aerodynamic drag affects the dust, the gas must feel the backreaction (i.e., feedback) from the dust as well, and the momentum equation of the gas is modified to
\begin{gather}
\frac{\pa{\mb v}'_g}{\pa t}+({\mb v}'_g\cdot\nabla){\mb v}'_g-\frac{3}{2}\Omega_0x\frac{\pa{\mb v}'_g}{\pa y}
=-\frac{\nabla P}{\rho_g}+\nu\nabla^2{\mb v}'_g + \nonumber \\
\quad 2\Omega_0 v'_{gy}{\mb e}_x-\frac{1}{2}\Omega_0 v'_{gx}{\mb e}_y + \frac{\rho_d}{\rho_g}\frac{{\mb v}'_d-{\mb v}'_g}{t_s}+f_0(x){\mb e}_y\ . \label{eq gas mom}
\end{gather}

The force difference between gas and dust mainly results from the fact that gas is subject to its own pressure whereas dust is not. It can be shown that when the dust is strongly coupled to gas, meaning $St\ll1$, the dust reaches a terminal velocity given by \citep{Jacquet2011,Laibe2014}
\begin{equation}
{\mb v}_d={\mb v}_g + \bm v_{\rm dif} + t_s\frac{\nabla P}{\rho_g}\ , \label{eq vd vg vdif}
\end{equation}
which describes that dust always drifts towards pressure maxima. Note that the original derivation ignores dust diffusion and external forcing. We supplement the right-hand side with $\bm v_{\rm dif}$ in response to our implicit incorporation of this term in $\bm v_d$, %Work by Lovascio \& Paadekooper (2019) showed that the difference between using ${\mb v}$ vs ${\mb v}_g$ in the viscous term is on the order of St$^2$ and can be ignored for tightly coupled particles. 
while our earlier analysis shows that the forcing term should be negligible compared with the pressure gradient term; see argument following Equation~(\ref{eq forcing definition}). Therefore, this expression is valid for our applications.

In the presence of a pressure maxima in the gas, dust would drift indefinitely into the pressure bump, leading to infinite concentration. However, this is prevented by turbulent diffusion. If feedback is ignored, then gas and dust dynamics are decoupled, and the dust distribution simply achieves an equilibrium profile whose width is set by a balance between concentration and diffusion. When considering feedback, however, the situation is much more involved.
%and an analytical solution is unlikely.
%\xb{which we will seek for numerical solutions in Section \ref{sec equilibrium}.}
As a first investigation, we reduce the mathematical complexity by considering a single-fluid formalism below.

\subsection{Single-fluid Formalism}

%The equations above combined give a two-fluid formalism of the problem. However, 
When assuming dust particles are strongly coupled with $St\ll1$, the problem can be cast into a one-fluid framework \citep{Laibe2014,Lin2017}, where the single-fluid density and velocity are defined as
\bgeq
\rho=\rho_g+\rho_d\ ,\quad {\mb v}=\frac{\rho_g{\mb v}_g+\rho_d{\mb v}_d}{\rho_g+\rho_d}\ .
\edeq
Since dust is pressureless, total pressure is still gas pressure
\begin{equation}
P=\rho_gc_s^2=\rho c_s^2f_g\ . \label{eq one fluid pre def}
\end{equation}
This equation relates the gas fraction $f_g$ to the equation of state.

%The single-fluid formalism describes the dust-gas mixture, with mutual drag reflected as a source term in the energy equation that behaves as nonlinear thermal conduction. The physics behind this description is the following. Higher dust concentration increases density without changing pressure, thus resulting in lower (effective) temperature. As dust drifts towards higher pressure, and hence attempts to enhance concentration there, it induces an effective conductive flux towards lowering system temperature at regions with higher pressure.

We derive the equations of the single-fluid system as follows. Firstly, the addition of Equation~(\ref{eq gas den}) and (\ref{eq dust den}) gives the continuity equation:
\begin{equation}
    \frac{\pa \rho}{\pa t} + \nabla\cdot(\rho\bm v') - \frac{3}{2}\Omega_0x\frac{\pa \rho}{\pa y} = 0\ . \label{eq one fluid den}
\end{equation}
As for the momentum equation, we simplify the derivation by assuming ${\bm v}'_g\sim{\bm v}'_d$. We 
multiply both sides of Equations~(\ref{eq gas mom})(\ref{eq dust mom}) by $\rho_g$ and $\rho_d$ respectively and then directly add them up, finally arriving at
\begin{gather}
\frac{\pa \bm v'}{\pa t} + (\bm v'\cdot\nabla)\bm v' - \frac{3}{2}\Omega_0x\frac{\pa \bm v'}{\pa y} = -\frac{1}{\rho}\nabla P + \nu f_g\nabla^2\bm v' +  \nonumber\\ 
\quad\frac{1}{\rho}\nabla\cdot(\rho_d\bm v_{\rm dif}\bm v_{\rm dif}) + 2\Omega_0v'_y\bm e_x - \frac{1}{2}\Omega_0v'_x\bm e_y + f_g f_0(x) {\bm e}_y \ . \label{eq one fluid mom}
\end{gather}
Despite the simplification, we maintain Equation~(\ref{eq vd vg vdif}) to account for the dust-gas drag.

%The left-hand side of the equation has been simplified to the familiar material derivative.
We further represent the equation of state in the form of the pressure diffusion equation, derived from Equations~(\ref{eq gas den})(\ref{eq vdif})(\ref{eq vd vg vdif})(\ref{eq one fluid pre def}):
\begin{equation}
    \frac{\pa P}{\pa t} + \nabla\cdot(P\bm v') - \frac{3}{2}\Omega_0x\frac{\pa P}{\pa y} = c_s^2t_s\nabla\cdot(f_d\nabla P) + \nabla\cdot(DP\nabla\ln f_g) \ . \label{eq one fluid pre}
\end{equation}
Here, dust drift behaves as nonlinear thermal conduction (first term on the right hand side), as pointed out in \cite{Lin2017}. The additional dust concentration diffusion flux gives rise to the last (nonlinear) term.

%Adopting this formalism here is not without caveats. In particular, dust concentration diffusion also leads to additional dust momentum flux, which is not accounted for (see Tominaga et al. 2019), though these terms are commonly ignored in multi-fluid dust formulations.

\section{Equilibrium states of the single-fluid system}
\label{sec equilibrium}

In this section, we numerically solve the single-fluid equations above for an equilibrium state of the system, on top of which we will further conduct linear perturbation analysis.
%derive the equations for the system in equilibrium and solve the states numerically. 
%The solutions are preparations for the perturbation analysis where we can exactly subtract the equilibrium state. 
From now on, we add a subscript $``_0"$ to quantities in the steady state. We have introduced this subscript in Section~\ref{subsec pressure bump}, and the notations in two places are consistent with each other.

\begin{figure*}
    \centering
    \includegraphics[width=\textwidth]{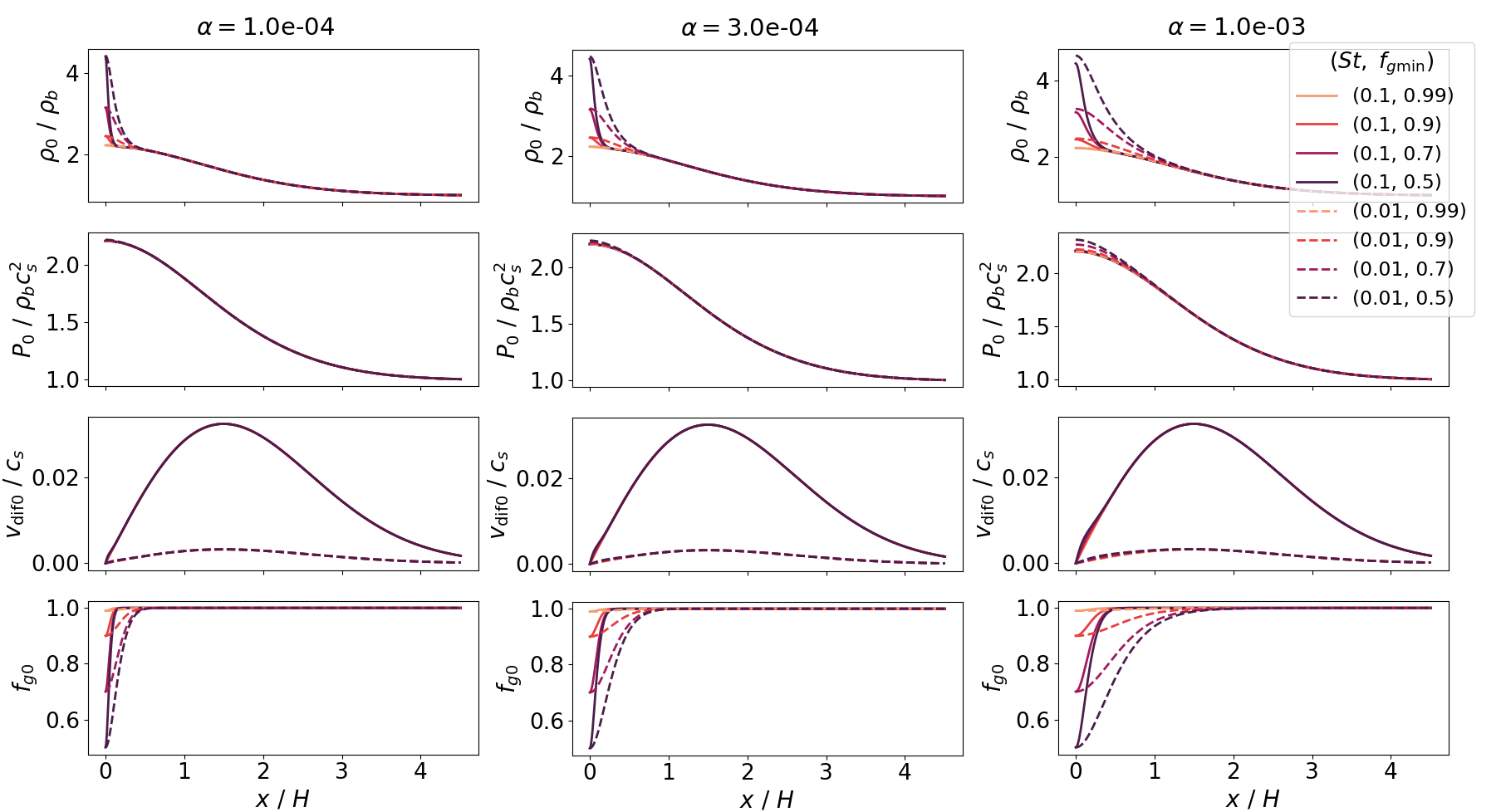}
    \caption{Equilibrium solutions for different combination of parameters. Only the $x>0$ region is plotted; $\rho_0$, $P_0$ and $f_{g0}$ are even functions in $x$ whereas $v_{\rm dif0}$ is odd. Each row presents $\rho_0$, $P_0$, $v_{\rm dif0}$ and $f_{g0}$ respectively. For all plots here, $A=0.8$ and $\Delta w/H=1.5$. Three columns correspond to different $\alpha$, as noted on the top. Colors represent different $f_{g{\rm min}}$ while line styles represent different $St$ values.}
    \label{fig steady profiles}
\end{figure*}

\subsection{Derivation of steady-state equations}

In equilibrium, we expect axisymmetry and therefore no dependence on $y$. The radial velocity of the single fluid is also zero: the gas stays at rest, while the bulk velocity of the dust, which drifts towards pressure maxima, is balanced by outward diffusion, giving an overall effect of $\bm v'_{d0}=0$. 
The absence of any bulk radial motion reflects the equilibrium state with an isolated dust trap, which is unlike the conventional scenario with dust/gas drifting inward/outward in the presence of a background pressure gradient.
%The absence of any bulk radial motion comes from our neglect of dust supply from the outer boundary, although a vanishing background gradient and hence symmetry about $x=0$ provides a stronger condition.
Therefore, after separating all vectors into $x$ and $y$ components, we drop the partial derivatives of $t$ and $y$ as well as terms involving $\bm{v}'_0$ from Equations~(\ref{eq one fluid den})(\ref{eq one fluid mom})(\ref{eq one fluid pre}) to obtain
%\begin{gather}
%    \frac{\pa (\rho_0 v'_{0x})}{\pa x} = 0\ , \label{eq continuity}\\
%    \frac{\pa (P_0v'_{0x})}{\pa x} = c_s^2t_s\frac{\pa}{\pa x}\left(f_{d0}\frac{\pa P_0}{\pa x}\right) + \frac{\pa}{\pa x}\left(D_0P_0\frac{\pa \ln f_{g0}}{\pa x}\right)\ , \\
%    v'_{0x}\frac{\pa v'_{0x}}{\pa x} = -\frac{1}{\rho_0}\frac{\pa P_0}{\pa x} + 2\Omega_0v'_{0y} + \frac{1}{\rho_0}\frac{\pa}{\pa x}\left[D_0^2\frac{\rho_0^2}{\rho_{d0}}\left(\frac{\pa \ln f_{g0}}{\pa x}\right)^2\right]\ , \label{eq x momentum}\\
%    v'_{0x}\frac{\pa v'_{0y}}{\pa x} = -\frac{1}{2}\Omega_0v'_{0x} + \nu f_{g0}\frac{\pa^2v'_{0y}}{\pa x^2} + f_{g0}f_0(x) \ . \label{eq y momentum}
%\end{gather}

%The radial velocity of the single fluid is also zero: the gas stays at rest, while the bulk velocity of the dust, which drifts towards pressure maxima, is balanced by outward diffusion, giving an overall effect of $\bm v'_{d0}=0$. Therefore, in steady state, one has
\begin{gather}
    v'_{0x} = 0 \ , \label{eq steady state vx}\\
    c_s^2t_sf_{d0}\frac{\pa P_0}{\pa x} + D_0P_0\frac{\pa \ln f_{g0}}{\pa x} = 0 \ , \label{eq steady state 1}\\
    -\frac{1}{\rho_0}\frac{\pa P_0}{\pa x} + 2\Omega_0v'_{0y} + \frac{1}{\rho_0}\frac{\pa}{\pa x}\left[D_0^2\frac{\rho_0^2}{\rho_{d0}}\left(\frac{\pa \ln f_{g0}}{\pa x}\right)^2\right] = 0 \ , \label{eq steady state 2}\\
    \nu f_{g0}\frac{\pa^2v'_{0y}}{\pa x^2} + f_{g0}f_0(x) = 0 \quad\Rightarrow\quad v'_{0y} = -\frac{c_s^2A}{2\Omega_0(\Delta w)^2}x{\rm e}^{-x^2/2(\Delta w)^2} \ , \label{eq steady state vy}
\end{gather}
where we use the fact that $P_0(x)$ and $f_{g0}(x)$ have no spatial gradient far from the pressure bump in obtaining Equation~(\ref{eq steady state 1}), which is derived from Equation~(\ref{eq one fluid pre}). Note that if there were no forcing, the equilibrium solution would become trivial, where all velocities would vanish and the
dust-to-gas ratio would become uniform. 

\subsection{Numerical solution}

\label{subsec equilibrium numerical solution}
We solve the equilibrium equations as an initial value problem by specifying conditions at $x=0$.
%evolving the system as $x$ increases. 
Due to the symmetry of our setting, $\rho_0(x)$ and $P_0(x)$ are even functions with respect to $x=0$ and thus 
%we directly obtain the solutions for $x<0$. 
we only solve the equations for $x>0$.
Our methods are described in detail in Appendix~\ref{append equilibrium}. We obtain equilibrium solutions with five dimensionless % non-dimensionalised
physical parameters, of which the information, fiducial values and the ranges explored in this work are summarised in Table~\ref{tab fiducial}. 

Among the parameters, $St=0.1$, corresponding to mm- to cm-sized dust for typical disc models in the outer disc,
%corresponding to cm-sized dust particles in the outer disc, 
would be an upper bound for our single-fluid formalism to remain approximately valid, which assumes strongly coupled dust and gas. The minimum gas fraction $f_{g{\rm min}}$ is the gas mass fraction at the center of the bump.
%always lies at the pressure maximum. 
We choose this parameter rather than a global dust-to-gas ratio by convention because $f_{g{\rm min}}$ is numerically easier to control. %and also likely to be better constrained in observation. 
Our lower bound of $f_{g{\rm min}}=0.5$ is chosen to correspond to the extreme situation with a $1:1$ gas-to-dust mass ratio in the bump center, which may be the case in some systems such as
%comparable to the extreme estimated dust-to-gas ratio of 1.7 
in HD 142527 (being 1.7, \citealp{Boehler17}). 

We show in Figure~\ref{fig steady profiles} the equilibrium solution in terms of the radial profiles of $P_0(x)$, $f_{g0}(x)$ and $v_{\rm dif0}(x)$ for a bump with $A=0.8$ and $\Delta w/H=1.5$. In all combination of parameters below, the density and pressure form a bump close to $x=0$ and quickly approaches the background value as $x$ exceeds $\Delta w$. The density excess close to $x=0$ and the minimum in the gas fraction profile indicate significant increase of dust concentration at the pressure maximum. We call this region a "dust bump" as opposed to %against
the wider gas bump. As can be seen from either $\rho_0$ or $f_{g0}$, the width of the equilibrium dust bump depends heavily on $St$ and $\alpha$: the dust bump is considerably narrower than the gas bump if the gas and dust are not well-coupled and/or if the concentration diffusion that balances the dust drift is weak. Tightly coupled systems give a slightly higher maximal pressure, which reflects the dust feedback to the gas.

It is of interest to translate $f_{g{\rm min}}$ to an averaged gas fraction through a ring. We define the mean gas and dust mass fraction $\overline{f_g}$ and $\overline{f_d}$ as
\begin{equation}
    \overline{f_g} = 1-\overline{f_d} = \frac{\int_{-x_B}^{x_B}P(x)dx}{c_s^2\int_{-x_B}^{x_B}\rho(x)dx}\ , \label{eq average gas fraction}
\end{equation}
where $x_B$ specifies the radial range of interest. A minimum gas fraction of 0.7 in our fiducial setting corresponds to $\overline{f_g}=0.980$ integrated from $x/\Delta w=-4$ to $+4$. In other words, if a bump quasi-statically evolved from a uniform mixture of gas and dust with $\overline{f_g}=0.98$ ($\overline{f_d}=0.02$), which is a reasonable condition, and if the bump could attract all the dust in a range of $\pm4\Delta w=\pm6H$, the equilibrium $f_{g{\rm min}}$ would be equal to our fiducial value. For $f_{g\rm min}=0.7$, a combination of large dust particles and low viscosity ($St=0.1, \alpha=3\times10^{-5}$) gives $\overline{f_g}=0.996$, while $St=0.003, \alpha=1\times10^{-3}$ gives a rather low $\overline{f_g}=0.89$ (we find no ``interesting'' instability anyway with this configuration or a more realistic $\overline{f_g}$). More values of $\overline{f_g}$ are annotated later in Figures~\ref{fig max gamma grid 0.7}, \ref{fig max gamma grid 0.5} with notes in Section~\ref{subsec parameter space}.
%This is twice the concentration of dust in the typical interstellar medium, a condition reasonable in dusty rings \citep{Jin16,Isella16}.

% Keeping $\partial_xf_g|_{x=0}=0$ constant, we vary $f_g|_{x=0}$ from zero to one and solve the IVP until $f_g(x)$ converges at $f_{gb}$.
\begin{table}
    \centering
    \begin{tabular}{cccc}
        \hline
        Parameter & Symbol & Fiducial Val. & Range \\\hline
        Gas bump magnitude & $A$ & 1.2 & 0.4--1.8\\
        Gas bump width & $\Delta w/H$ & 1.5 & 1.0--2.0\\
        Stokes number & $St$ & 0.03 & 0.003--0.1\\
        Viscous parameter & $\alpha$ & $3\times10^{-4}$ & $3\times10^{-5}$--$1\times10^{-3}$\\
        Minimum gas fraction & $f_{g{\rm min}}$ & 0.7 & 0.5--0.99\\\hline
    \end{tabular}
    \caption{Parameters of the dust-trapping ring.}%steady-state ODE.}
    \label{tab fiducial}
\end{table}

\section{Formulation of the perturbation equations}
\label{sec formulation of perturbation}
\subsection{Linearised system of equations}
Based on the equilibrium results, now we proceed to obtain the perturbation equations to investigate potential instabilities. Using the subscript ``$_1$" to denote perturbation variables, we consider a plane wave perturbation of the form
\begin{gather}
    \rho_1(x,y,t) = {\rm Re}[\rho_1(x){\rm e}^{i(ky-\omega t)}] \nonumber \ ,\\
    P_1(x,y,t) = {\rm Re}[P_1(x){\rm e}^{i(ky-\omega t)}] \nonumber \ ,\\
    \bm v_1(x,y,t) = {\rm Re}[\bm v_1(x){\rm e}^{i(ky-\omega t)}]\ ,
    \label{eq perturb var def}
\end{gather}
where $k$ (being a real number) is the $y$-direction wavenumber, $\omega$ is the complex frequency, $i^2=-1$, and Re[$\cdot$] takes the real part. The perturbation variables $\rho_1(x)$, $P_1(x)$ and $\bm{v}_1(x)$  throughout this paper represent the complex 1D functions on the right hand side in Equation~(\ref{eq perturb var def}) unless otherwise stated to denote the real 2D waveform. % whose real parts represent corresponding physical properties. 
Note that we do not impose (anti-)symmetry here but solve the perturbation equations over the full domain of $x$. %Both $k$ and $\omega$ are complex, their real parts representing 
The real part of $\omega$ represents oscillation and the imaginary part implies temporal growth or damping in the perturbation magnitude. We write $\omega=\omega_r+i\gamma$, where $\omega_r$ and $\gamma$ are real. An unstable perturbation with its magnitude growing with time has $\gamma>0$.

We introduce the following notation of perturbation ratios $\mathfrak{p}_1(x)\equiv P_1(x)/P_0(x)$ and $\mathfrak{f}_{g1}(x)\equiv f_{g1}(x)/f_{g0}(x)$. We substitute the perturbations into Equations~(\ref{eq vdif})(\ref{eq one fluid den})(\ref{eq one fluid mom})(\ref{eq one fluid pre}) to obtain the linearised system of equations. The derivation and detailed form of the system are lengthy and involve considerable algebra, which we outline in Appendix~\ref{append perturbation}. We only show the compact form here, expressed as a matrix of linear operators acting on the perturbation variables:

\begin{equation}
    \begin{bmatrix}
        \mathcal M_{00} & \mathcal M_{01} & \mathcal M_{02} & \mathcal M_{03}\\
        \mathcal M_{10} & \mathcal M_{11} & \mathcal M_{12} & \mathcal M_{13}\\
        \mathcal M_{20} & \mathcal M_{21} & \mathcal M_{22} & \mathcal M_{23}\\
        \mathcal M_{30} & \mathcal M_{31} & \mathcal M_{32} & \mathcal M_{33}
    \end{bmatrix}
    \begin{bmatrix}
        \mathfrak{p}_1(x)\\
        \mathfrak{f}_{g1}(x)\\
        v'_{1x}(x)\\
        v'_{1y}(x)
    \end{bmatrix} = 0\ . \label{eq perturb matrix}
\end{equation}

This is a system of four second-order linear ordinary differential equations in four functional variables, $\mathfrak{p}_1(x)$, $\mathfrak{f}_{g1}(x)$, $v'_{1x}(x)$, and $v'_{1y}(x)$. The matrix $\mathcal{M}(x,\omega,k)$ consists of block coefficients $\mathcal{M}_{ij} (i,j=0,1,2,3)$, which are differential operators of order at most two and may be functions of $x$, the already known equilibrium variables, and the yet-undetermined perturbation parameters $\omega$ and $k$. For a given $k$, we view the system of equations as an eigenproblem and solve for the eigenvalue $\omega=\omega_m$ with the corresponding eigenfunction. 
%The physical interpretation is a stationary vibration mode with an $x$-direction profile specified by the eigenfunction, a $y$-direction sine wave of spatial frequency $k$, an oscillatory angular frequency $\omega_r$, and a transient evolution rate $\gamma$ which can be either growth or decay depending on the sign of $\gamma$. 
The system allows for numerous modes,
%but we are only interested in the few unstable ones, 
but only a handful of modes are unstable, which we will focus on.
%The system allows for more than one mode, but we are primarily interested in the most unstable one. Therefore, for a given system \xb{setting} (i.e., with parameters listed in Table~\ref{tab fiducial} and $k$ fixed), we only solve for the eigenvalue with the maximum $\gamma$. 
% Note that, unless otherwise specified, the term ``eigenvalue'' of the perturbation problem consistently refers to $\omega_m$ in this paper.  %\textbf{It is numerically close but not equal to an eigenvalue of the finite difference matrix $\mathcal{M}(x,0,k)$, the difference due to the nonlinear boundary conditions specified below.} %\textbf{Due to the nonlinear boundary conditions specified below, $\omega_m$} should not be confused with and is not equal to an eigenvalue of the finite difference matrix.
%We have put $\omega$ on the diagonal.

\subsection{Boundary conditions}
\label{subsec boundary conditions}
Boundary conditions are required for a complete eigenproblem. As can be observed from Figure~\ref{fig steady profiles}, $\rho_0, P_0$ and $f_{g0}$ quickly approach background values as $|x|$ increases, while $v_{\rm dif0}$ shows a slower decay. Therefore, we set $\rho_0 = \rho_b, P_0 = P_b = c_s^2\rho_b, f_{g0}=1$, and $v'_{0y}=0$ at the boundary, while still using nonzero $v_{{\rm dif}0x}(x)$ from the equilibrium solution. Since dust is depleted here, $\mathfrak{f}_{g1}=0$. The perturbation equations at the boundary $x=\pm x_B$ can therefore be reduced to three equations in three variables $\mathfrak{p}_1$, $v'_{1x}$, and $v'_{1y}$ (the second perturbation equation becomes trivial).

Now, %assuming that physical quantities in equilibrium vary in a much larger spatial scale compared to the radial wavelength, 
we apply the WKBJ approximation, i.e., to take $\mathfrak{p}_1, v'_{1x},$ and $v'_{1y}$ as a plane wave proportional to $\exp{(ik_xx)}$, %respectively, 
where $k_x$, the asymptotic radial wavenumber shared by the three perturbation variables, is yet to be determined. This form is motivated by the fact that physical quantities in equilibrium vary slowly with $x$ near the boundaries; similar methods have been used by \cite{Li2000,Ono16}. We stress that $k_x$ is only used to specify the asymptotic relation at the boundaries, i.e., we do not assume that the perturbation variables constitute a plane wave everywhere. The boundary perturbation equations are therefore reduced to a linear system, whose coefficient matrix must have a vanishing determinant for a nontrivial solution. The boundary perturbation equations before and after the WKBJ approximation, as well as the form of the determinant, can be found in Appendix~\ref{append boundary}.

The zero determinant condition yields a dispersion relation $k_x=k_x(\omega,k,x)$, which is a polynomial equation of fourth degree in $k_x$. %We stress that $k_x$ is only used to specify the asymptotic relation at the boundaries.
%This relation, or rather the concept of $k_x$, is used only at the boundary. 
Two of the four complex solutions unphysically go to infinity both in real and imaginary parts as $\nu\to0$. %, indicating high-frequency, highly transient viscous waves that are not the focus here. 
Between the remaining two, one and only one has a positive real part if $k$ is not too close to zero. %in regions in the parameter space where we expect unstable modes (needs further explanation). 
We obtain this root $k_x$ numerically at the outer and inner boundaries respectively and adopt it as the outgoing boundary condition: $d\mathfrak{p}_1/dx=ik_x\mathfrak{p}_1, d\mathfrak{f}_{g1}/dx=ik_x\mathfrak{f}_{g1}, dv'_{1x}/dx=ik_xv'_{1x},$ and $dv'_{1y}/dx=ik_xv'_{1y}$.

\subsection{Numerical treatment}
We solve the eigenproblem numerically by discretising the differential equations in $x$ and representing all coefficients with matrix elements, similar to \cite{Ono16}. We perform calculations over a range of $-4\Delta w<x<4\Delta w$, which sufficiently covers the pressure bump region, over a uniform grid of $N=1001$ nodes. Doubling the node number would give eigenvalues that agree with our fiducial resolution within three to four digits. To construct the matrix, we use \texttt{findiff} \citep{findiff}, a Python package for finite difference numerical derivatives and partial differential equations in any number of dimensions. We then solve $\omega$ with a positive imaginary part such that the determinant of the matrix goes to zero. Once the desired eigenvalue $\omega_m=\omega_{rm}+i\gamma_m$ is found, we substitute it for $\omega$ in the matrix and calculate the eigenfunction, which we denote as a vector function in $x$ with parameter $\omega=\omega_m$, namely, $\Vec{u}_{1m}(x,\omega_m)=(\mathfrak{p}_{1m}(x), \mathfrak{f}_{g1m}(x), v'_{1xm}(x), v'_{1ym}(x))^\top|_{\omega=\omega_m}$. We use the subscript ``$_m$'' to denote eigenmodal quantities. We normalise the eigenfunction in magnitude and phase such that $\Vec{u}_{1m}(x,\omega_m)$ has length unity and $\mathfrak{p}_{1m}(0)|_{\omega=\omega_m}$ is real. The original perturbation variables $P_{1m}(x)$ and $f_{g1m}(x)$ are then recovered from $\mathfrak{p}_{1m}(x)$ and $\mathfrak{f}_{g1m}(x)$. Finally, the physically meaningful waveform in 2D, as later displayed in Figures~\ref{fig Type1 eigenfunction 0.99}--\ref{fig Type2 eigenfunction 0.70}, is obtained from Equation~(\ref{eq perturb var def}), where we arbitrarily take $t=0$. The phases of these 2D waveforms depend on both $x$ and $y$ as the 1D perturbation variables are complex. We describe details of matrix construction and determination of $\omega_m$ and $\Vec{u}_{1m}(x,\omega_m)$ in Appendix~\ref{append perturb}.

\section{Results of the linear analysis}
\label{sec results}

\begin{figure}
    \centering
    \includegraphics[width=0.5\textwidth]{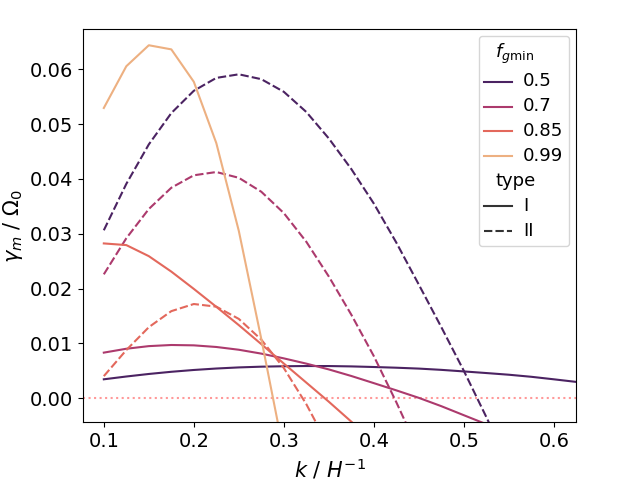}
    \caption{Dispersion relation of a sharp bump, showing $\gamma_m$ as a function of $k$, for different dust content measured in $f_{g\rm min}$. Two types of DRWI are plotted using solid and dashed lines respectively. The Type \RNum{2} curve for $f_{g{\rm min}}=0.99$ is below the lower limit of this figure. Parameters are set as $A=1.2$, $\Delta w/H=1.5$, $St=0.03$, $\alpha=3\times10^{-4}$. The value of $\overline{f_d}$ corresponding to each level of $f_{g\rm min}$ is measured as $0.038,0.020,0.009,$ and $6\times10^{-4}$ respectively.}
    \label{fig dispersion sharp bump}
\end{figure}

We identify solutions of the eigenproblem by a broad search on the $\gamma$--$\omega_r$ plane (Appendix~\ref{append perturb}). Only two modes are unstable among the numerous solutions. We term them the Type~\RNum{1} and Type~\RNum{2} DRWI.
%We discover two types of the DRWI in the dusty bump that have distinct characteristics. We refer to them as Type \RNum{1} and \RNum{2}. 
For reasons to be discussed below, we believe that Type \RNum{1} is a direct generalisation from the classical RWI, % \textbf{in a vortensity minimum} %in a pressure gradient of pure gas, 
whereas Type \RNum{2} is first identified in this work whose origin is closely related to the presence of dust. %and is closely associated with the dust content.

The two types are distinguished by the value of $\omega_{rm}$. Type \RNum{1} features $\omega_{rm}=0$, i.e., the mode has no phase velocity at $x=0$ in the rotating frame and therefore is stationary at the pressure maximum. Its eigenfunctions are symmetric or antisymmetric about $x=0$. In contrast, Type \RNum{2} has nonzero $\omega_{rm}$, indicating a co-rotation radius off the peak, and the perturbation profiles do not have the (anti-)symmetry as Type \RNum{1} does. 

\subsection{Two types of the DRWI: dispersion relation and eigenmodes %eigenvalues and eigenfunctions
}
\label{subsec sharp bump}

\begin{figure*}
    \centering
    \includegraphics[width=0.81\textwidth]{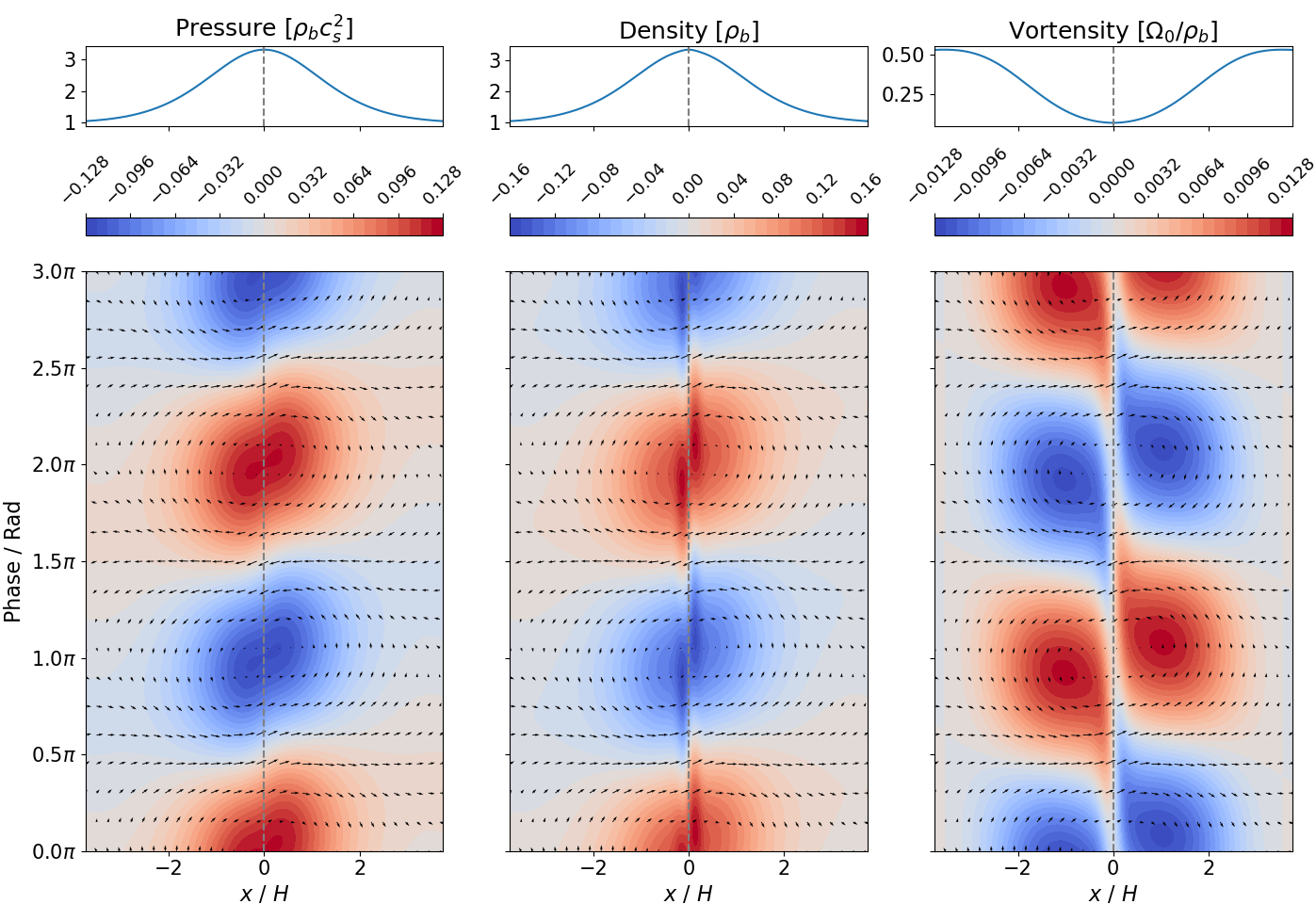}
    \caption{Eigenfunctions of Type \RNum{1} DRWI for a system with low dust content ($f_{g{\rm min}}=0.99$). The top panels present the $y$-independent equilibrium profiles $P_0$, $\rho_0$, and $q_0$, and the bottom panels show the perturbation functions $P_1$, $\rho_1$, and $q_1$ respectively. The eigenfunctions are drawn for one and half a wavelength in the $y$-axis, indicated by the phase. Arrows in the bottom panels denotes the perturbed velocity field $\bm v'_1$, with the arrow length proportional to the perturbed speed magnitude. In all panels, a grey dashed line marks the $x$-location of the co-rotation radius. Here $k=0.2 H^{-1}$ and $\omega_m=(0+0.0577i)\Omega_0$ and other parameters are the same as Figure~\ref{fig dispersion sharp bump}.}
    \label{fig Type1 eigenfunction 0.99}
\end{figure*}
\begin{figure*}
    \centering
    \includegraphics[width=0.81\textwidth]{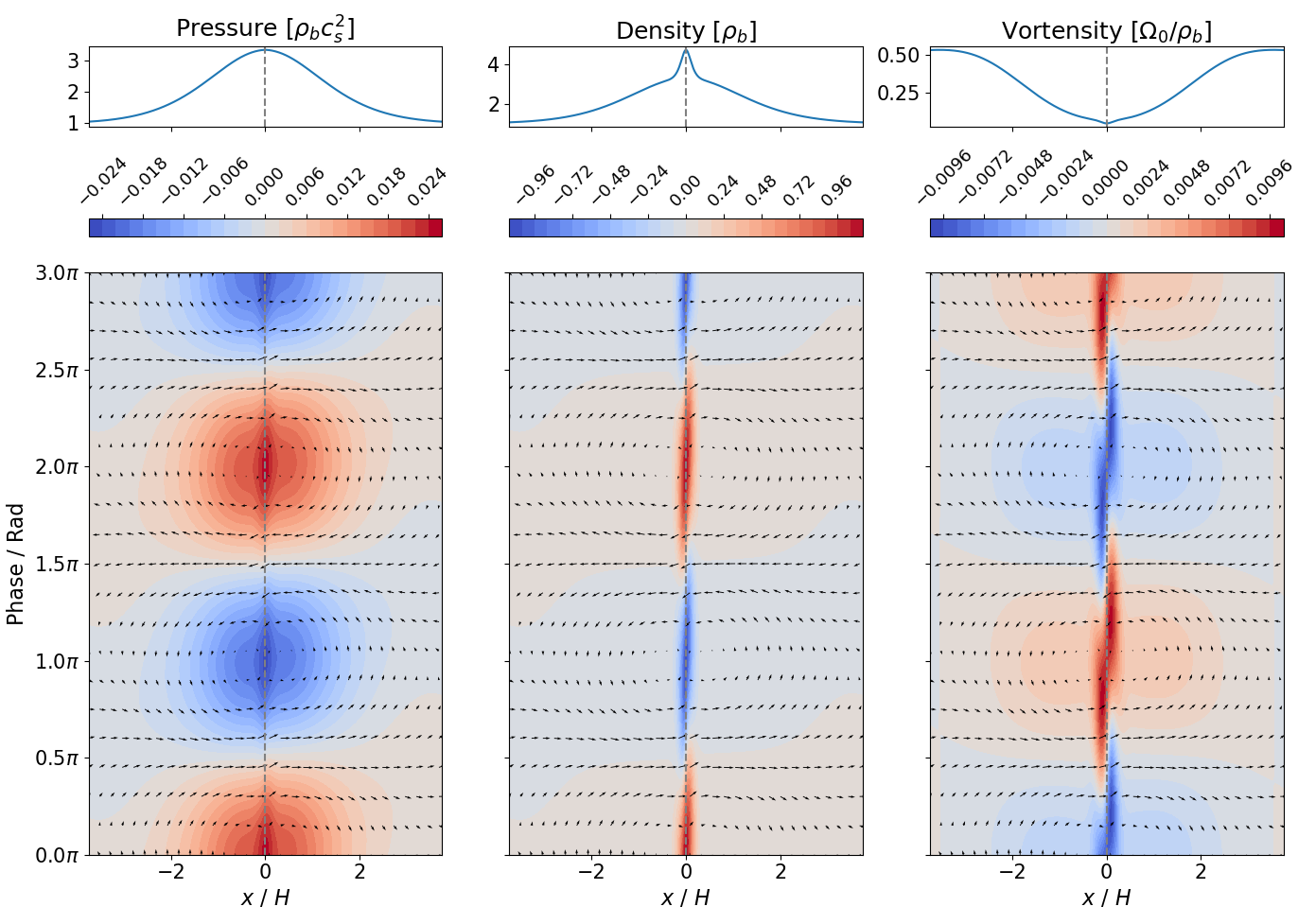}
    \caption{Eigenfunction of Type \RNum{1} DRWI for a system with moderate dust content ($f_{g{\rm min}}=0.7$). Here $k=0.2H^{-1}$ and $\omega_m=(0+0.0096i)\Omega_0$. Other details are the same as Figure~\ref{fig Type1 eigenfunction 0.99}.}
    \label{fig Type1 eigenfunction 0.70}
\end{figure*}

\begin{figure*}
    \centering
    \includegraphics[width=0.81\textwidth]{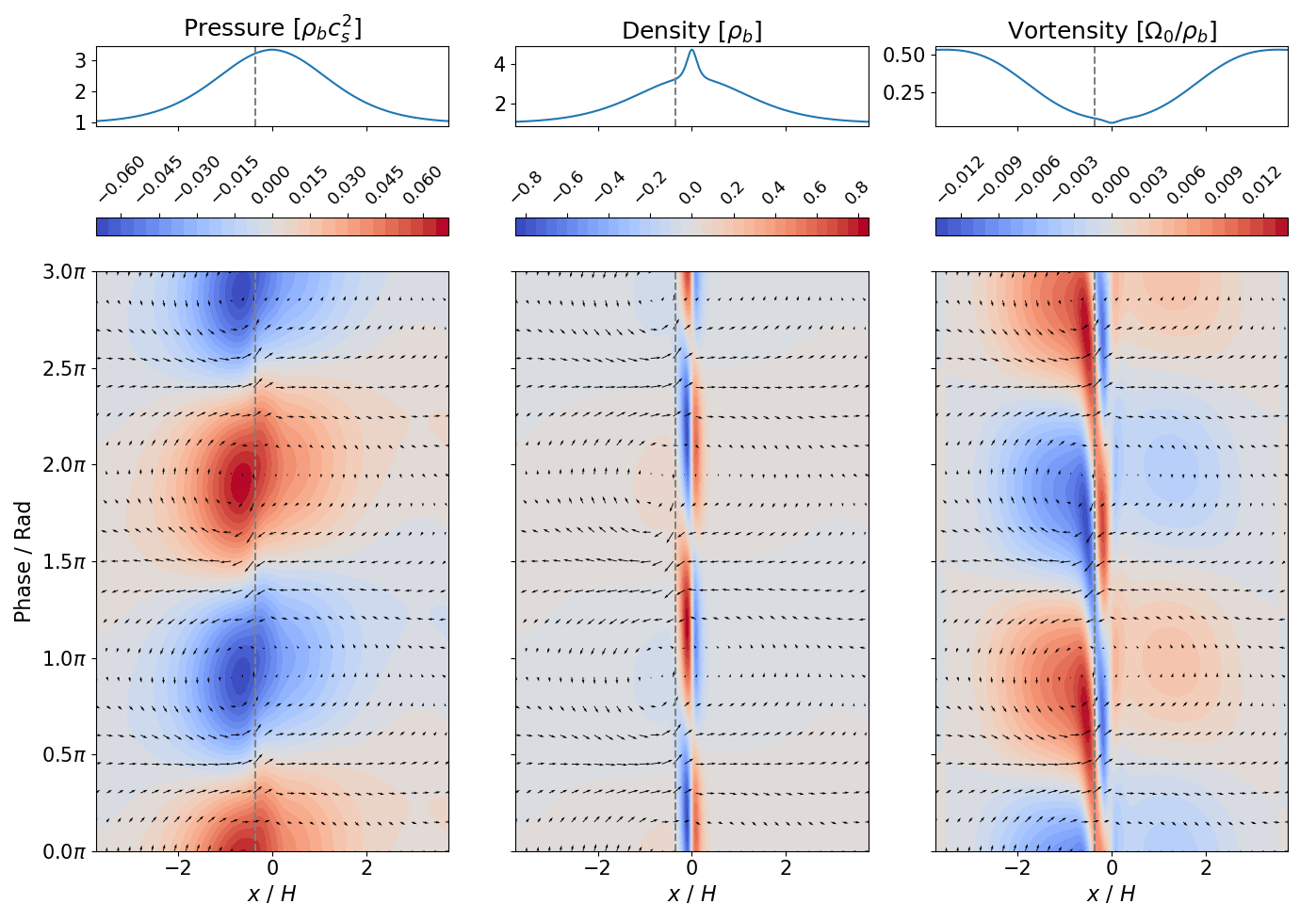}
    \caption{Eigenfunction of Type \RNum{2} DRWI for a system with moderate dust content ($f_{g{\rm min}}=0.7$). Here $k=0.2H^{-1}$ and $\omega_m=(0.1244+0.0407i)\Omega_0$. %Note that this figure, similar to Figure~\ref{fig Type1 eigenfunction 0.99} and Figure~\ref{fig Type1 eigenfunction 0.70}, are drawn in the rotating frame at $x=0$, instead of in the rotating frame at $x=x_c$.  
    Other details are the same as Figure~\ref{fig Type1 eigenfunction 0.99}.}
    \label{fig Type2 eigenfunction 0.70}
\end{figure*}

Now, for the purpose of illustration, we demonstrate the main properties of the two DRWI types by a representative result in Figure~\ref{fig dispersion sharp bump}. With fiducial settings of $A=1.2, \Delta w/H=1.5, St=0.03,$ and $\alpha=3\times10^{-4}$, various values of $f_{g\rm min}=0.5,0.7,0.85,0.99$ are chosen, with the corresponding $\overline{f_d}$ measured to be $0.038,0.020,0.009,$ and $6\times10^{-4}$.
%that respectively correspond to $\overline{f_d}=0.038,0.020,0.009,6\times10^{-4}$.
%In our context, a "sharp bump" refers to large $A$ and small $\Delta w$ %such that a purely gaseous bump ($f_{g{\rm min}}\to1$) is unstable to classical RWI. 
%such that our system is unstable to both types of DRWI. For a sharp bump, the dispersion relations of the two types follow different trends with a range of $f_{g{\rm min}}$, as shown in Figure~\ref{fig dispersion sharp bump}. 
Our calculation starts from $k=0.1$, which corresponds to a wavelength of the disc circumference if the local pressure scale height satisfies $H/R_0=0.1$. 
%In all cases, $\gamma_m(k)$ reaches a most unstable level (possibly before $k=0.1$) before reducing to negative for large $k$. However, the peak of $\gamma_m(k)$ is influenced very differently by dust concentration: for Type \RNum{1}, higher level of dust significantly suppresses the maximum of $\gamma_{m}$, whereas the opposite is true for Type \RNum{2}. 
We see that as dust concentration increases, the dispersion relation for the type I DRWI extends to larger $k$ (shorter wavelength), while the fastest growth rates decreases. On the other hand, the fastest growth rate of type II DRWI increases with dust concentration.

In the next, we show the eigenfunctions of the two modes to further examine the underlying physics. Apart from $P_1, f_{g1}, v'_{1x},$ and $v'_{1y}$, the vortensity $q$ is known to be vital for the mechanism of the RWI and also of interest here.
%We will present and discuss the perturbed vortensity $q_1$ along with other perturbation variables.
The vortensity is defined by
\begin{equation}
    q \equiv \frac{(2-3/2)\Omega_0+(\nabla\times \bm v')_z}{\rho} = \frac{1}{\rho}\left(\frac{1}{2}\Omega_0+\frac{\pa v'_y}{\pa x}-\frac{\pa v'_x}{\pa y}\right)
\end{equation}
as proper for pure gas in a Keplerian-rotating shearing sheet (see Appendix~\ref{append vortensity}). Its linear perturbation is then
\begin{equation}
    q_1 = -\frac{1}{\rho_0^2}\left(\frac{1}{2}\Omega_0+\frac{dv'_{0y}}{dx}\right)\rho_1 + \frac{1}{\rho_0}\left(\frac{dv'_{1y}}{dx} - ikv'_{1x}\right)\ .
\end{equation}

Starting from the Type~\RNum{1} DRWI, we first look at the case with $f_{g{\rm min}}=0.99$, which is close to the dust-free scenario, and the corresponding eigenfunctions of perturbed pressure, density and vortensity are shown in Figure~\ref{fig Type1 eigenfunction 0.99}.
%We use $f_{g{\rm min}}=0.99$ to approximate a pressure bump of pure gas. 
In this limit, Type \RNum{1} is strongly unstable with a maximal $\gamma_m$ on the order of $10^{-2}$ to $10^{-1}\Omega_0$, in agreement with the RWI investigated in \cite{Ono16} with similar gas bump profiles.
%Moreover, the eigenfunctions of the perturbed problem shown in Figure~\ref{fig Type1 eigenfunction 0.99} manifest patterns matching the classical RWI. 
The pressure and density perturbations show alternate peaks and troughs along the $\hat{y}$ direction accompanied respectively by anti-cyclonic and cyclonic velocity perturbations. The vortensity perturbations show patterns of two Rossby waves along $\hat{y}$ on the two sides of the background vortensity minimum ($x=0$), with a phase difference.
%shows a perturbed pattern of two parallel Rossby waves at both sides of the equilibrium vortensity minimum ($x=0$) with a phase difference. 
The growth of this instability, as explained in \cite{Ono16}, can be ascribed to $v'_{1x}$ advecting large background vortensity towards a positive vortensity perturbation and vice versa (e.g., along the horizontal line of phase $0.5\pi$ in Figure~\ref{fig Type1 eigenfunction 0.99}). On the other hand, for a higher $k$ such that $\gamma_m<0$, the vortensity begins to show an opposite phase difference that suppresses the perturbation. Based on these reasons, we recognise Type \RNum{1} as the RWI loaded with dust.
%We caution that the interpretation of the vortensity flow is inexact in our system because the vortensity is no longer conserved \xb{due to the presence of dust}, although we expect any deviation to vanish in the pure gas limit.
\footnote{Although vortensity is no longer strictly conserved due to the presence of dust, the deviation is expected to be small with only mild dust mass loading ($f_{g{\rm min}}=0.99$, or $\overline{f_d}=6\times10^{-4}$).}

In the eigenfunction described above, the magnitude of the density perturbation is %substantially 
strongly
enhanced within the dust bump. %a narrow range of roughly $|x|<0.3$, where the dust concentration is non-negligible. We call this region a "dust bump" as against the wider gas bump.
The phenomenon becomes more pronounced for a system with higher total dust content. In Figure~\ref{fig Type1 eigenfunction 0.70} where $f_{g\rm min}=0.7$ or $\overline{f_d}=0.020$, both the density and the vortensity perturbations are mainly concentrated in the dust bump. %This is the morphological reason of the name "dusty vortex instability". 
Vortensity sources are no longer negligible here, while the pattern remains
%although the pattern looks 
similar to the $f_{g\rm min}=0.99$ case outside the dust bump, where the vortensity-flow explanation of the instability still applies. We will further discuss the instability mechanism in Section~\ref{subsec vor sources}. %In particular, the "baroclinity" $\nabla P\times\nabla(1/\rho)$, which generates vortensity in non-isothermal gas dynamics (ref), 

The Type \RNum{2} DRWI shows essentially different eigenfunctions (Figure~\ref{fig Type2 eigenfunction 0.70}). %We still call it the dusty vortex instability as the dust content is necessary for the growth of the vortex patterns here. 
The non-zero $\omega_{rm}$ implies a $y$-direction phase velocity in the co-rotating frame at $x=0$. Therefore, the patterns in Figure~\ref{fig Type2 eigenfunction 0.70}, where $\omega_{rm}>0$, should be understood %imagined 
as travelling up along the $y$-axis with time. Another viewpoint is that %, causing
the co-rotation radius $x_c$, defined implicitly by $v_{0y}(x_c) = v'_{0y}(x_c)-(3/2)\Omega_0x_c = \omega_{rm}/k$, deviates from the pressure maximum towards approximately the edge of the dust bump. For $\omega_{rm}>0$, we have $x_c<0$. The pressure perturbation appears distorted across $x_c$ and reaches maximum/minimum at $x<0$. The density perturbation forms
%$y$-parallel 
periodic patterns along $\hat{y}$, with a positive patch on one side of $x=0$ matched with negative on the other side and vice versa. The perturbed vortensity patterns outside the dust bump still resemble two Rossby waves, but now the vortensity advection does not effectively contribute to the growth of the instability in the interval $x_c<x<0$. 
%with a phase difference close to, although not exactly, $\pi$. the vortensity flow fails to  account for the instability in the interval $x_c<x<0$. where $q_1$ tends to be neutralised, rather than enhanced, by vortensity advection either from the left or the right. \textbf{For example, the rightward flow at phase 0.5$\pi$ advects large background vortensity onto the negatively perturbed region where $x_c<x<0$.} 
We will elaborate on this observation quantitatively in %defer the analysis on the mechanism to
Section~\ref{subsec vor sources}, where we point out that the Type~\RNum{2} DRWI requires vortensity sources in the dust bump to be unstable at all.

As expected from the symmetry of our formulation, Type \RNum{2} DRWI modes always come in pairs: if $\omega_{rm}+i\gamma_m$ is an eigenvalue, then so is $-\omega_{rm}+i\gamma_m$. The 2D waveform of $-\omega_{rm}+i\gamma_m$ can be obtained from that of $\omega_{rm}+i\gamma_m$ by mapping $x\mapsto-x$ and $y\mapsto-y$, i.e., by reflection over the origin. The pair of modes likely coexist in real bumps, which implies a complicated mixture of their travelling patterns. Still, one may expect the $P_1$ patterns to appear to travel up along the $\hat{y}$ direction on the inner side of the bump $(x<0)$, where a mode with positive $\omega_{rm}$ has much stronger pressure perturbation than its negative-$\omega_{rm}$ counterpart; the opposite is expected on the outer side $(x>0)$. Our simulation in Section~\ref{subsec mild bump run} confirms this prediction.

While the two types of instabilities follow distinct trends in Figure~\ref{fig dispersion sharp bump}, their relation becomes more complicated where the dust bump is sharper. We observe bifurcation phenomena, where Type \RNum{2} merges into or forks from Type \RNum{1}, which we describe in further detail in Appendix~\ref{append bifurcation}. While the discussion above on eigenfunction remains valid, the bifurcation implies a smooth transition between the two types of the DRWI and hence between symmetric and asymmetric perturbation patterns in strongly unstable regions of the parameter space.

\subsection{Effect of viscosity on the RWI}
\label{subsec viscosity}

\begin{figure}
    \centering
    \includegraphics[width=0.5\textwidth]{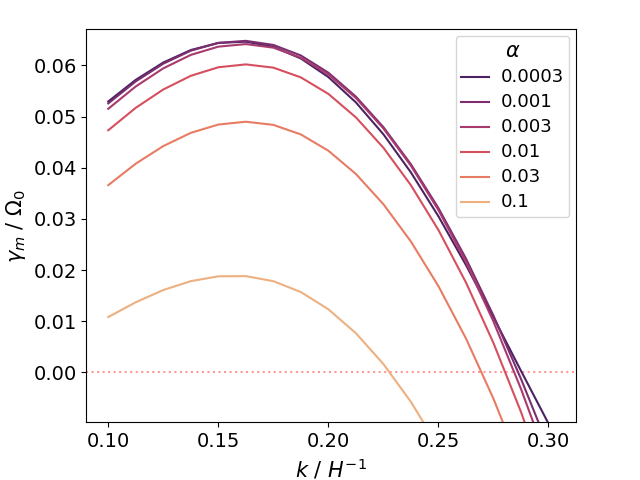}
    \caption{Dispersion relation of the classical RWI with different viscosity. A low gas content of $f_{g\rm min}=0.99$ is considered to approximate the pure-gas limit. Other parameters are set as $A=1.2$, $\Delta w/H=1.5$, $St=0.03$.}
    \label{fig classical RWI viscosity}
\end{figure}

We have shown that the dispersion relation and eigenfunction patterns of the Type~\RNum{1} DRWI approach those of the classical RWI in the limit of pure gas. However, the DRWI incorporates the turbulent viscosity, a physical process neglected in most previous studies on the RWI. As a short digression, our formulation can naturally be used to calculate the linear behavior of the RWI in the presence of gas turbulence.

%Our formulation readily applies to the classical RWI in the presence of gas turbulence. 
In Figure~\ref{fig classical RWI viscosity}, we compare a wide range of $\alpha$ for an approximately dust-free bump ($f_{g{\rm min}}$=0.99). While high viscosity suppresses the instability, $\alpha\leq3\times10^{-3}$ hardly influences the dispersion relation. The $y$-direction wavenumber corresponding to the maximal $\gamma_m$ also stays almost invariant. Our linear analysis here is consistent with simulations that the RWI in the linear regime is largely unaffected by realistic disc viscosity settings \citep{Lin14}. Analytical work by \cite{Gholipour14} gave similar results, on which we improve by properly setting up the background equilibrium state.
%formulating the viscosity from general equations and deriving relevant perturbations consistently.

\begin{figure*}
    \centering
    \includegraphics[width=\textwidth]{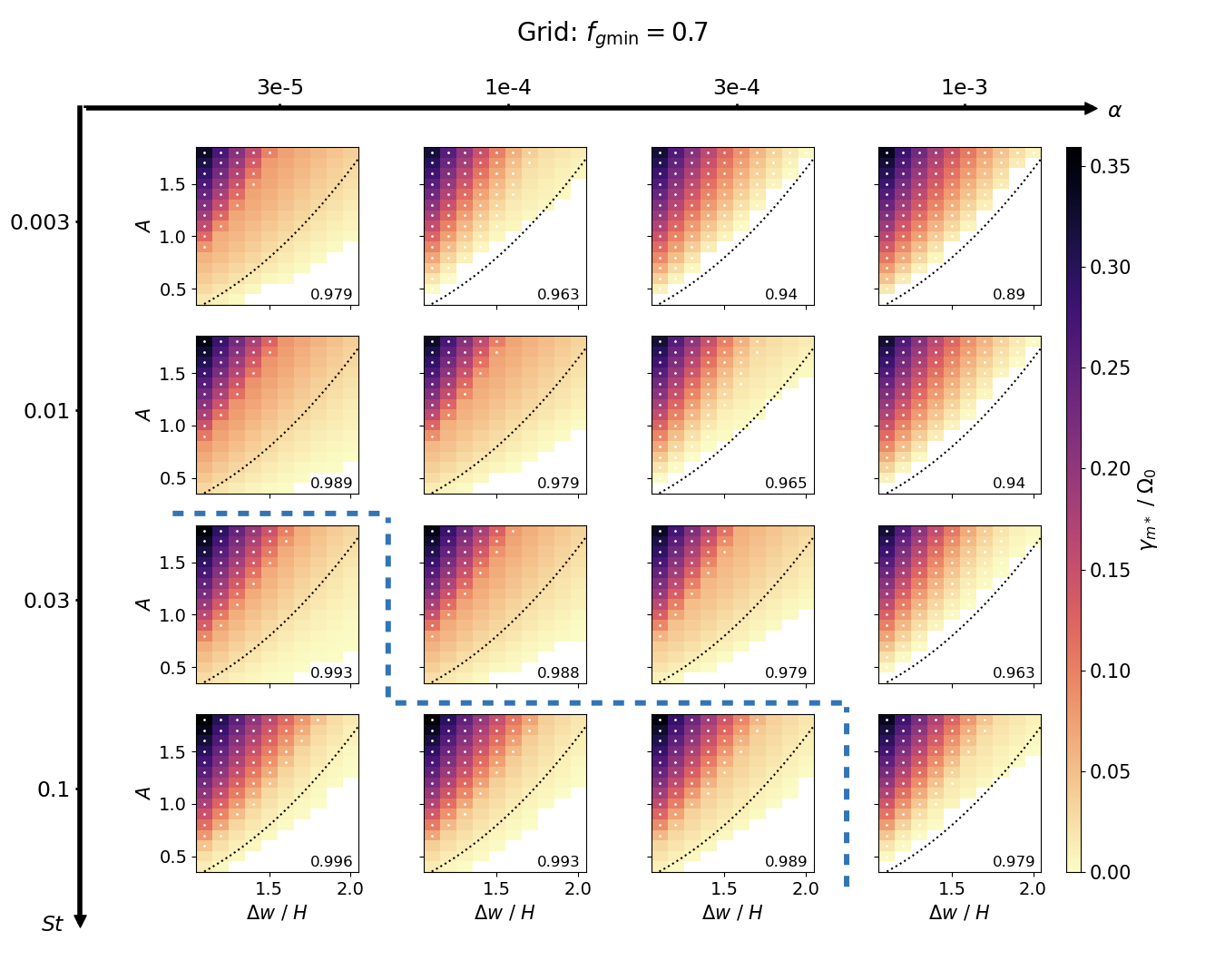}
    \caption{Grid search of $\gamma_{m*}$ %maximal  
    for $f_{g{\rm min}}=0.7$. Each panel is covered by a linearly uniform grid of step size 0.1 along both axes. The values of $St$ and $\alpha$ for each panel are annotated on the axes of the entire figure. Colors represent the value of $\gamma_{m*}$ for a given set of parameters, listed in Table~\ref{tab fiducial}, with darker shades denoting higher $\gamma_{m*}$ and white denoting $\gamma_{m*}<0$. 
    In each panel, white dots denote pixels where Type~\RNum{1} is more unstable than Type~\RNum{2}, the black dotted curve denotes where the dust-free bump is marginally stable to the standard RWI, and the number at the bottom right denotes $\overline{f_g}$ averaged over all pixels on the panel. The blue dashed line separates four bottom left panels from the rest, as the former deviate from the trend related to $St$ and $\alpha$ (see discussion in Section~\ref{subsubsec gamma m*}). %The color fades completely where the maximal $\gamma_m\leq0$.  %\protect\footnotemark
    All panels here and in Figure~\ref{fig max gamma grid 0.5} are colored in one single scale for comparison.}
    \label{fig max gamma grid 0.7}
\end{figure*}

\begin{figure*}
    \centering
    \includegraphics[width=\textwidth]{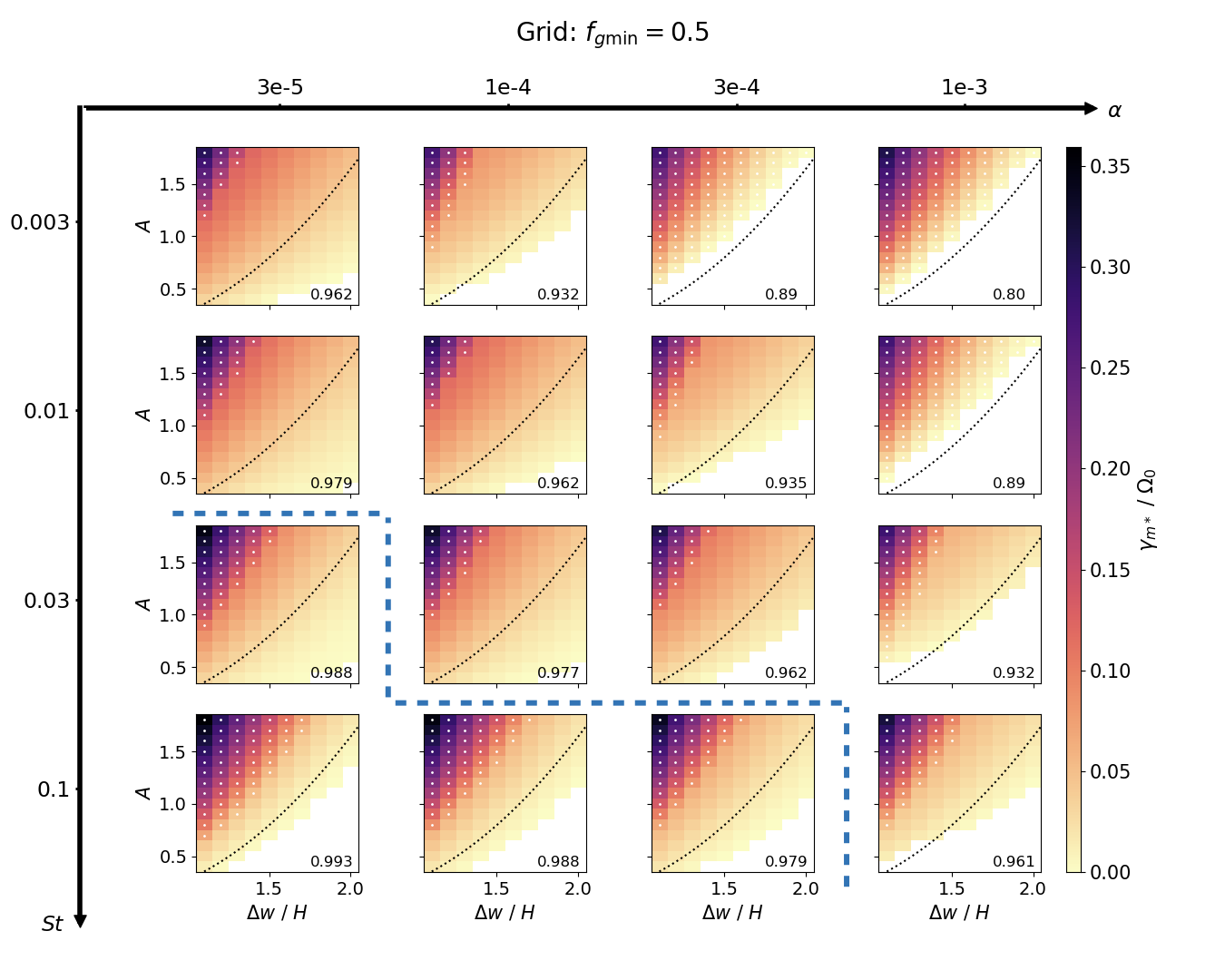}
    \caption{Grid search of $\gamma_{m*}$ % maximal $\gamma_m$ 
    for $f_{g{\rm min}}=0.5$. This figure is similar to Figure~\ref{fig max gamma grid 0.7}. All panels here and in Figure~\ref{fig max gamma grid 0.7} are colored in one single scale for comparison.}
    \label{fig max gamma grid 0.5}
\end{figure*}

\begin{figure*}
    \centering
    \includegraphics[width=\textwidth]{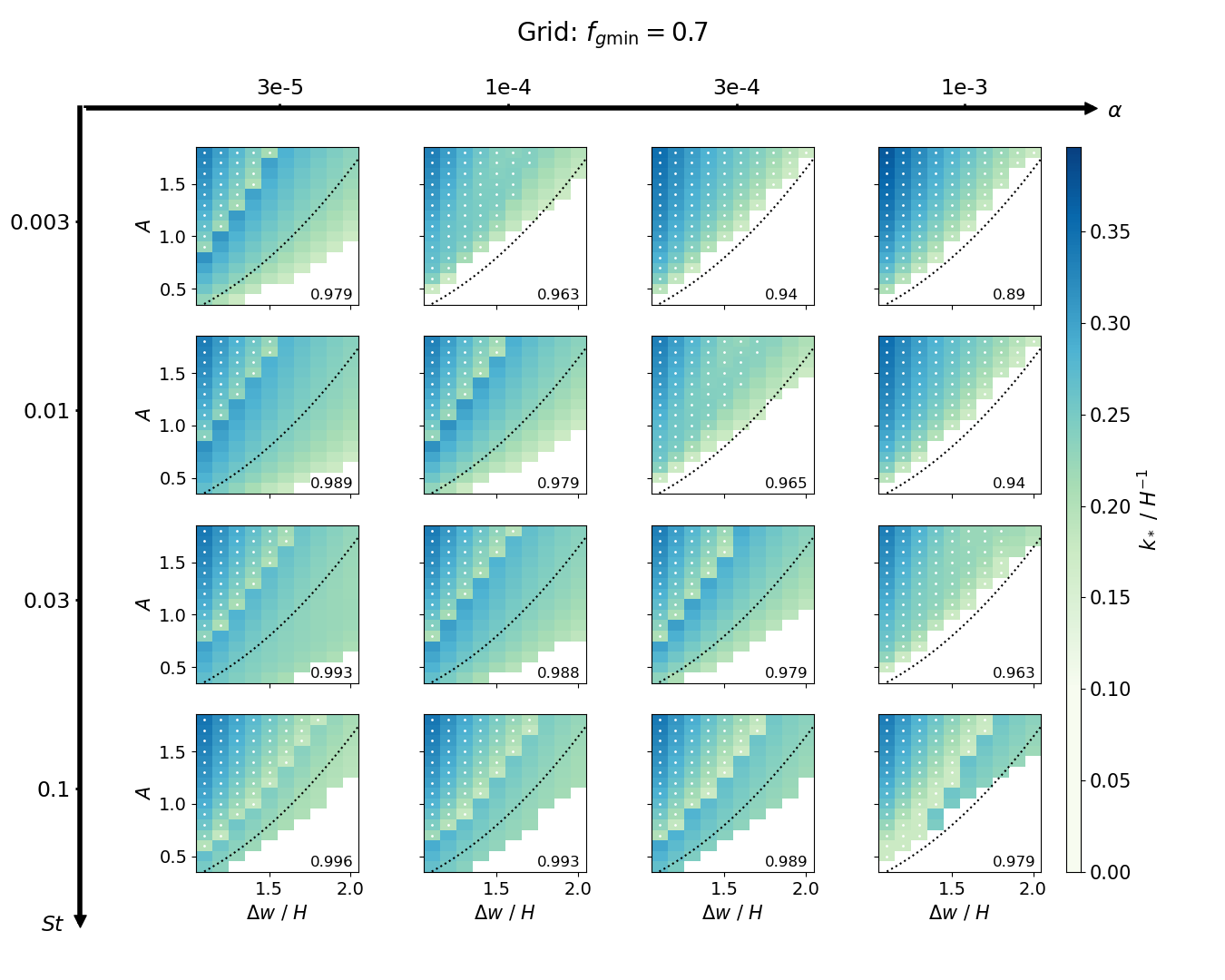}
    \caption{Grid search of the most unstable wavenumber $k_*$ for $f_{g{\rm min}}=0.7$. Darker shades denote higher $k_*$ and white pixels have no unstable mode. We take $k_*=0.1H^{-1}$ for monotonically decreasing $\gamma_m(k)$ curves. All panels here and in Figure~\ref{fig max k grid 0.5} are colored in one single scale for comparison. For other details, see the caption of Figure~\ref{fig max gamma grid 0.7}.}
    \label{fig max k grid 0.7}
\end{figure*}

\begin{figure*}
    \centering
    \includegraphics[width=\textwidth]{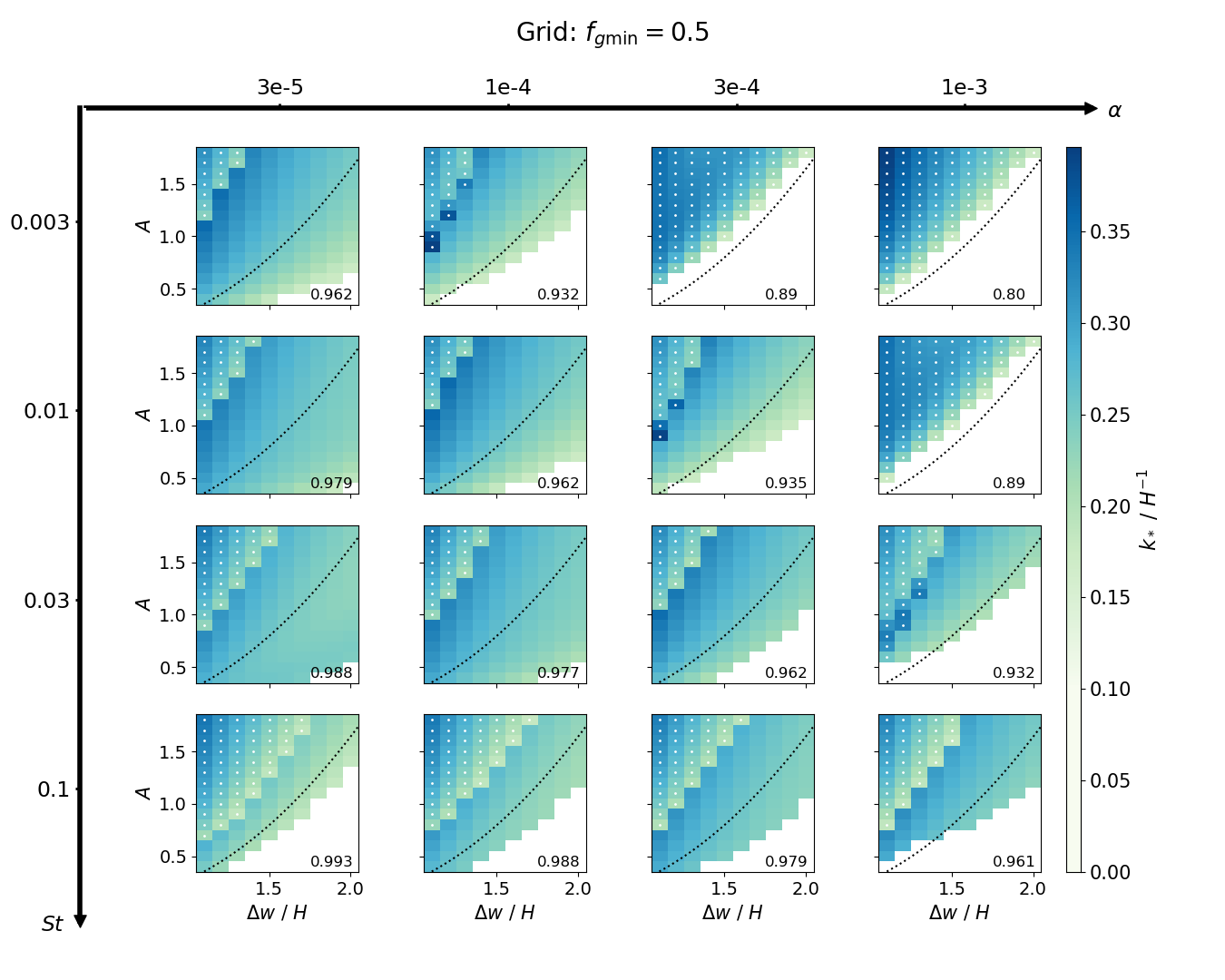}
    \caption{Grid search of the most unstable wavenumber $k_*$ for $f_{g{\rm min}}=0.5$. This figure is similar to Figure~\ref{fig max k grid 0.7}. All panels here and in Figure~\ref{fig max k grid 0.7} are colored in one single scale for comparison.}
    \label{fig max k grid 0.5}
\end{figure*}

\subsection{Parameter space of the most unstable DRWI}
\label{subsec parameter space}
To understand the parametric dependence of the two types of the DRWI, we perform a grid search in the parameter range listed in Table~\ref{tab fiducial} except that only two levels of dust content, $f_{g{\rm min}}=0.7$ and 0.5, are selected. The maximum growth rates $\gamma_{m*}$ with regard to different $k$ and the corresponding wavenumbers $k_*$ are shown in Figures~\ref{fig max gamma grid 0.7}, \ref{fig max gamma grid 0.5} and Figures~\ref{fig max k grid 0.7}, \ref{fig max k grid 0.5} respectively. For all these figures, each panel represents one particular $(St,\alpha)$, while each pixel in each panel gives one $(A, \Delta w)$. White dots denote pixels where Type~\RNum{1} is more unstable than Type~\RNum{2}. Black dotted curves denote where the dust-free bump is marginally stable to the standard RWI. %with $f_{g{\rm min}}=0.99$ is marginally stable. 
\footnote{We only calculate the black dotted curve accurate to the ($\Delta w, A$) pixel size for $f_{g{\rm min}}=1-10^{-8}$, $St=0.03$ and $\alpha=3\times10^{-4}$ and duplicate it to all panels. Section~\ref{subsec viscosity} and additional tests on different $St$ verify that the line does not change location significantly across panels.} We also calculate the mean gas mass fraction $\overline{f_g}$ for each pixel. As the bump sharpens, it concentrates dust more vigorously, but the total amount of gas also increases, making $\overline{f_g}$ %This value 
barely dependent on $A$ or $\Delta w$. We show the panel-wise averaged results at the bottom right of each panel, where the significant digits reflect the magnitude of pixel-wise deviation. %Section~\ref{subsec viscosity} and additional tests on different $St$ verify that the line does not change location significantly across panels.}
%The highest colored pixels roughly along the diagonal in each panel trace the boundary between a sharp and a mild bump: for $f_{g{\rm min}}=0.99$ and different $(St, \alpha)$, Type \RNum{1} DRWI maximally gives $0<\gamma_m\lesssim10^{-2}$ for each of these pixels and $\gamma_m<0$ one pixel below (here we confirm the known fact that the classical RWI is insensitive to viscosity). We tested that more concentrated dust only further stabilised Type \RNum{1} DRWI, consistent with our results in Section~\ref{subsec sharp bump}. Pixels above the diagonal are not explored as they belong to the sharp bump case. Starting from the diagonal in each panel, we move downwards and rightwards and calculate the eigenvalue of Type \RNum{2} DRWI for each pixel until the mode becomes stable. Some panels on the upper right are left blank because all mild bumps there do not have unstable Type \RNum{2} DRWI. We also calculated $\alpha=3\times10^{-3}$ but found that all Type \RNum{2} DRWI had $\gamma_m<0$ for both $f_{g{\rm min}}=0.7$ and 0.5 in the mild bump. Presenting this result would add a blank column of panels on the right of the figures.

\subsubsection{Maximum growth rate $\gamma_{m*}$}
\label{subsubsec gamma m*}
We first focus on the maximum growth rate (among different $k$). %with regard to different $k$. 
Three levels of inspection reveal the effect of different parameters: between pixels within one panel for $A$ and $\Delta w$, between the sixteen panels within one figure for $St$ and $\alpha$, and between Figures~\ref{fig max gamma grid 0.7} and \ref{fig max gamma grid 0.5} for $f_{g\rm min}$. In the following, we will start from the first and third levels, where the trends are relatively straightforward, before elaborating on the second level of comparison.

Each panel shows similar pixel-level trends: sharper pressure bumps (large $A$, small $\Delta w$) induce faster growth rates for both types %either type
of the DRWI. While the Type~\RNum{1} DRWI shows a steep %ascending 
slope with respect to $A$ or $\Delta w$ and prevails for very sharp bumps, Type~\RNum{2} dominates in a broad range of realistic parameters and even renders the pressure bump %system 
unstable when it is stable to the classical (dust-free) RWI (i.e., colored pixels below the black dotted curve). %renders unstable some pressure bumps that are 

Regarding the comparison between Figures~\ref{fig max gamma grid 0.7} and \ref{fig max gamma grid 0.5}, % \textbf{white-dotted pixels in the former figure has slightly higher maximal growth rates than the pixels at the same location in the latter figure (e.g., compare the pixel with $St=0.03,\alpha=3\times10^{-4},A=1.8,\Delta w/H=1.4$ in each figure). The opposite is true for pixels without white dots (e.g., the pixel with $St=0.01,\alpha=1\times10^{-4},A=0.5,\Delta w/H=1.5$). This demonstrates that} 
a higher dust concentration tends to stabilise the Type~\RNum{1} but destabilise the Type~\RNum{2} DRWI, consistent with Section~\ref{subsec sharp bump}. For example, the pixel with $St=0.03,\alpha=3\times10^{-4},A=1.8,\Delta w/H=1.4$ has a darker color in Figure~\ref{fig max gamma grid 0.7} than in Figure~\ref{fig max gamma grid 0.5} (a Type~\RNum{1}-dominant case, $\gamma_{m*}=0.16\Omega_0$ versus $0.12\Omega_0$), whereas the opposite is true for the pixel with $St=0.01,\alpha=1\times10^{-4},A=0.5,\Delta w/H=1.5$ (Type~\RNum{2}-dominant).

% As displayed in Figures~\ref{fig max gamma grid 0.7}, \ref{fig max gamma grid 0.5}, while the Type~\RNum{1} DRWI dominates for very sharp bumps, Type~\RNum{2} grows faster in a broad range of realistic parameters and even %renders unstable some pressure bumps that are 
% renders the system unstable when it is stable to the \xb{classical} RWI. A comparison between the two figures reveals that a higher dust concentration tends to stabilise the Type~\RNum{1} but destabilise Type \RNum{2} DRWI, consistent with Section~\ref{subsec sharp bump}. %: $f_{g{\rm min}}=0.5$ may have unstable modes for configurations where $f_{g{\rm min}}=0.7$ always gives $\gamma_m<0$ (e.g., the panel $\alpha=1\times10^{-3}$ and $St=0.03$), and the former tends to have a larger $\gamma_m$ (darker color) when both are unstable. 

Now, we compare different panels within Figure~\ref{fig max gamma grid 0.7} to explain how $\gamma_{m*}$ changes with $St$ and $\alpha$. The comparison also applies to Figure~\ref{fig max gamma grid 0.5}. The similar colors on the upper left corner of each panel demonstrate that the Type~\RNum{1} DRWI, when dominant, is insensitive to $St$ or $\alpha$. Conversely, the trend of the Type~\RNum{2} DRWI is most clearly seen from the lower right region of each panel. For most panels (those above the blue dashed line), a combination of small $\alpha$ and large $St$ shows the broadest range of colored pixels. For example, a bump with $A=0.8$ and $\Delta w/H=2.0$ is unstable to the Type~\RNum{2} DRWI for $St=0.03$ and $\alpha=1\times10^{-4}$, which is not true for $St\leq0.01, \alpha=1\times10^{-4}$ or for $St=0.03, \alpha\geq3\times10^{-4}$. The total area of colored pixels on the panel $(St,\alpha)=(0.03,1\times10^{-4})$ or $(0.01,3\times10^{-5})$ is larger than that on the panel on its upper right, i.e., $(St,\alpha)=(0.03,3\times10^{-4})$, $(0.01,1\times10^{-4})$, or $(0.003,3\times10^{-5})$. In other words, the susceptibility of the system to the Type~\RNum{2} DRWI largely varies along the diagonal of the figure from moderately high $St$ and low $\alpha$ (most unstable) to low $St$ and high $\alpha$ (least unstable). %However, the four panels below the blue dashed line deviate from the diagonal trend as they show a shrinkage of the unstable range compared to adjacent panels on their upper right.
% Therefore, the maximum growth rate of the Type~\RNum{2} DRWI is largely positively correlated with $St$ but negatively correlated with $\alpha$ and $f_{g\rm min}$. 
We have seen that a sharp pressure bump promotes both types of the DRWI; the correlation here likely similarly points to the Type~\RNum{2} DRWI favoring a sharp dust bump in addition to a sharp gas bump (see Figure~\ref{fig steady profiles} for how the dust bump profile changes with $St, \alpha$ and $f_{g\rm min}$). Notably, here we believe that $St$ and $\alpha$ only indirectly influence the Type~\RNum{2} DRWI by modifying the equilibrium bump profile instead of directly involving in the mechanism of the instability, a point we will argue more rigorously after describing the $k_*$ trends. 

However, the four panels below the blue dashed line deviate from the diagonal trend as they show a shrinkage of the unstable range compared to adjacent panels on their upper right.
%This is open to various explanations.
%However, the opposite tendency below the blue dashed line is open to various explanations. 
%Possibly excessively sharp dust \textbf{density gradients} %bump 
%become suboptimal for the Type~\RNum{2} DRWI. Here, one may concern that the dust bump is worse resolved, but doubling the resolution only further marginally reduce $\gamma_{m*}$. 
This corner corresponds to very low $\alpha$ and relatively large $St$, leading to a very sharp dust bump. We confirm that our resolution is adequate for resolving the dust bump, and speculate on the potential causes that reverse the trend. First,
%Alternatively, 
the sharp dust bump implies a very low average dust mass fraction $\overline{f_d}$,
%which may become inadequate for Type~\RNum{2} since the latter requires the presence of dust to operate.
where dust feedback likely becomes too spatially restricted for the Type~\RNum{2} DRWI to operate.
%since the presence of dust is a necessary condition of this instability, \textbf{a very low average dust mass fraction $\overline{f_d}$ may adversely affect its operation.}
Also,
%weak dust-gas coupling with $St\gtrsim0.1$ may interfere with the instability, although here our one-fluid formalism might not capture the physics correctly. 
weak dust-gas coupling with $St\gtrsim0.1$ may be subject to two-fluid effects not fully captured in our one-fluid formalism.
Later we find that the mechanism of the Type~\RNum{2} DRWI does not necessitate streaming motion and thus refrain from further analysis of the marginally coupled system.

\subsubsection{Most unstable wavenumber $k_*$}
\label{subsubsec k*}
We show in Figures~\ref{fig max k grid 0.7}, \ref{fig max k grid 0.5} the most unstable wavenumber $k_*$ in the sense that $\gamma_m(k)$ reaches maximum at $k=k_*$. The apparent discontinuous transition in $k_*$ in most of the panels reflect a switch from Type~\RNum{1} (with white dots) to Type~\RNum{2} (no white dots) regimes. Generally, for both types of the DRWI, higher maximum growth rates correspond to higher $k_*$, as is also seen from Figure~\ref{fig dispersion sharp bump}. For $f_{g\rm min}=0.5$, a few Type~\RNum{1} cases have abnormally large $k_*$ (e.g., the deep blue pixel at $St=0.003,\alpha=1\times10^{-4},A=1.0,\Delta w/H=1.1$). %which is likely caused by the suppressive effect of dust that may also 
%\textbf{as high dust content in equilibrium may shift the peak of the dispersion relation to a higher wavenumber} (\textbf{cf.} Figure~\ref{fig dispersion sharp bump}).
This is related to the fact that $\gamma_m(k)$ is quite flat in the full dispersion relation of Type~\RNum{1} DRWI when $\gamma_{m*}$ is low (cf. Figure~\ref{fig dispersion sharp bump}) and thus $k_*$ can be parameter-sensitive.
%Remarkably, the wavelength here is 
Interestingly, typical unstable wavelengths in $\hat{y}$ are
comparable to the disc size and much longer than typical length scales of the SI, the latter being only a fraction of the disc scale height. 
We find no unstable mode in high $k\gtrsim1H^{-1}$ (Appendix~\ref{append perturb}), except when we reduce the turbulence level to $\alpha<10^{-6}$. This is
likely due to the turbulent diffusion that strongly suppresses small-scale instabilities.
%Indeed, unstable modes with $k>1$ emerge only 

Also, we find that the most unstable wavenumber of the Type~\RNum{2} DRWI is insensitive to $St$. %This has implications on its mechanism, which we elaborate on in Section~\ref{subsec mechanism}.
The important implication is that this instability is unlikely related to the dust streaming motion, the mechanism used to explain the SI and more generally the resonant drag instabilities \citep[RDI;][]{Squire2018}, where the outcome sensitively depends on the Stokes number.
%contrary to our results in Figure~\ref{fig max k grid 0.7} and \ref{fig max k grid 0.5}.
To further verify this, we conducted another series of calculations, gradually reducing $St$ and $\alpha$ simultaneously until $St\sim10^{-5}$ and $\alpha\sim10^{-7}$, %calculation setting $St\sim10^{-5}$ and $\alpha\sim10^{-7}$, 
so that the dust bump remains similar to that in our fiducial setting (Table~\ref{tab fiducial}) but the dust is tightly coupled with the gas. We find that the dispersion relation remains similar for $k\lesssim 1H^{-1}$, %between the two cases, 
pointing to the fact that it is mainly the dust mass loading, rather than dust-gas streaming that shape the properties of this instability.
This conclusion is further strengthened by examining the perturbed relative kinetic energy of the dust and the gas with regard to the center of mass (Appendix~\ref{append relative Ek}). In the fiducial eigenfunction with $k=0.2H^{-1}$, it is found to be only 0.06\% of the perturbed kinetic energy of the single fluid.
The finding sets the stage for our understanding of the Type~\RNum{2} DRWI as tightly coupled motion of gas and dust in the following section.

\subsection{Physical ingredients of the DRWI}
\label{subsec vor sources}

We have identified Rossby waves in the morphology of both types of the DRWI. A Rossby wave is characterised by periodic votensity perturbation patterns with a background vortensity gradient normal to its travelling direction. The vortensity perturbations imply velocity perturbations and hence vortensity advection along the gradient, which solely governs the vortensity budget if the vortensity is conserved (e.g., pure isotropic gas). The RWI, then, is a result of two Rossby waves on each side of a background vortensity minimum positively feeding back to each other \citep[for a detailed interpretation of the Rossby wave and the RWI, see][]{Ono16}.

The physical picture 
%of both 
behind the Rossby waves
%and the RWI 
requires conservation of the vortensity. However, the dust bump introduces vortensity sources, which modifies the RWI to become the Type~\RNum{1} DRWI and brings the Type~\RNum{2} DRWI into existence. In the following, we
%will outline the theory of 
analyse
the vortensity budget in the dust-trapping ring in Section~\ref{subsubsec vor theory}, which is followed by a discussion in Section~\ref{subsubsec vor analysis} on the evolution of vortensity in and outside the dust bump and the properties of vortensity sources. 
We aim at identifying the governing physical ingredients of the DRWI, where
we demonstrate that
%our recognising Rossby waves away from the dust bump is still valid and that the misalignment between gas and dust density 
the Rossby waves still
play important roles in both types of the DRWI, with dust playing a damping/driving role in Type~\RNum{1}/Type~\RNum{2}.
We further tentatively explain in Appendix~\ref{append propagation mechanism} the propagation process of the Type~\RNum{2} DRWI, but a comprehensive investigation of the instability mechanism is beyond the scope of this paper.

%\textbf{Here, we only aim to identify the governing physical ingredients of the DRWI rather than to investigate its mechanism comprehensively. We only  tentatively explain in Appendix~\ref{append propagation mechanism} the propagation process of the Type~\RNum{2} DRWI, where we focus on how the Rossby waves and the perturbation patterns in the dust bump harmonise their phase velocity to meet the global eigenvalue $\omega_m$. The mechanism of the growth of the instability is beyond the scope of this paper.}

%\textbf{The dispersion relation has demonstrated the definitive and constructive role of the dust in the Type~\RNum{2} DRWI. What remains to be shown is how the dust bump and the broader gas bump interact and facilitate the propagation and growth of the instability.  In Section~\ref{subsec mechanism}, we interpret the Type~\RNum{2} DRWI first from the pressure-density wave perspective and then from the vortensity wave perspective, with an emphasis on how the different physical processes in the system harmonise their phase velocity and growth rates to meet the global eigenvalue $\omega_m$. The proposed mechanism broadly explains the representative eigenfunction patterns and parametric dependence of the Type~\RNum{2} DRWI.}

\begin{figure*}
    \centering
    \includegraphics[width=0.9\textwidth]{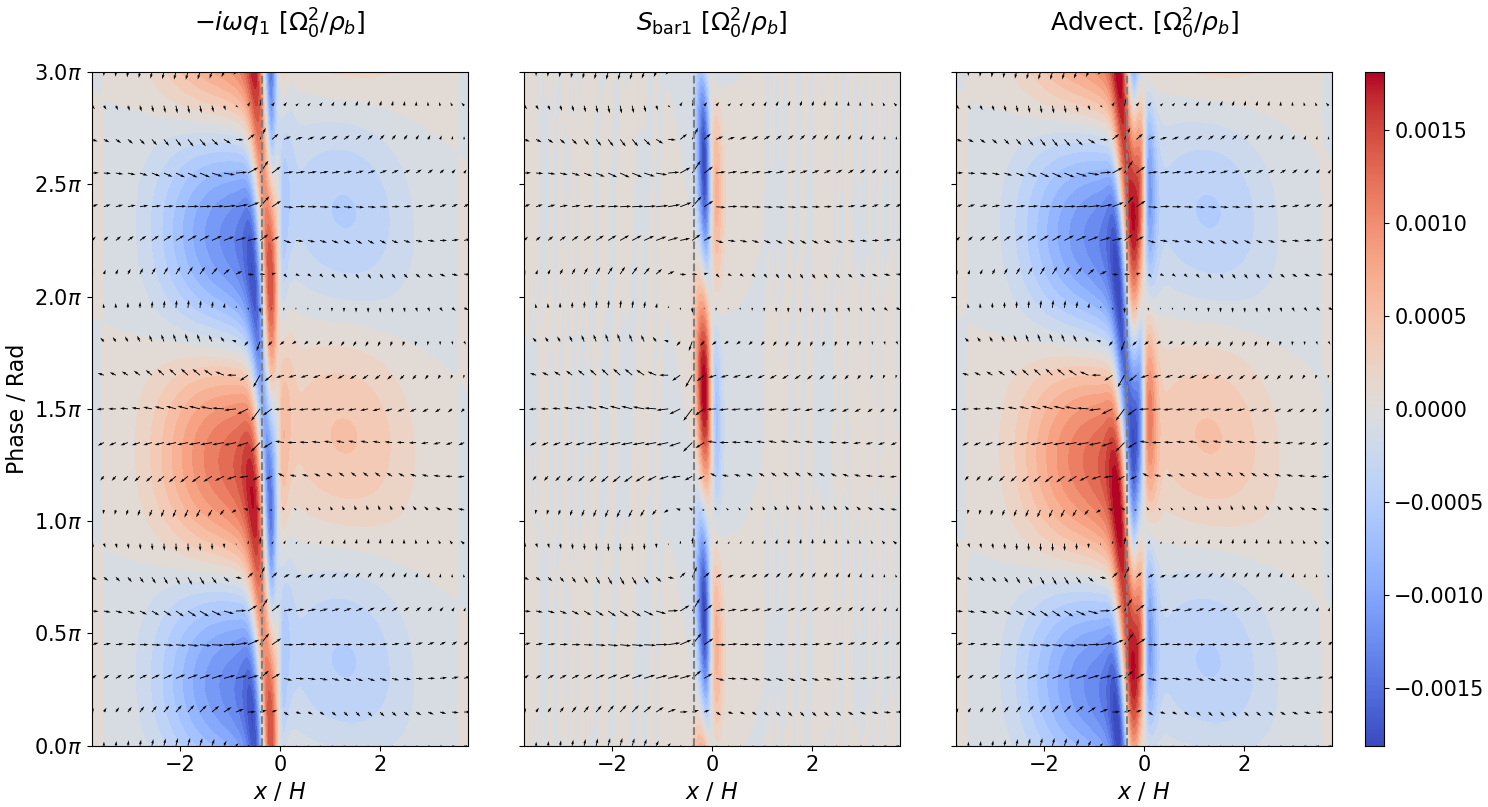}
    \caption{Terms related to the evolution of vortensity of Type \RNum{2} DRWI for a system with moderate dust content ($f_{g{\rm min}}=0.7$). The eigenmode is identical to that shown in Figure~\ref{fig Type2 eigenfunction 0.70}. The three panels show the time derivative, the baroclinic source, and the advection term of the vortensity perturbation. All panels are colored in one single scale for comparison. The $y$ range, the perturbed velocity field and the co-rotation radius are similarly plotted as in Figure~\ref{fig Type1 eigenfunction 0.99}.}
    \label{fig vor sources}
\end{figure*}

\subsubsection{The vortensity budget}
\label{subsubsec vor theory}

The DRWI involves a mixture of gas and dust that violates the conservation of vortensity. Specifically, the vortensity equation derived from Equations~(\ref{eq one fluid den})(\ref{eq one fluid mom}) takes the following form:
\begin{equation}
    \left(\frac{\pa}{\pa t}+\bm v'\cdot\nabla-\frac{3}{2}\Omega_0x\frac{\pa}{\pa y}\right)q = S \ , \label{eq one fluid vor}
\end{equation}
where the source $S$ satisfies
\begin{gather}
    S{\bm e}_z = \frac{1}{\rho}\nabla P\times\nabla\left(\frac{1}{\rho}\right) + \nu\frac{1}{\rho}\nabla\times(f_g\nabla^2\bm v') + \nonumber\\
    \frac{1}{\rho}\nabla\times\left[\frac{1}{\rho}\nabla\cdot(\rho_d\bm v_{\rm dif}\bm v_{\rm dif})\right] + \frac{1}{\rho}\frac{\pa [f_gf_0(x)]}{\pa x}{\bm e}_z\ . \label{eq vor source}
\end{gather}
Derivation of Equation~(\ref{eq one fluid vor}) can be found in Appendix~\ref{append vortensity}. The first term on the right-hand side of Equation~(\ref{eq vor source}) is usually known as the ``baroclinic"
%contribution for fluids whose density depends on a variable temperature besides pressure. It vanishes in the case of a barotropic flow, where $\rho$ is only a function of $P$, and hence the conservation of vortensity in the absence of diffusion or external forcing. 
term that arises when the fluid is not barotopic ($\rho$ being only a function of $P$). %However,
In our system, the dependence of $\rho$ on $f_g$ %in our formulation 
implies that vortensity may be created or consumed by any misalignment between the density and pressure gradients. The second and third terms might be crudely understood as vortensity diffusion due to gas viscosity and dust concentration diffusion respectively, and the fourth term emerges from the external forcing. 

The source terms have zero net %make no total
contribution in equilibrium: the baroclinic and dust diffusion terms vanish, while the gas diffusion term is balanced by the external torque. In perturbation, though, the vortensity equation will become
\begin{equation}
    (-i\omega+ikv'_{0y}-\frac{3}{2}ik\Omega_0x)q_1 + \frac{dq_0}{dx}v'_{1x} = S_1 = S_{\rm bar1} + S_{\nu1} + S_{\rm dif1} + S_{\rm ext1} \ , \label{eq one fluid vor perturb}
\end{equation}
where we express the perturbed source as a sum of the baroclinic $S_{\rm bar1}$, viscous $S_{\nu1}$, dust diffusion $S_{\rm dif1}$, and external forcing $S_{\rm ext1}$ terms, in parallel with the four in Equation~(\ref{eq vor source}). In particular,
\begin{equation}
    S_{\rm bar1} = \frac{ik}{\rho_0^3}\left(\frac{d\rho_0}{dx}P_1 - \frac{dP_0}{dx}\rho_1\right) = -\frac{ikP_0v_{{\rm dif}0x}}{\rho_0^2}\left(\frac{1}{c_s^2t_s}\mathfrak{f}_{g1}+\frac{f_{d0}}{D_0}\mathfrak{p}_1\right) \ . \label{eq Sbar1}
\end{equation}
This term is dominant among the sources and plays significant roles in the two types of the DRWI, as shown in the following sub-subsection. 

\subsubsection{Vortensity analysis}
\label{subsubsec vor analysis}
%The $S_{\rm bar1}$ term \textbf{in Equation~(\ref{eq Sbar1})} makes a substantial contribution to the Type \RNum{2} DRWI, as shown in Figure~\ref{fig vor sources}. 
We first analyse the vortensity budget of the Type~\RNum{2} DRWI, shown in Figure~\ref{fig vor sources}.
Here, we compare the time derivative of the vortensity perturbation, $-i\omega q_1$, the baroclinic source, $S_{\rm bar1}$, and the advection term, $-[ikv'_{0y}-(3/2)ik\Omega_0x]q_1-(dq_0/dx)v'_{1x}$.
%The latter two would add up equal to the time derivative in the absence of other sources, as indicated by Equation~(\ref{eq one fluid vor perturb}). Indeed, 
We find that the combination of the middle and right panels in Figure~\ref{fig vor sources}, representing the latter two terms,
%looks very similar to 
largely account for the total vortensity perturbation, as shown in the left panel.
%although the combined result shows a slightly larger magnitude and fine structural differences presumably explained by the gas or the dust diffusion.
We have also examined that the contributions from other terms, primarily from gas and dust diffusion, are relatively minor and only yield certain small-amplitude fine-scale features.
%\textbf{Here we note that $S_{\nu1}$ is proportional to $\nu$, which implies that weak turbulence hardly violates conservation of vortensity. This explains why, in Section~\ref{subsec viscosity}, the dispersion relation of the classical RWI is only disturbed when $\alpha>10^{-2}$.}

The baroclinity barely appears in the interval $x<x_c$ and is relatively weak in $x>0$. Advection dominates the evolution of $q_1$ in these regions, supporting our interpretation of classical Rossby waves based on the conservation of vortensity. However, in the narrow interval in between, $S_{\rm bar1}$ is stronger and one observes a discrepancy between $-i\omega q_1$ and the advection. %Positive $S_{\rm bar1}$ (e.g., $1.0\pi$--$2.0\pi$) roughly matches negative advection, therefore annulling the effect of the latter. But as we argued in Section~\ref{subsec sharp bump}, here the advection term discourages the vortensity perturbation. 
To quantify the effects of the baroclinity and the advection, we select the region $x_c\leq x\leq0, 0\leq ky<2\pi$ and calculate the cross-correlation between $q_1$ and the three terms shown in Figure~\ref{fig vor sources} along the $y$-axis with circular boundary conditions. The results are all sinusoidal as expected. Measuring the phase of the sinusoids, we find that the time derivative of $q_1$ has a phase lead of $71.9^\circ$ over $q_1$ itself, which is plainly equal to $-{\rm arg}(-i\omega_m)$. %\footnote{In a plane wave, a temporal phase lag corresponds to a spatial phase lead. The cross-correlation reveals the latter, hence the minus sign before the argument.}
The angle is less than $90^\circ$ (a positive imaginary part of $\omega_m$), indicating instability. $S_{\rm bar1}$ lags behind $q_1$ by $24.3^\circ$, a small angle compared to $90^\circ$, thus significantly enhancing $q_1$. In contrast, the advection term leads $q_1$ by $89.7^\circ$. The instability in the interval $x_c<x<0$, then, may be %qualitatively 
interpreted as the baroclinic source driving the growth of the vortensity perturbation, whereas the advection only serves to propagate the $q_1$ patterns. %against advection. 

%\textbf{Instabilities of dusty pressure bumps are sometimes thought to be associated to the Kelvin-Helmholtz Instability (KHI), as the dust backreaction induces a sharp radial gradient in gas azimuthal velocity \citep[e.g.,][]{Yang20}. The Type~\RNum{2} DRWI probably extracts some kinetic energy from such dust-induced velocity shear, but the classical KHI does not consider contribution from the baroclinity.}

We also perform similar calculations on the Type~\RNum{1} DRWI exemplified in Figure~\ref{fig Type1 eigenfunction 0.70}. In the region $-0.3H\leq x\leq0,0\leq ky<2\pi$ (roughly the left half of the dust bump), the advection is completely in phase with $q_1$ whereas the baroclinic source lags behind by $162.2^\circ$. Now, the advection encourages the growth of $q_1$ even in the dust bump, %(Section~\ref{subsec sharp bump})
but the baroclinic source still works against the advection. This explains why the dust tends to suppress the Type~\RNum{1} DRWI: the dust saps the perturbed vortensity from the positive feedback loop involving the Rossby waves. %\textbf{consumes the perturbed vortensity that the Rossby waves build up and rely on.} %acts as a parasite of 
%the Rossby waves by consuming the perturbed vortensity that the latter build up. 
In this sense, the same mechanism of instability underlies the Type~\RNum{1} DRWI and the classical RWI. This concludes %terminates 
our analysis on the interaction between the dust bump and the gaseous Rossby waves in the linear regime of the DRWIs. %Type~\RNum{1} DRWI.

\section{Numerical test and the nonlinear regime}
\label{sec simulation}
\begin{figure*}
    \centering
    \includegraphics[width=\textwidth]{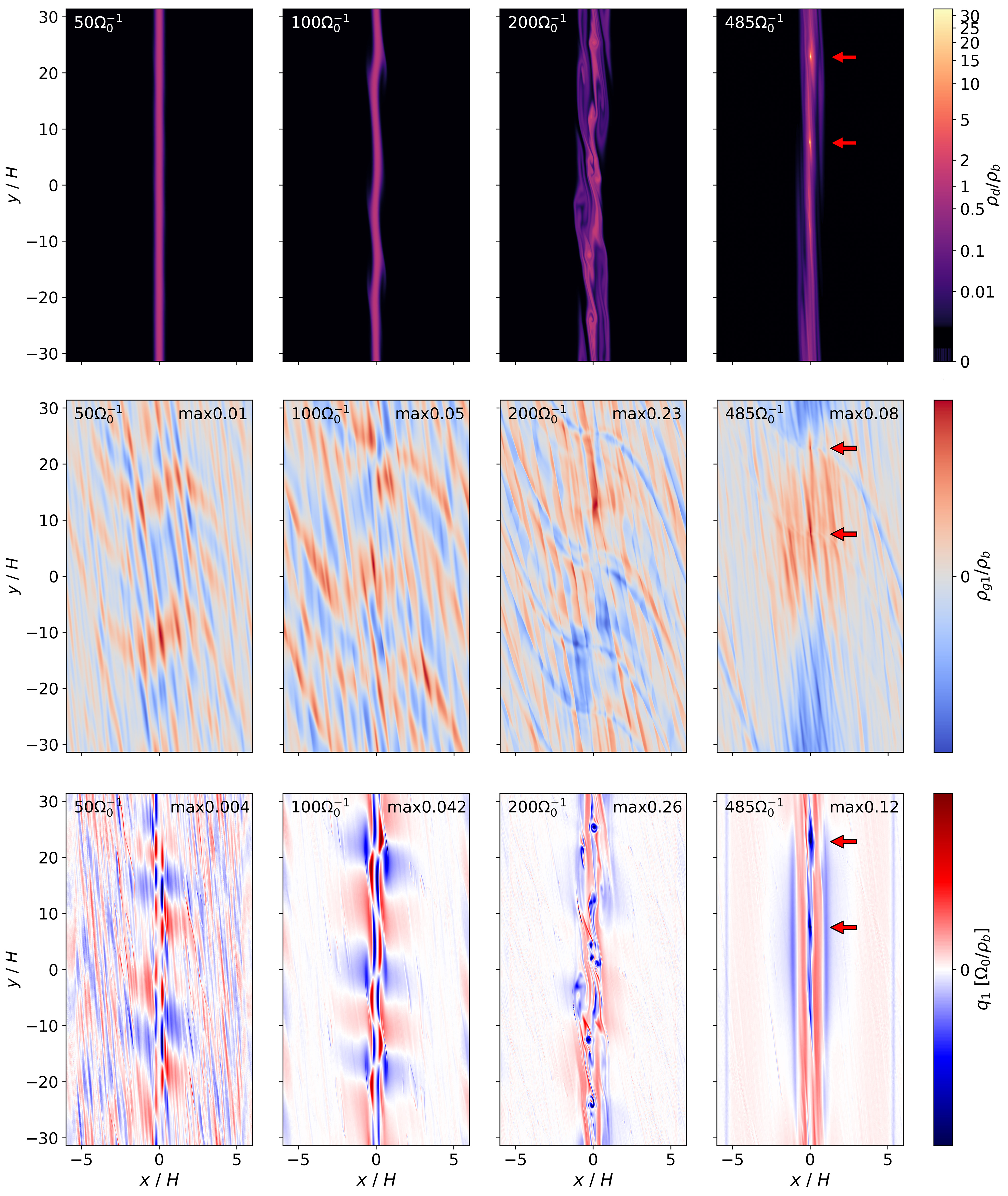}
    \caption{Snapshots in the ``mild bump'' run of the dust density, perturbed gas density and perturbed one-fluid vortensity (Equation~(\ref{eq one fluid vor perturb})) after inserting the perturbation. Note that the aspect ratio is not drawn to scale. The time is annotated on the top left of each panel. Panels in the top row are colored in the same power-law scale. Panels in the middle and bottom rows are colored linearly and not in the same scale: the maximum $|\rho_{g1}|$ or $|q_1|$ of each is annotated on the top right. Red arrows in the last column point to dust density maxima.}
    \label{fig simulation}
\end{figure*}

\begin{figure*}
    \centering
    \includegraphics[width=\textwidth]{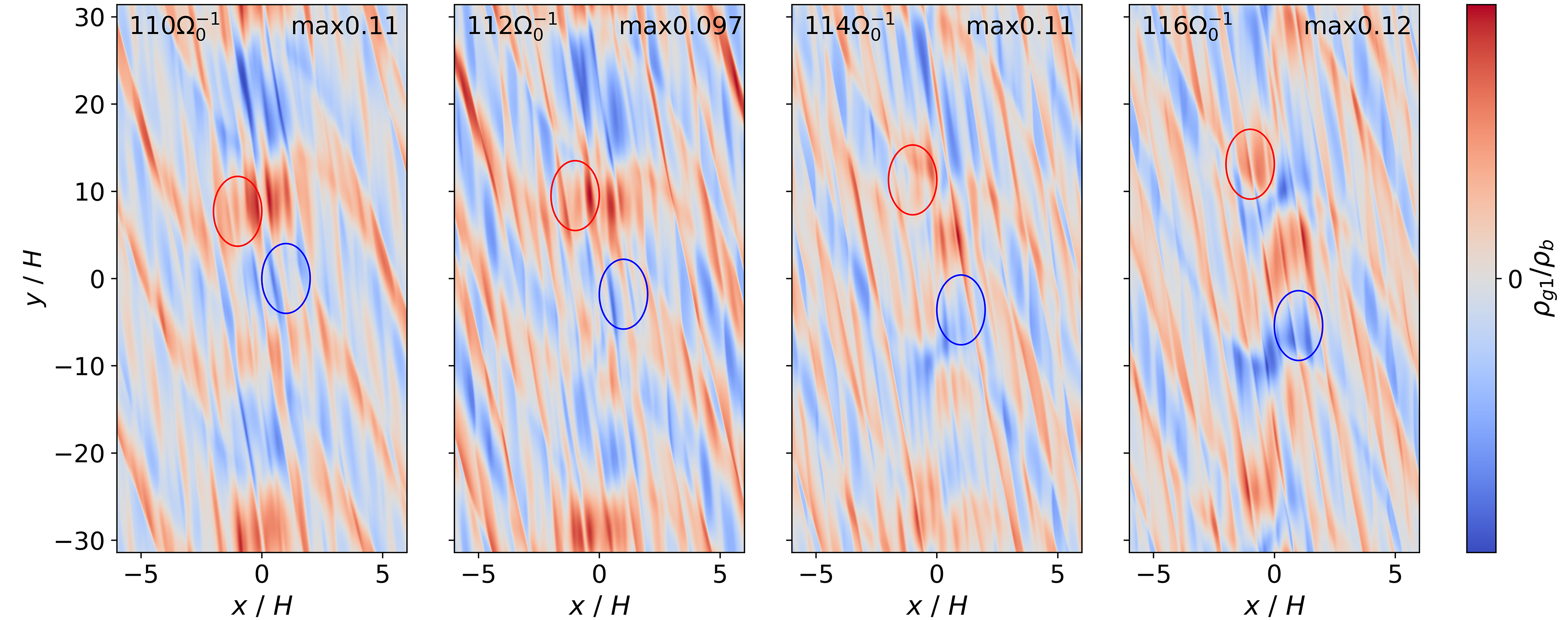}
    \caption{Same as the middle row in Figure~\ref{fig simulation}, but in a shorter time-scale. Red and blue circles ($v_y=\pm0.9c_s$) indicate azimuthally travelling patterns.}
    \label{fig simulation_travel}
\end{figure*}

\begin{figure*}
    \centering
    \includegraphics[width=\textwidth]{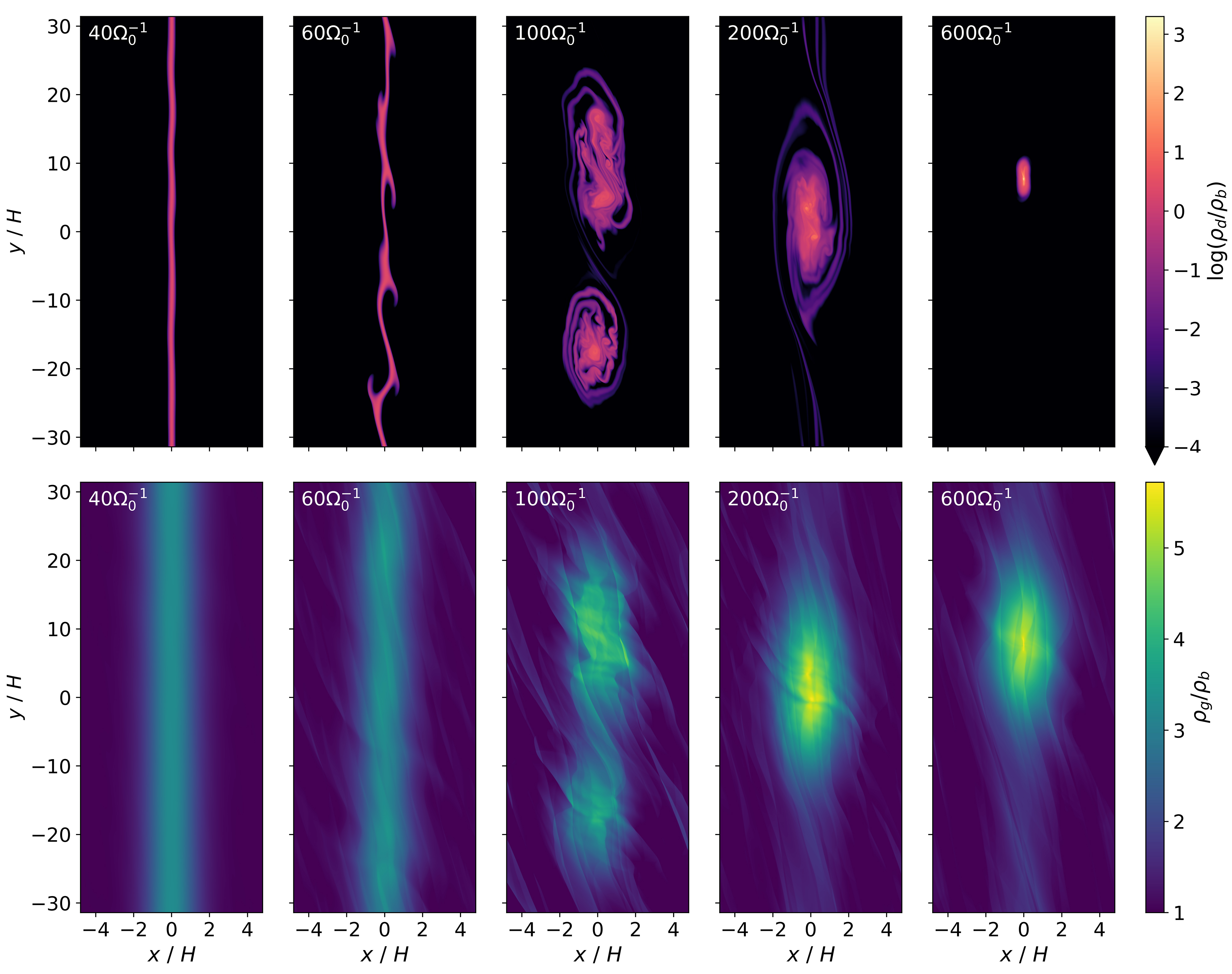}
    \caption{Snapshots in the ``sharp bump'' run (the Type~\RNum{1} DRWI being unstable) of the dust and gas density after inserting the perturbation. The time is annotated on the top left of each panel. The top row is colored in the same logarithmic scale and the bottom in the same linear scale.}
    \label{fig simulation_Type1}
\end{figure*}

In this section, we qualitatively verify the two types of DRWI and investigate their evolution in the nonlinear regime. We use the multifluid dust module in \texttt{Athena++} \citep{Stone2008,Huang22}. Our numerical setup keeps the formulation in Section~\ref{subsec pressure bump} and \ref{subsec dust diffusion and concentration}, treating the gas and dust as two fluids in a local shearing sheet and establishing the external forcing to maintain the pressure bump. Differently, though, we adopt the standard Navier-Stokes viscosity in \texttt{Athena++}. The external forcing is then modified to satisfy the new equilibrium equation in place of Equation~(\ref{eq f0 v0y}):
\begin{equation}
    f_0(x) + \frac{1}{\rho_{g0}}\frac{\partial}{\partial x}\left[\rho_{g0}\nu\frac{\partial}{\partial x}\left(v'_{0y}-\frac{3}{2}\Omega_0x\right)\right] = 0\ ,
\end{equation}
which gives the form of $f_0(x)$ implemented in the simulation:
\begin{gather}
    f_0(x) = -\frac{A}{2}\alpha c_s\Omega_0\left(\frac{H}{\Delta w}\right)^3\left(-\frac{x^3}{\Delta w^3}+\frac{3x}{\Delta w}\right)\exp\left(-\frac{x^2}{2\Delta w^2}\right) + \nonumber \\ -\frac{A^2}{2}\alpha c_s\Omega_0\left(\frac{H}{\Delta w}\right)^3\left(-\frac{x^3}{\Delta w^3}+\frac{x}{\Delta w}\right)\exp\left(-\frac{x^2}{\Delta w^2}\right) + \nonumber \\
    -\frac{3A}{2}\alpha c_s\Omega_0\left(\frac{H}{\Delta w}\right)\left(\frac{x}{\Delta w}\right)\exp\left(-\frac{x^2}{2\Delta w^2}\right)\ .
\end{gather}
Also, we use the conventional treatment that includes the dust concentration diffusion in the continuity equation and does not absorb ${\bm v}_{\rm dif}$ into ${\bm v}_d$ \citep[][Equation (A1)]{Huang22}. 

We expect no substantial deviation in terms of linear evolution where viscosity and dust diffusion processes are unimportant. However, the equilibrium profile %becomes different 
is slightly influenced by the different setups in the simulation compared to our analytical derivation (mainly due to the use of two-fluid instead of single-fluid formalism), and we reach the steady state by a preliminary axisymmetric run. For a given set of parameters, we use a sheet size of $x\times y = 8\Delta w\times0.078125\pi H$ with $1024\times12$ cells. Initially, we set $\rho_g$ as in Equation~(\ref{eq pressure bump profile}), $v'_{gy}$ and $v'_{dy}$ as in Equation~(\ref{eq steady state vy}), and $v'_{gx}=v'_{dx}=0$. The initial dust density is set as a Gaussian whose height satisfies $f_{g\rm min}$ and whose width ensures that the total dust weight equals to that calculated in Section~\ref{sec equilibrium}. After the equilibrium is reached, we scale up the simulation with a sheet size of $x\times y = 8\Delta w\times20\pi H$ with $1024\times3072$ cells, which has the same resolution as the preliminary run and is enough to capture a linear wave of $k=0.1H^{-1}$. 

To verify the Type~\RNum{2} DRWI, we use the parameter $A=0.8, \Delta w/H=1.5, St=0.03, \alpha=1\times10^{-4},$ and $\overline{f_g}=0.980$ (or equivalently $f_{g\rm min}=0.536$). Our linear calculations predict that this system is stable to the Type~\RNum{1} DRWI while $\gamma_{m*}=0.03\Omega_0$ for Type~\RNum{2}. We preliminarily run this system for 10000 $\Omega_0^{-1}$, after which the time derivative of the dust density is below $10^{-6}\rho_b\Omega_0$. Then, we insert random noise of amplitude $0.01c_s$ into the gas velocity and run the full-scale simulation. We term it the ``mild bump'' run. This run with the dust turned off is tested to be stable to the RWI. We also study a ``sharp bump'' run where the Type~\RNum{1} DRWI dominates. %operates. 
The parameter is $A=1.2, \Delta w/H=1.2, St=0.03, \alpha=1\times10^{-4},$ and $\overline{f_g}=0.980$ (or equivalently $f_{g\rm min}=0.541$), to which our linear calculations predict that $\gamma_{m*}=0.12\Omega_0$  and $0.10\Omega_0$ for the Type~\RNum{1} and \RNum{2} DRWI respectively. This run follows the same procedures as described above. We will describe the two runs separately in the following subsections.

\subsection{The mild bump run: development of the Type~\RNum{2} DRWI %only the Type~\RNum{2} DRWI unstable
}
\label{subsec mild bump run}
The evolution of the ``mild bump'' run ($A=0.8, \Delta w/H=1.5$) %dust density
is shown in Figure~\ref{fig simulation} and Figure~\ref{fig simulation_travel}. % As expected, whereas the same pressure bump \textbf{with the dust turned off} is stable to the RWI \textbf{in simulations}, 
The dust-gas instability starts to evolve into the nonlinear regime when $t\gtrsim100\Omega_0^{-1}$. Before that, the dominant azimuthal wavenumber is approximately $k=0.3$. The gas and dust density perturbations are anti-correlated. Moreover, Figure~\ref{fig simulation_travel} shows azimuthally travelling gas density perturbation patterns with $v_y\simeq\pm0.9c_s$. These are characteristic of the Type~\RNum{2} DRWI. The upward-moving patterns at $x<0$ correspond to %major Rossby wave of 
the Type~\RNum{2} DRWI mode with $\omega_{rm}>0$, while the downward-moving patterns at $x>0$ the mode %Rossby wave 
with $\omega_{rm}<0$.
% in accord with the Type~\RNum{2} DRWI. The perturbation patterns appear to be standing because the Type~\RNum{2} DRWI with $\omega_{rm}>0$ and $\omega_{rm}<0$ develops simultaneously.
% Asymmetric noise (e.g., larger initial perturbation amplitude for $x>0,y>0$), or even numerical noise at MeshBlock\footnote{The MeshBlock is the unit subdomain for parallel computation in \texttt{Athena++}. A ghost zone of two cells is used at the MeshBlock boundaries to minimise, although not eliminate, numerical noise.} boundaries) 
% would give a mixture of standing and travelling patterns of the Type~\RNum{2} DRWI. 
%In fact, if we evolve the full-scale system without inserted noise and with the $y$-direction decomposed into multiple MeshBlocks, the unit subdomain for parallel computation in Athena++, numerical noise at the MeshBlock boundaries may accumulate and grow into travelling patterns of the Type~\RNum{2} DRWI.

Upon entering the nonlinear regime after $t\geq200\Omega_0^{-1}$, the dust ring is deformed into anticyclonic vortices, where dust becomes more and more concentrated into scales $\lesssim H$, with maximum $\rho_d$ constantly increasing. These dust-gathering vortices correspond to negative one-fluid vortensity perturbation seen in the bottom panels of Figure~\ref{fig simulation}. Interestingly, their locations %The location of these dust-gathering vortices 
seem unrelated to the sign of the pressure perturbation: they may stay in either positive or negative pressure extrema or no extremum at all. More precisely, whereas the $\rho_{g1}$ patterns are still travelling at an azimuthal velocity comparable to the sound speed\footnote{The pattern of $\rho_{g1}$ at $t=485\Omega_0^{-1}$ that appears as an extended density maxima in fact consists of two traveling waves, %The apparent vortex pattern of $\rho_{g1}$ at $t=485\Omega_0^{-1}$ is still non-stationary, 
with the left and right halves to separate soon.}, the dust vortices become almost stationary. %Meanwhile, the gas density perturbations show \textbf{a tendency to merge into one large anticyclone, as is usually observed in the RWI, although} %both large-scale vortices, resembling the linear instability, e.g., the negative $\rho_{g1}$ region for $-25H<y<0$ at $200\Omega_0^{-1}$ corresponding to a gauze of positive $q_1$ and 
%fine spiral waves %of low pressure stemming from the dusty vortices are also seen.
Moreover, the gas density perturbation gradually decays in magnitude (compare the maximum $|\rho_{g1}|$ at $t=200\Omega_0^{-1}$ and $t=485\Omega_0^{-1}$),
%\xb{saturating at a rather low level with $\rho_{g1}/\rho_b\lesssim10\%$}, in contrast
reaching a characteristic level of $\rho_{g1}/\rho_b\lesssim10\%$, in contrast
%contrary 
to the still-concentrating dust vortices.
In the meantime, the system is accompanied by numerous fine-scale density waves, presumably triggered by local dust concentrations.

The dust is continuously gathered and dusty vortices merge into each other.  Several hundred $\Omega_0^{-1}$ after we insert perturbation, one or two dust-loaded vortices are left with maximum dust density $\rho_{d\rm max}/\rho_b$ ranging from several tens to more than one hundred. Although this slightly falls short of the density threshold for gravitational collapse \citep[e.g., $\rho_{d\max}/\rho_b\gtrsim200$ in typical outer disc conditions; see Equation (16) and Section 5.1 in][]{Xu22a}, dust vertical settling is not included in this work. The equilibrium dust scale height can be estimated by $H_d/H=\sqrt{\alpha/St}=0.06$ \citep{Dubrulle1995}. Under the assumption that the vertical dimension does not impact adversely on dust concentration in 2D, this %dust settling of a factor of several
will suffice to lead to planetesimal formation by clumping even if dust mass loading itself does not further promote settling \citep[which could be observed in 2D axisymmetric and 3D simulations; see][]{Lin2019,Xu22b}. Moreover, the dust in each of these vortices is likely massive enough for the self-gravity to overcome turbulent diffusion, for which \cite{Klahr2020} derived a critical minimum dust cloud radius $l_c/H=\sqrt{\delta/9St}$, where $\delta\equiv D/c_sH$ is the dimensionless diffusivity. In our problem ($\delta<\alpha=1\times10^{-4}, St=0.03$), $l_c<0.02H$. In comparison, the typical length scale of the dust vortices at $t=485\Omega_0^{-1}$, measured in regions with $\rho_d\geq12\rho_b$ (so that the actual dust-to-gas density ratio reaches $\sim200$ after accounting for dust settling), reaches $\sim0.06H$ in $x$ and $\sim0.6H$ in $y$. 

The total dust mass in each vortex that is gravitationally bound is estimated as
\begin{equation}
    m_{d,\rm vortex} \simeq 0.2\left(\frac{R_0}{30{\rm au}}\right)^2\left(\frac{H/R_0}{0.1}\right)\left(\frac{\rho_b}{1\rm g/cm^2}\right) {\rm M_\oplus}\ .
\end{equation}
This is larger than the mass of typical
%upper limits of the initial mass function of 
planetesimals and may already be considered to be planetary embryos if it collapses into a single object.
%but the amount of dust directly involved in the clumping process remains uncertain as the vertical dimension is unresolved.
On the other hand, we caution that our study lacks the vertical dimension and does not include self-gravity, and thus the fate of such dust clumps remains to be revealed. In the absence of self-gravity
%gravitational collapse, 
they are quickly dissipated after several tens of $\Omega_0^{-1}$, although they re-emerge $\sim400\Omega_0^{-1}$ later when the dust is spread back into the ring and then triggers a new round of the Type~\RNum{2} DRWI. Moreover, the nonlinear outcome of the DRWI also likely depends on the nature of disc turbulence where our treatment is highly simplified. We speculate that the system may instead form multiple planetesimals (as suggested in \citealp{Xu22b}), especially as there is no strong gas vortex that may tend to gather all nearby dust towards a common collapse site at its center. %to enforce dust concentration.
%since dust in the entire bump is mostly gathered in very few dust vortices, each of such vortices is \xb{likely} massive enough for the self-gravity to overcome turbulent diffusion \citep{Klahr2020}.

One important characteristic of dust clumping in the Type~\RNum{2} DRWI is that the dust ring is retained. This is primarily because of the weak density perturbations in the gas (as opposed to the Type~\RNum{1} case to be discussed next). As a result, dust concentration may not be easily identified observationally, especially when the dust ring is optically thick. On the other hand, the Type~\RNum{2} DRWI does induce certain level of azimuthal asymmetries in the form of non-uniform dust distribution and/or corrugation. For example, in the last column in Figure~\ref{fig simulation}, the large-scale dust mass azimuthal contrast (estimated as the dust density within the most massive quarter of the $y$ range divided by that within its opposite quarter) is $\sim3$. %comparable to the observed intensity contrast of weakly asymmetric disc rings \citep{vanderMarel21} and may be detectable by ALMA if the dust is not completely optically thick.} 
Azimuthal assymetries in dust rings up to similar levels of contrast have been seen in a number of systems such as DM Tau \citep{Hashimoto21} and LkCa 15 \citep{Facchini2020,Long2022}, and they are suggested to be common \citep{vanderMarel21}. Such azimuthal asymmetries could serve as indirect evidence for the presence of Type~\RNum{2} DRWI and hence dust clumping. Our results further suggest that peaks in the azimuthal brightness profile in dust ring are not necessarily co-spatial with the azimuthal gas pressure maxima.

%Overall, our results imply that ringed outer Class \RNum{2} discs \citep[e.g.][]{Pinte2016,Isella16,Villenave2022} may witness robust dust clumping via the Type~\RNum{2} DRWI even in the presence of turbulence, although details of this scenario await 3D simulations. %may provide sites where, even in the presence of turbulence, dust clumping occurs robustly via the Type~\RNum{2} DRWI.

\subsection{The sharp bump run: dominance of the Type~\RNum{1} DRWI} %the Type~\RNum{1} DRWI more unstable than Type~\RNum{2}}
%Sharper pressure bump profiles may render the Type~\RNum{1} DRWI more unstable than Type~\RNum{2} and more readily induce dust overdensity. %Changing the pressure bump amplitude to $A=1.5$ while controlling other parameters (keeping $\overline{f_g}=0.98$ but modifying $f_{g\rm min}$ to 0.55), 
In the ``sharp bump'' run ($A=1.2, \Delta w/H=1.2$), shown in Figure~\ref{fig simulation_Type1}, we observe that dust and gas vortices develop and merge, forming one single anticyclonic gas vortex at saturated state approximately $200\Omega_0^{-1}$ after the initial perturbation.
%saturated gas anticyclone.
The overall evolution process closely resembles the development of the standard RWI in dusty discs \citep{Meheut2012,Zhuetal2014}, and as a result, all the dust in the ring concentrates towards the gas vortex center.
This is clearly distinct from the mild bump run where
%the gas density perturbations reduce to a rather low level
%recede prematurely 
%the dust vortices do not follow the pressure maxima. 
the dust ring is retained thanks to low levels of gas perturbation while developing dust concentration and clumping within the ring. In our sharp bump run,
%Then, 
the contraction of the dust in the vortex continues to develop %even as the gas vortex stops developing, 
(while the gas vortex has already saturated),
eventually saturating at $t\simeq600\Omega_0^{-1}$ with $\rho_{d\rm max}/\rho_b\simeq2\times10^3$. %$\rho_{d\rm max}/\rho_b\simeq700$. 
%This density does not even need any settling to collapse and lead to planetesimal formation \textbf{or even directly towards a planet} (again, assuming that the vertical dimension does not hinder the development of the compact dust vortex).
% \xb{We observe that the gas vortex has} long lifetimes of at least $1000\Omega_0^{-1}$, although the peak dust density fluctuates \textbf{between $10^1\rho_b$ and $10^3\rho_b$ later on}.
This peak dust density is significantly higher than the Type-\RNum{2} case, and we can also estimate the total dust mass gathered in the vortex to be
%Dust in the entire bump is
%mostly gathered, with a substantial total mass estimated as
\begin{equation}
    m_{d,\rm vortex} \simeq 4\pi R_0\Delta w \overline{\rho_d} = 4\left(\frac{R_0}{30{\rm au}}\right)^2\left(\frac{H/R_0}{0.1}\right)\left(\frac{\rho_b}{1\rm g/cm^2}\right) {\rm M_\oplus}\ ,
\end{equation}
where $\rm M_\oplus$ is the Earth mass.
This is also much higher than the Type~\RNum{2} counterpart, and is in the mass range of planetary embryos. Again, future 3D studies including self-gravity is needed to reveal the fate of such dust clump. Also, we observe that the gas vortex has long lifetimes of at least $1000\Omega_0^{-1}$, although the peak dust density fluctuates between $10^1\rho_b$ and $10^3\rho_b$ after saturation. The lifetime of dust-laden vortices in 2D has been studied extensively \citep{Chang2010,Fu2014b,Crnkovic-Rubsamen2015,Lovascio2022} and depends on factors such as the initial dust-to-gas mass ratio, viscosity, dust feedback and dust grain size. The vortex in our sharp bump run is consistent with \cite{Lovascio2022} with similar spatial scale, dust size and total dust-to-gas mass ratio (lifetime $\sim10^3\Omega_0^{-1}$ there). %Works on planet-induced  vortices with dust feedback \citep{Li2020,Hammer2021} suggested similar longevity.
Works on planet-induced \citep{Li2020,Hammer2021} or 3D \citep{Lyra2018,Hammer2023} vortices with dust feedback suggested similar longevity, although dust settling could disturb the midplane vortex structure.
%Different from \cite{Fu2014b}, the vortex in our simulation survives in spite of dust feedback, presumably because the compact dust spatial distribution barely influences the gas dynamics away from the vortex center. 
Since the dust is well confined in the long-lived gas vortex, we speculate that the stronger dust clump resulting from Type~\RNum{1} DRWI is more likely to form massive planetesimals/planetary embryos.

% Recently, \cite{Regaly2023} investigated the excitation of the RWI at the edge of the dead zone. Obviating the conventional requirement of a steep viscosity gradient, the RWI was found to operate in a smooth transition of viscosity if dust concentration reduces local turbulence. The RWI-induced vortices then collects dust efficiently, sometimes leading to dust overdensity capable of gravitational collapse. Such "planetary nurseries" are very similar to the dust vortices formed via the DRWI in a pressure bump and are presumably governed by analogous mechanisms.

One uncertainty in our scenario is that maximum dust density in dusty vortices is very sensitive to the prescription %formulation
of dust diffusivity $D$, which is currently given as a function of the local dust and gas density. If we assume no weakening of turbulent diffusion due to the dust mass loading, $\rho_{d\rm max}$ will reduce by approximately one order of magnitude for both the mild and sharp bump runs. % (this also applies to the Type~\RNum{1} DRWI). %Previous works \citep[e.g.][]{Sano2000,Okuzumi2009,Dzyurkevich2013} found that dust weakens MRI turbulence by capturing free electrons and ions on its surface. Tiny grains are expected to have the strongest recombination effect due to their large surface area. Independent from this, 
In previous works, enhanced dust mass loading with dust feedback is found to reduce turbulent diffusion in the magneto-rotational instability (MRI) turbulence %\textbf{from non-ideal MHD effects}
\citep{Xu22b}. SI-induced turbulent diffusivity was also found to be sensitive to the dust-to-gas ratio
%and the domain size 
\citep{Schreiber2018}.
%More realistic modelling of the dust diffusivity must be based on a deeper understanding of the interaction between MRI turbulence and dust particles across various length scales. 
Further investigations in dust diffusivities within dust clumps are needed that incorporates more realistic background gas turbulence.
%The real scenario, then, is complicated by the dust size distribution and our insufficient understanding of the interaction between MRI turbulence and dust particles in non-ideal MHD. 

%For whichever formulation of $D$, the dusty vortices are inevitably destructed and the dust spread over the pressure maximum roughly $100\Omega_0^{-1}$ after reaching $\rho_{d\rm max}$. This is expected in view of the very large dust concentration gradient and hence strong diffusion at the vortices. This implies that, if a dusty vortex fails to reach the collapsing threshold, the dust can start over from a ring and attempt another round of mass loading. Such recurrence is confirmed in simulations, which suggests robust probability of dust clumping in turbulent dust-trapping rings.

% is simplified from \cite{Xu22b} in that we treat the gas and dust as two fluids in a local shearing sheet setting and mimic the non-ideal MHD effect with viscosity. 

%\subsection{Planetesimal/embryo formation: big or small?}

%\subsection{From dust-trapping rings to dust asymmetry}

\section{Summary and discussion}
\label{sec conclusions}
We introduce a physically-motivated local shearing sheet%two-dimensional 
model of turbulent dust-trapping rings in PPDs. 
%To mimic realistic disc conditions where the pressure bump forms from planet-induced torque or magnetic flux concentration in the presence of MRI turbulence, we establish an equilibrium between a generic external forcing on the gas and turbulent viscosity parameterized by $\alpha$. The dust, modelled with the gas as a single fluid and with feedback considered, drifts towards the pressure maximum but meanwhile experiences concentration diffusion, therefore forming an equilibrium dust bump with a finite width. %We self-consistently establish the pressure maximum and the dust ring in equilibrium, the former via a balance of external forcing versus viscosity and the latter via dust drift versus turbulent diffusion.
We establish a pressure bump by implementing a forcing term that mimics torques that drive ring formation (e.g., by planets, or magnetic flux concentration), balanced by viscosity that mimics disc turbulence. The dust is modeled as a fluid including backreaction, which also evolves into an equilibrium dust bump profile by balancing radial drift towards pressure maxima and turbulent diffusion.
We aim to identify linear instabilities that operate and potentially lead to planetesimal formation in this realistic setting.

We find two types of instabilities, which we term the DRWI. Type~\RNum{1} is generalised from the standard RWI while Type~\RNum{2} is first identified here. The Type~\RNum{1} DRWI, characterised by a vanishing phase velocity and (anti-)symmetric eigenfunction patterns, dominates in relatively sharp pressure bumps and/or bumps with low dust content. In contrast, the Type~\RNum{2} DRWI travels along the $y$ axis, has different perturbation magnitudes on either side of the pressure bump, and operates in relatively mild and dusty bumps. Its maximum growth rate is largely determined by the equilibrium gradients of the gas and dust density. %fraction. %We also show that the linear RWI is insensitive to turbulence for $\alpha\lesssim3\times10^{-3}$. 

The standard RWI is understood in terms of conservation of vortensity. However, our vortensity source analysis highlights the effective baroclinity in the dust bump, which only consumes the vortensity budget in the Type~\RNum{1} DRWI but mainly contributes to the vortensity growth in Type~\RNum{2}. Therefore, we believe that vortensity advection, the incentive of the classical RWI, also accounts for the growth of the Type~\RNum{1} DRWI, while both the advection and baroclinity drive the Type~\RNum{2} DRWI. %We further interpret the mechanism of the \textbf{latter} %Type~\RNum{2} DRWI
%as two travelling Rossby waves coupled with a \textbf{``dust wave''. The propagation of the major and minor Rossby waves is synchronised by the combined factors of vortensity advection, differential background shear and the different perturbation magnitudes, while the concurrence between the Rossby waves and the dust wave can be understood in terms of the density flow or the vortensity budget. The vortensity advection and the baroclinity are mainly responsible for the growth of the Rossby waves and the dust wave respectively.} %wave in the dust bump.  and further speculate the existence of resonance between the gas and dust waves. 

The two types of DRWI are qualitatively verified in simulations, and they show distinct nonlinear outcomes with major observational implications. 
%A sharp bump enables the 
In general, Type~\RNum{1} DRWI dominates in the presence of a sharp bump. This yields a standard gas vortex characterized by a pressure maximum in the center, and it traps and concentrates all the dust originally in the ring. On the other hand, in a mild bump, the Type~\RNum{2} DRWI operates and develops into sub-$H$-sized dust anticyclones, whereas the gas density 
only shows weak perturbations. This allows the dust ring to be largely preserved, while exhibiting azimuthal asymmetries.
%perturbation does not form nonlinear vortices. %which become largely disengaged from the gas perturbation patterns that remain linear. 
In both cases, %saturated gas vortex, which traps and concentrates dust in its center. In either case,
the non-linear evolution of the DRWI triggers significant dust mass loading in the form of dust vortices, which hold potential for dust clumping and hence planetesimal formation % and their non-linear evolution promises compact dust vortices that likely trigger dust clumping. 
or direct formation of planetary embryos.

\subsection{Discussion}
\label{subsec discussion}
%\textbf{After the summary above, we discuss the prospect of the two types of the DRWI towards a variety of azimuthally asymmetric dust substructures from the axisymetric dust-trapping ring.}
The DRWIs are likely closely related to certain instability phenomena in previous simulations of pressure bumps with dust feedback. For example, they provide a potential explanation to simulations in \cite{Xu22b}, where the ring could be broken into dusty non-axisymmetric filaments, qualitatively similar to the nonlinear patterns of the Type~\RNum{2} DRWI. Also, at the outer edge of a dead zone, the steep increase of turbulent viscosity leads to radial local gas overdensity. While a sufficiently narrow transition width induces formation of large-scale gas vortices with dust concentrating inside \citep[ascribed to the RWI,][with the dust content found to impede vortex formation and dust concentration]{Miranda2017}, a smoother transition produces no large-scale gas vortices but dust clumps of scales $\lesssim H$ \citep{Huang20}. The edge of a gap opened by a massive planet is subject to similar instabilities, with large-scale gas vortices emerging only in the absence of dust backreaction while non-negligible dust concentration instead encourages formation of small dust vortices \citep{Yang20}. 3D simulations in VSI-turbulent pressure bumps also found a tendency of dusty vortex formation towards axisymmetric rings for increasing average dust-to-gas mass ratio or the Stokes number \citep{Lehmann2022}.
%While the link between the DRWI and these findings deserves further investigation,
While further investigation is needed, the Type~\RNum{2} DRWI offers a viable physical explanation of these findings.
%dust-trapping vortices in general serve as a potential path \xb{to form structures that range} from axisymmetric dust distribution to large-scale azimuthal concentration of solid material.
%dust-trapping rings likely serve as a potential avenue to form %a range of 
%smaller-scale substructures in general that range from relatively low-contrast azimuthal asymmetries to dust-laden vortices with \xb{strong} dust clumping.

%\textbf{The largest-scale substructures discussed above may be directly observable by ALMA. 

It is worth considering how our local analyses and simulations of the DRWI %on an isolated radial pressure maximum 
fit in realistic global disc structures, which has geometric curvature as well as a background pressure gradient. We expect the instability to be qualitatively robust in the presence of the disc curvature since we recover the classical RWI. \citet{Pan2020} noted that sustaining Rossby waves requires the presence of the second derivative of the background shear (of order $\Omega_0/R_0$), which the standard shearing sheet does not capture in background equilibrium. This is resolved as we form a pressure bump that provides strong radial structure ($\partial^2v'_{0y}/\partial x^2\sim\Omega_0/H\gg \Omega_0/R_0$). Further, our physical ingredient analysis suggests that the DRWI is probably insensitive to the particular bump shape with or without a background pressure gradient, as long as the pressure maximum concentrates dust to serve as the vortensity source and the two bump flanks provide equilibrium vortensity slopes. %(although perhaps with more radially asymmetric waveforms if the two slopes are different). 
On the other hand, a background pressure gradient can induce a net dust radial flux if no other dust trap exists outside the bump in question. The dust drift could trigger the two-dimensional SI in small scales \citep{Pan2020} that may coexist and/or interact with the DRWI. % We also note that, while \cite{Pan2020} pointed out that Rossby waves in a 2D, incompressible, homogeneous disc require the second derivative of the background shear (of order $\Omega_0/R_0$) which the shearing sheet does not capture, our pressure maximum provides strong radial structure ($\partial^2v'_{0y}/\partial x^2\sim\Omega_0/H\gg \Omega_0/R_0$) and thus is not subject to this issue. }

The ubiquity of dust-trapping rings and the relatively rarer occurrence of high-contrast asymmetries such as arcs and crescents \citep{Andrews20} suggest that %such asymmetries is either hard to form or short-lived. The first possibility suggests that
most of the rings are likely moderate in sharpness: they must trap dust effectively in the presence of background radial drift while still stable to the vortex-forming Type~\RNum{1} DRWI. This implication is related to the recent global study by \cite{Chang2023}, % on the stability of axisymetric dust traps to the classical RWI, 
which showed that isothermal axisymetric pressure maxima remain (classical-) RWI-stable for a reasonably large range of bump widths. %, while somewhat tighter constraints are placed on adiabatic bumps or discs with large background aspect ratios $H/R$. 
On the other hand, weak-to-modest level of azimuthal asymmetry in dust rings appears to be common \citep{vanderMarel21}. This is suggestive that the Type~\RNum{2} DRWI likely operates and leads to dust clumping while preserving the overall morphology of the dust rings. Another possibility is that dust-laden vortices do form but quickly die out, although the exact lifetime is model-sensitive \citep[e.g.,][]{Fu2014a,Fu2014b,Rometsch2021,Hammer2023}.
%Our analysis indicates that dust content likely considerably changes the ring stability, with the further complexity that dust asymmetry produced by the Type~\RNum{2} DRWI may or may not be detected.

The dust-trapping ring rests in a broader context of spatial and size evolution of solids in PPDs. The Type~\RNum{2} DRWI favors relatively large particles with $St=10^{-2}$--$10^{-1}$, consistent with upper bounds of drift-limited dust size in typical conditions in outer PPDs \citep{Birnstiel2012}. The ring is found to further enhance the average dust size by alleviating drift and fragmentation barriers \citep{Li2019,Laune2020}, thus likely encouraging the onset of the DRWI. It is conceivable that the pressure bump gathers and nurses the dust progressively over drift and coagulation time-scales until mature for the instability.

%\textbf{While we did not find short-wavelength instabilities in the ring, it remains possible that the DRWI triggers subsequent instabilities locally. In particular, the SI potentially operates in or near the dust clumps in the nonlinear stage of the DRWI, which is accessible to reduced turbulence (due to low $f_g$) and local pressure gradient (due to perturbation). In our mild bump run, for example, the clumpy region $(0<x<0.05H, 22.5H<y<23.3H)$ at $t=485\Omega_0^{-1}$ has $f_g\sim0.1$ (hence $D/c_sH\sim10^{-5}$) and is subject to a local pressure gradient $\eta\sim10^{-2}$ where $\eta\equiv-\partial P/\partial R\cdot1/(2\rho_g\Omega_0^2R)$ \citep{Youdin2007}.}

The formation of planetesimals/embryos in the pressure bump bears on their later evolution paths. For instance, formation models built out of a self-interacting planetesimal ring (regardless of their origin) can be compatible with the formation scenario of terrestrial planets and super-Earths \citep{Woo2023,Batygin2023}. A dust-trapping ring also likely allows pebble accretion to operate efficiently that leads to rapid planet assembly \citep{Jiang2023}.
%Efficient pebble accretion and hence planet formation is expected in light of the radial dust concentration. Recently, \cite{Jiang2023} showed that the ring could serve as a factory assembly where one planet after another grew in and then migrated off the ring, with the latter being replenished with dust via inward radial drift. 
The fact that Type~\RNum{2} DRWI leaves the pressure bump largely intact
%and is compatible with this scenario, which may develop into a system of multiple planets inside the ring. 
likely favors the production of a planetesimal ring and/or direct formation of embryos which fit into the scenarios above.
Our study bears a number of simplifications and caveats that deserve future studies. Among them includes the local treatment of the isolated pressure bump, as discussed above.
Moreover, our 2D study also neglects vital 3D processes such as dust settling and vertical gas flow, which may alter the linear DRWI and its non-linear evolution. %and the vertical dimension may alter the morphology and linear/non-linear evolution of the DRWI. 
We approximate the dust-gas mixture with a single fluid, although two-fluid simulations largely agree with the calculations. Self-gravity is ignored throughout this work, and thus planetesimal/embryo formation is only inferred. Also, our treatment of the MRI turbulence as a diffusive process and of the dust diffusivity as a simplistic function %of the viscous parameter and the gas fraction 
calls for first-principle insights in modelling the MRI and/or other forms of turbulence. We intend to generalise our work to 3D in the future, with a more realistic and thorough consideration of physical processes. Despite current limitations, our work pioneers a rigorous effort to uncover fundamental dynamical scenarios that bridge widespread observed dust structures to the crucial evolutionary stage of solid material towards future planets.

% The DRWI produces low-wavenumber dust and gas structures, which do not adequately explain the small-scale dust patterns in 3D pressure bump simulations with non-ideal MHD turbulence \citep{Xu22b}. 

% Caveats: shearing sheet vs disc geometry (polar, vertical); one-fluid formulation; turbulence mimicked by the viscosity

% Future work

%Vortensity in a shearing sheet is defined by (useful derivation: http://www.damtp.cam.ac.uk/user/gio10/dad_notes2.pdf, p.39, although we need a more authoritative reference)

\section*{Acknowledgements}

We thank the anonymous referee for detailed comments and suggestions that helped improve the clarity of this paper. We thank Pinghui Huang for instructions on the multi-fluid dust module in Athena++, and Cong Yu, Min-Kai Lin and Marius Lehmann for useful discussions. We also acknowledge the Chinese Center of Advanced Science and Technology for hosting the Protoplanetary Disk and Planet Formation Summer School in 2022 where part of this work is conducted.
This work is supported by the National Science Foundation of
China under grant No. 12233004, and the China Manned Space
Project, with No. CMS-CSST-2021-B09. We acknowledge the Tsinghua Astrophysics High-Performance Computing platform at Tsinghua University for providing computational and data storage rsources that have contributed to the research results reported within this paper.

Software: \texttt{NumPy} \citep{harris2020numpy}, \texttt{SciPy} \citep{virtanen2020scipy}, \texttt{Matplotlib} \citep{hunter2007matplotlib}, \texttt{Findiff} \citep{findiff}, \texttt{Athena++} \citep{Stone2020,Huang22}

% The Acknowledgements section is not numbered. Here you can thank helpful colleagues, acknowledge funding agencies, telescopes and facilities used etc. Try to keep it short.

%%%%%%%%%%%%%%%%%%%%%%%%%%%%%%%%%%%%%%%%%%%%%%%%%%
\section*{Data Availability}
Data of the linear analyses and simulation in this paper are available upon request to the authors.
 
% The inclusion of a Data Availability Statement is a requirement for articles published in MNRAS. Data Availability Statements provide a standardised format for readers to understand the availability of data underlying the research results described in the article. The statement may refer to original data generated in the course of the study or to third-party data analysed in the article. The statement should describe and provide means of access, where possible, by linking to the data or providing the required accession numbers for the relevant databases or DOIs.


\bibliographystyle{mnras}
\bibliography{bib} % if your bibtex file is called example.bib

%%%%%%%%%%%%%%%%%%%% APPENDIX %%%%%%%%%%%%%%%%%%%%

\appendix

\section{Derivations}
\subsection{Perturbation equations}
\label{append perturbation}

To derive the system of perturbation equations (\ref{eq perturb matrix}), we first use the definitions of $f_d$, $\bm{v}_{\rm dif}$, $D$ and $\rho$ in Equations~(\ref{eq vdif})(\ref{eq D def})(\ref{eq one fluid pre def}) to obtain their perturbed forms, also denoted with a subscript $``_1"$:
\begin{gather}
    f_{d1} = -f_{g0}\mathfrak{f}_{g1}\ , \\
    v_{{\rm dif}1x} = \left(\frac{D_0}{f_{d0}}\frac{d}{dx} + \frac{f_{g0}}{f_{d0}}v_{{\rm dif}0x} + \frac{D_0'f_{g0}v_{{\rm dif}0x}}{D_0}\right)\mathfrak{f}_{g1} \ , \\
    v_{{\rm dif}1y} = \frac{ikD_0}{f_{d0}}\mathfrak{f}_{g1} \ ,\\
    D_1 = D'_0f_{g0}\mathfrak{f}_{g1} \ , \\
    \frac{\rho_1}{\rho_0} = \mathfrak{p}_1 - \mathfrak{f}_{g1} \ ,
\end{gather}
where $D_0'\equiv[dD(f_g)/df_g]_{f_g=f_{g0}}$. 

We will heavily use equilibrium solutions in the detailed form of the perturbation equations. We avoid explicitly using $x$-derivatives of $P_0$, $f_{g0}$ and $\rho_0$ to circumvent numerical errors, instead substituting them with $v_{{\rm dif}0x}$:
\begin{gather}
    \frac{1}{P_0}\frac{dP_0}{dx} = -\frac{v_{{\rm dif}0x}}{c_s^2t_s}\ , \\
    \frac{1}{f_{g0}}\frac{df_{g0}}{dx} = \frac{f_{d0}v_{{\rm dif}0x}}{D_0}\ , \\
    \frac{1}{\rho_0}\frac{d\rho_0}{dx} = -(\frac{1}{c_s^2t_s}+\frac{f_{d0}}{D_0})v_{{\rm dif}0x} \ .
\end{gather}
These representations come from Equations~(\ref{eq vdif})(\ref{eq one fluid pre def})(\ref{eq steady state 1}). 

The next step is to linearise the single fluid equations. Substitution of the perturbation variables into Equations~(\ref{eq one fluid den})(\ref{eq one fluid mom})(\ref{eq one fluid pre}) gives
\begin{gather}
    -i\omega\rho_1 + \frac{d(\rho_0v'_{1x})}{dx} + ik(v'_{0y}\rho_1 + \rho_0v'_{1y} -\frac{3}{2}\Omega_0x\rho_1) = 0 \ ,
\end{gather}
\begin{gather}
    -i\omega P_1 + \frac{d(P_0v'_{1x})}{dx} + ik(v'_{0y}P_1 + P_0v'_{1y} - \frac{3}{2}\Omega_0xP_1) = \nonumber \\
    c_s^2t_s\frac{d}{dx}\left(f_{d0}\frac{dP_1}{dx}\right) - c_s^2t_sk^2f_{d0}P_1 + c_s^2t_s\frac{d}{dx}\left(\frac{dP_0}{dx}f_{d1}\right) + \frac{d(f_{d0}P_0v_{{\rm dif}1x})}{dx} + \nonumber \\ ikf_{d0}P_0v'_{{\rm dif}1y} + \frac{d(f_{d1}P_0v_{{\rm dif}0x})}{dx} + \frac{d(f_{d0}P_1v_{{\rm dif}0x})}{dx} \ ,
\end{gather}
\begin{gather}
    -i\omega v'_{1x} + v'_{0y}ikv'_{1x} - \frac{3}{2}ik\Omega_0xv'_{1x} = -\frac{1}{\rho_0}\frac{dP_1}{dx} + \frac{\rho_1}{\rho_0^2}\frac{dP_0}{dx} + \nonumber \\  2\Omega_0v'_{1y} + \nu f_{g0}(\frac{d^2v'_{1x}}{dx^2} - k^2v'_{1x}) + \frac{1}{\rho_0}\frac{d(\rho_{d1}v_{{\rm dif}0x}^2)}{dx} + \nonumber\\ \frac{2}{\rho_0}\frac{d(\rho_{d0}v_{{\rm dif}0x}v_{{\rm dif}1x})}{dx} + \frac{ik}{\rho_0}\rho_{d0}v_{{\rm dif}0x}v_{{\rm dif}1y} - \frac{\rho_1}{\rho_0^2}\frac{d(\rho_{d0}v_{{\rm dif}0x}^2)}{dx}\ , \label{eq begin derive perturb 3}
\end{gather}
\begin{gather}
    -i\omega v'_{1y} + \frac{dv'_{0y}}{dx}v'_{1x} + ikv'_{0y}v'_{1y} - \frac{3}{2}ik\Omega_0xv'_{1y} = -\frac{ik}{\rho_0}P_1 - \frac{1}{2}\Omega_0v'_{1x} +\nonumber\\ \nu f_{g0}(\frac{d^2v'_{1y}}{dx^2}-k^2v'_{1y}) + %\nu\frac{d^2v'_{0y}}{dx^2}f_{g1} + 
    \frac{1}{\rho_0}\frac{d(\rho_{d0}v_{{\rm dif}0x}v_{{\rm dif}1y})}{dx} \ . \label{eq begin derive perturb 4} %+ f_{g1}f_0(x)
\end{gather}
Assuming perturbation variables to be much less than corresponding background values, we have ignored all terms that contain perturbation variables of order higher than one. We also applied the force equilibrium, Equation~(\ref{eq steady state vy}), to cancel out the second $x$-derivative term of $v'_{0y}$ and the external force $f_0$ from the last equation above. 

Then comes substantial work of substitution, expansion of derivatives of products, and algebra. We did not see any shortcut ahead. During these manipulations, we divide the first and second perturbation equations by $P_0$ to simplify the expression. They are cast into
\begin{gather}
    (-i\omega+ikv'_{0y}-\frac{3}{2}ik\Omega_0x)(\mathfrak{p}_1-\mathfrak{f}_{g1}) + \nonumber \\ \left[\frac{d}{dx}-(\frac{1}{c_s^2}+\frac{f_{d0}}{D_0})v_{{\rm dif}0x}\right]v'_{1x} + ikv'_{1y} = 0 \ , \label{eq derive perturb 1}
\end{gather}
\begin{gather}
    \left[-c_s^2t_sf_{d0}\frac{d^2}{dx^2} + (c_s^2t_s\frac{f_{d0}f_{g0}v_{{\rm dif}0x}}{D_0} + f_{d0}v_{{\rm dif}0x})\frac{d}{dx} + \right. \nonumber\\ \left. -i\omega + ikv'_{0y} - \frac{3}{2}ik\Omega_0x + k^2c_s^2t_sf_{d0}\right]\mathfrak{p}_1 + \nonumber \\
    \left[-D_0\frac{d^2}{dx^2} + \left((\frac{D_0}{c_s^2t_s} - f_{g0})v_{{\rm dif}0x} - \frac{2D_0'}{D_0}f_{d0}f_{g0}v_{{\rm dif}0x}\right)\frac{d}{dx} + \right. \nonumber \\ \left. (-f_{g0})\frac{dv_{{\rm dif}0x}}{dx} + (\frac{1}{c_s^2t_s}-\frac{f_{d0}}{D_0})f_{g0}v_{{\rm dif}0x}^2 + k^2D_0 - \frac{D_0'f_{d0}f_{g0}}{D_0}\frac{dv_{{\rm dif}0x}}{dx} + \right. \nonumber \\ \left. \frac{D_0'f_{d0}f_{g0}v_{{\rm dif}0x}^2}{D_0^2}(\frac{D_0}{c_s^2t_s}+2f_{g0}-1) - f_{d0}^2f_{g0}^2v_{{\rm dif}0x}^2\frac{D_0''D_0-D_0'^2}{D_0^3}\right]\mathfrak{f}_{g1} + \nonumber \\
    (\frac{d}{dx} - \frac{v_{{\rm dif}0x}}{c_s^2t_s})v'_{1x} + ikv'_{1y} = 0 \ . \label{eq derive perturb 2}
\end{gather}

Equation~(\ref{eq derive perturb 1}) at present explicitly contains $\omega$ in the coefficients of both $\mathfrak{p}_1$ and $\mathfrak{f}_{g1}$. To calculate eigenvalues $\omega_m$ efficiently (see Appendix~\ref{append perturb}), we put Equation~(\ref{eq derive perturb 2}) in the first line among the four lines in the system (\ref{eq perturb matrix}) and set the second line as the difference between Equation~(\ref{eq derive perturb 2}) and Equation~(\ref{eq derive perturb 1}). This arrangement ensures that $\omega$ only appears explicitly in the diagonal of the system. The third and fourth lines in Equation~(\ref{eq perturb matrix}) are simply %the fully derived 
Equations~(\ref{eq begin derive perturb 3})(\ref{eq begin derive perturb 4}). The final expression condensed into the perturbation system (\ref{eq perturb matrix}) is given below:

\begin{gather}
    \left[-c_s^2t_sf_{d0}\frac{d^2}{dx^2} + \right. \nonumber \\ \left. (c_s^2t_s\frac{f_{d0}f_{g0}v_{{\rm dif}0x}}{D_0} + f_{d0}v_{{\rm dif}0x})\frac{d}{dx} - i\Delta\omega + k^2c_s^2t_sf_{d0}\right]\mathfrak{p}_1 + \nonumber \\
    \left[-D_0\frac{d^2}{dx^2} + \left((\frac{D_0}{c_s^2t_s} - f_{g0})v_{{\rm dif}0x} - \frac{2D_0'}{D_0}f_{d0}f_{g0}v_{{\rm dif}0x}\right)\frac{d}{dx} + k^2D_0 + \right. \nonumber \\ \left. (-f_{g0})\frac{dv_{{\rm dif}0x}}{dx} + (\frac{1}{c_s^2t_s}-\frac{f_{d0}}{D_0})f_{g0}v_{{\rm dif}0x}^2 - \frac{D_0'f_{d0}f_{g0}}{D_0}\frac{dv_{{\rm dif}0x}}{dx} + \right. \nonumber \\ \left. \frac{D_0'f_{d0}f_{g0}v_{{\rm dif}0x}^2}{D_0^2}(\frac{D_0}{c_s^2t_s}+2f_{g0}-1)  - f_{d0}^2f_{g0}^2v_{{\rm dif}0x}^2\frac{D_0''D_0-D_0'^2}{D_0^3}\right]\mathfrak{f}_{g1} + \nonumber \\
    (\frac{d}{dx} - \frac{v_{{\rm dif}0x}}{c_s^2t_s})v'_{1x} + ikv'_{1y} = 0 \ , \label{eq perturb1}
\end{gather}
\begin{gather}
    \left[-c_s^2t_sf_{d0}\frac{d^2}{dx^2} + (c_s^2t_s\frac{f_{d0}f_{g0}v_{{\rm dif}0x}}{D_0} + f_{d0}v_{{\rm dif}0x})\frac{d}{dx} + k^2c_s^2t_sf_{d0}\right]\mathfrak{p}_1 + \nonumber \\
    \left[-D_0\frac{d^2}{dx^2} + \left((\frac{D_0}{c_s^2t_s} - f_{g0})v_{{\rm dif}0x} - \frac{2D_0'}{D_0}f_{d0}f_{g0}v_{{\rm dif}0x}\right)\frac{d}{dx} + \right. \nonumber \\ \left. (-f_{g0})\frac{dv_{{\rm dif}0x}}{dx} + (\frac{1}{c_s^2t_s}-\frac{f_{d0}}{D_0})f_{g0}v_{{\rm dif}0x}^2 + k^2D_0  - i\Delta\omega + \right. \nonumber \\ \left. \frac{D_0'f_{d0}f_{g0}v_{{\rm dif}0x}^2}{D_0^2}(\frac{D_0}{c_s^2t_s}+2f_{g0}-1) - \frac{D_0'f_{d0}f_{g0}}{D_0}\frac{dv_{{\rm dif}0x}}{dx} -  \right. \nonumber \\ \left. f_{d0}^2f_{g0}^2v_{{\rm dif}0x}^2\frac{D_0''D_0-D_0'^2}{D_0^3}\right]\mathfrak{f}_{g1} + \frac{f_{d0}v_{{\rm dif0}x}}{D_0}v'_{1x} = 0 \ ,  \label{eq perturb2}
\end{gather}
\begin{gather}
    \left[\left(c_s^2f_{g0} - f_{d0}v_{{\rm dif}0x}^2\right)\frac{d}{dx}\right]\mathfrak{p}_1 + \nonumber \\ 
    \left[-2D_0v_{{\rm dif}0x}\frac{d^2}{dx^2} + \left(-2D_0\frac{dv_{{\rm dif}0x}}{dx} + (\frac{2D_0}{c_s^2t_s}+3-4f_{g0})v_{{\rm dif}0x}^2 - \right.\right. \nonumber \\ \left.\left.\frac{4D_0'}{D_0}f_{d0}f_{g0}v_{{\rm dif}0x}^2\right)\frac{d}{dx} + 2\Omega_0v'_{0y} + (1-2f_{g0})\frac{dv_{{\rm dif}0x}^2}{dx} - \right. \nonumber \\ \left. (\frac{1-2f_{g0}}{c_s^2t_s}+\frac{f_{d0}}{D_0})v_{{\rm dif}0x}^3 + k^2D_0v_{{\rm dif}0x} - \frac{2D_0'f_{d0}f_{g0}}{D_0}\frac{dv_{{\rm dif}0x}^2}{dx} + \right. \nonumber\\ \left. \frac{2D_0'f_{d0}f_{g0}v_{{\rm dif}0x}^3}{D_0^2}(\frac{D_0}{c_s^2t_s}+f_{g0}) - 2f_{d0}^2f_{g0}^2v_{{\rm dif}0x}^3\frac{D_0''D_0-D_0'^2}{D_0^3} \right]\mathfrak{f}_{g1} + \nonumber \\ \left[-\nu f_{g0}(\frac{d^2}{dx^2} - k^2) - i\Delta\omega \right]v'_{1x} - 2\Omega_0v'_{1y} = 0 \ , \label{eq perturb3}
\end{gather}
\begin{gather}
    ikc_s^2f_{g0}\mathfrak{p}_1 - ikD_0\left[v_{{\rm dif}0x}\frac{d}{dx} + \right. \nonumber\\ \left.  \frac{dv_{{\rm dif}0x}}{dx} - (\frac{1}{c_s^2t_s}+\frac{f_{d0}}{D_0} - \frac{f_{d0}f_{g0}D_0'}{D_0^2})v_{{\rm dif}0x}^2\right]\mathfrak{f}_{g1} + \nonumber\\ (\frac{dv'_{0y}}{dx} + \frac{1}{2}\Omega_0)v'_{1x} + \left[-\nu f_{g0}(\frac{d^2}{dx^2} - k^2) -i\Delta\omega \right]v'_{1y} = 0 \ . \label{eq perturb4}
\end{gather}
We have used the notation $\Delta\omega\equiv\omega-kv'_{0y}+(3/2)k\Omega_0x$. The term $2\Omega_0v'_{0y}\mathfrak{f}_{g1}$ in Equation~(\ref{eq perturb3}) was introduced via the equilibrium Equation~(\ref{eq steady state 2}). In Equation~(\ref{eq perturb matrix}), each of the four rows of the matrix represents Equation~(\ref{eq perturb1})(\ref{eq perturb2})(\ref{eq perturb3})(\ref{eq perturb4}) respectively, and each of the four rows represents the coefficients of the four functional
variables $\mathfrak{p}_1, \mathfrak{f}_{g1}, v'_{1x},$ and $v'_{1y}$. For example, the third block in the second row $\mathcal M_{12}$ denotes the coefficient of $v'_{1x}$ in Equation~(\ref{eq perturb2}), namely, $f_{d0}v_{{\rm dif}0x}/D_0$.

\subsection{Boundary conditions for the perturbation equations}
\label{append boundary}
The perturbation equations at the boundary $x=\pm x_B$ come in the following form:
\begin{gather}
    -i\Delta\omega\mathfrak{p}_1 + (\frac{d}{dx} - \frac{v_{{\rm dif}0x}}{c_s^2t_s})v'_{1x} + ikv'_{1y} = 0 \ ,\\
    c_s^2\frac{d}{dx}\mathfrak{p}_1 + \left[-\nu(\frac{d^2}{dx^2}-k^2) - i\Delta\omega \right]v'_{1x} - 2\Omega_0v'_{1y} = 0 \ ,\\
    ikc_s^2\mathfrak{p}_1 + (\frac{dv'_{0y}}{dx} + \frac{1}{2}\Omega_0)v'_{1x} + \left[-\nu(\frac{d^2}{dx^2}-k^2)-i\Delta\omega \right]v'_{1y} = 0 \ ,
\end{gather}
where $\Delta\omega$ is defined as in Appendix~\ref{append perturbation}.

These are in fact the perturbation equations in the limit of pure gas, applicable to, for example, a pressure bump without any dust. They are derived either by setting $f_{g0}=1$ and $f_{g1}=0$ in the general perturbation equations (\ref{eq perturb1})--(\ref{eq perturb4}) or by removing dust content from Equations~(\ref{eq one fluid den})--(\ref{eq one fluid pre}) and then linearizing them directly.
%These equations are verified by direct derivation from Equations~(\ref{eq one fluid den})(\ref{eq one fluid pre})(\ref{eq one fluid mom}) without dust and external force. 

Application of the WKBJ approximation means to take $\mathfrak{p}_1, v'_{1x},$ and $v'_{1y}$ proportional to a shared plane-wave form $\exp{(ik_xx)}$. %respectively. 
The equations above are therefore reduced to the following linear system:
\begin{gather}
    \frac{-i\Delta\omega}{\Omega_0}\mathfrak{p}_1 + \frac{ik_x - v_{{\rm dif}0x}/c_s^2t_s}{1/H}\frac{v'_{1x}}{c_s} + \frac{ik}{1/H}\frac{v'_{1y}}{c_s} = 0\ , \\
    \frac{ik_x}{1/H}\mathfrak{p}_1 + \left[\frac{1}{\Omega_0}\nu(k_x^2+k^2) - \frac{i\Delta\omega}{\Omega_0} \right]\frac{v'_{1x}}{c_s} - 2\frac{v'_{1y}}{c_s} = 0\ , \\
    \frac{ik}{1/H}\mathfrak{p}_1 + (\frac{1}{\Omega_0}\frac{dv'_{0y}}{dx} + \frac{1}{2})\frac{v'_{1x}}{c_s} + \left[\frac{1}{\Omega_0}\nu(k_x^2+k^2)-\frac{i\Delta\omega}{\Omega_0} \right]\frac{v'_{1y}}{c_s} = 0 \ ,
\end{gather}
where we non-dimensionalised all physical quantities. This implies the condition for the existence of nontrivial solution:
\begin{equation}
    \det\begin{bmatrix}
    \displaystyle \frac{-i\Delta\omega}{\Omega_0} & \displaystyle \frac{ik_x - v_{{\rm dif}0x}/c_s^2t_s}{1/H} & \displaystyle \frac{ik}{1/H} \\
    \displaystyle \frac{ik_x}{1/H} & \displaystyle \frac{1}{\Omega_0}\nu(k_x^2+k^2) - \frac{i\Delta\omega}{\Omega_0} & -2 \\
    \displaystyle \frac{ik}{1/H} & \displaystyle \frac{1}{\Omega_0}\frac{dv'_{0y}}{dx} + \frac{1}{2} & \displaystyle \frac{1}{\Omega_0}\nu(k_x^2+k^2) - \frac{i\Delta\omega}{\Omega_0}
    \end{bmatrix} = 0 \ .
\end{equation}
This algebraic equation gives the desired dispersion relation at the boundaries.

\subsection{Vortensity equation}
\label{append vortensity}
Here, we derive the expression of the vortensity $q$ in a Keplerian shearing sheet and its governing equation for the single fluid formulation. We start by taking the curl on both sides of Equation~(\ref{eq one fluid mom}). We use the vector calculus identities $(\bm v'\cdot\nabla)\bm v'=(\nabla\times\bm v')\times\bm v'+(1/2)\nabla(|\bm v'|^2)$ and $\nabla\times\nabla f=0$ to obtain
\begin{equation}
    \frac{\pa\bm w'}{\pa t} - \frac{3}{2}\Omega_0x\frac{\pa\bm w'}{\pa y} + \nabla\times(\bm w'\times\bm v') + \frac{1}{2}\Omega_0(\nabla\cdot\bm v'){\bm e}_z = \rho S{\bm e}_z\ ,
\end{equation}
where $\bm w'\equiv\nabla\times \bm v'$, and the source term has been put down in Equation~(\ref{eq vor source}). Then, we use the identity $\nabla\times(\bm A\times\bm B) = (\nabla\cdot\bm B)\bm A + (\bm B\cdot\nabla)\bm A - (\nabla\cdot\bm A)\bm B - (\bm A\cdot\nabla)\bm B$ to expand the curl term on the left-hand side, followed by the use of Equation~(\ref{eq one fluid den}) to substitute the velocity divergence term:
\begin{gather}
    \left(\frac{\pa}{\pa t}+\bm v'\cdot\nabla-\frac{3}{2}\Omega_0x\frac{\pa}{\pa y}\right)\bm w' = \nonumber\\
    \left(\frac{1}{2}\Omega_0{\bm e}_z+\bm w'\right)\frac{1}{\rho}\left(\frac{\pa}{\pa t}+\bm v'\cdot\nabla-\frac{3}{2}\Omega_0x\frac{\pa}{\pa y}\right)\rho + \rho S{\bm e}_z \ , \label{eq derive q}
\end{gather}
where the term $(\bm w'\cdot\nabla)\bm v'$ vanishes because $\bm w'$ only has the $z$-component.

All terms in the equation above are non-zero only in the $z$ direction. If we define the vortensity by
\begin{equation}
    q \equiv \frac{(1/2)\Omega_0+ w'_z}{\rho}
\end{equation}
and divide both sides of Equation~(\ref{eq derive q}) by $\rho$, the two differential operators can be combined into one acting on $q$. This shows the motivation of defining $q$ in the particular way here and the derivation of Equation~(\ref{eq one fluid vor}). The source term will vanish in the limit of pure gas, leaving the vortensity as a conserved material quantity.

\subsection{Relative kinetic energy}
\label{append relative Ek}
The perturbed relative velocity is given by
\begin{equation}
    {\bm v}_{d1}-{\bm v}_{g1} = {\bm v}_{\rm dif1} + (c_s^2t_s\frac{1}{P_0}\frac{dP_1}{dx} +v_{{\rm dif}0x}\frac{P_1}{P_0}){\bm e}_x + ikc_s^2t_s\frac{P_1}{P_0}{\bm e}_y
\end{equation}
from Equation~(\ref{eq vd vg vdif}). Then, the perturbed kinetic energy due to the gas-dust relative motion is the total perturbed kinetic energy of the gas and the dust subtracted by the kinetic energy of the center of mass, i.e., that of the single fluid:
\begin{gather}
    E_{k1,\rm rel} = \int_{-x_B}^{x_B}\frac{1}{2}(\rho_{g0}|{\bm v}_{g1}|^2+\rho_{d0}|{\bm v}_{d1}|^2-\rho_0|{\bm v}_1|^2)dx \nonumber \\ = \int_{-x_B}^{x_B}\frac{1}{2}\frac{\rho_{d0}\rho_{g0}}{\rho_{d0}+\rho_{g0}}|{\bm v}_{d1}-{\bm v}_{g1}|^2dx \ .
\end{gather}
The squared modulus of the complex perturbed velocity does not depend on the azimuthal phase, so it suffices to integrate $E_{k1,\rm rel}$ radially (with respect to $x$).

\section{Numerical techniques for solving the equilibrium equations}
\label{append equilibrium}
We solve the equilibrium equations (\ref{eq steady state vx})--(\ref{eq steady state vy}) as an initial value problem (IVP) by focusing on half of the domain $0<x<x_B$, exploiting the symmetry $P_0(x)=P_0(-x)$ and $f_{g0}(x)=f_{g0}(-x)$.
We formulate the variables as a three-component vector $(u_0, u_1, u_2)$ and specify the equations and initial values as follows:
\begin{gather}
    u_0 = P_0\ , \label{eq steady state numerical first}\\
    u_1 = f_{g0}\ ,\\
    u_2 = v_{\rm dif0} = \frac{D_0}{f_{d0}}\frac{\pa \ln f_{g0}}{\pa x}\ ,
\end{gather}
\begin{gather}
    \frac{du_0}{dx} = -\frac{1}{c_s^2t_s}u_0u_2\ ,\\
    \frac{du_1}{dx} = \frac{(1-u_1)u_1}{D_0}u_2\ ,\\
    \frac{du_2}{dx} = -\frac{u_1}{2t_s(1-u_1)} -\frac{\Omega_0v'_{0y}}{(1-u_1)u_2} + \left(\frac{1}{2c_s^2t_s}+\frac{1}{2D_0}\right)u_2^2\ , \label{eq steady state numerical singularity}
\end{gather}
\begin{gather}
    u_0|_{x=0} = P_{\rm max}\ ,\\
    u_1|_{x=0} = f_{g{\rm min}}\ ,\\
    u_2|_{x=0} = 0\ , \label{eq steady state numerical last}
\end{gather}
where $P_{\rm max}$ and $f_{g{\rm min}}$ are the pressure and gas fraction at $x=0$. As dust is most concentrated at the pressure maximum, we expect $f_{g{\rm min}}$ to be a minimum of the gas fraction profile. The maximal pressure $P_{\rm max}$ can in fact be taken arbitrarily, as we would normalise $P_0(x)$ by the background value $P_b=\lim_{x\to\infty}P_0(x)$ given by the numerical solution. We simply set $P_{\rm max}=1$. Here, treating the system as an IVP is much simpler and more robust than treating it as a boundary value problem in the domain $-x_B<x<x_B$ where the boundary conditions are not exactly known.

%One iteration suffices for the IVP approach because conditions at $x=0$ are known: arbitrary for $u_0$, a free parameter for $u_1$, and exactly zero for $u_2$. The third condition in particular relies on the symmetric setting of our problem: iterations are likely necessary for, e.g., a pressure bump with a background pressure gradient.

To solve the problem (\ref{eq steady state numerical first})--(\ref{eq steady state numerical last}), we use the Fortran package LSODA (via \texttt{scipy.integrate}), a highly optimised solver for initial value problems that can automatically switch between non-stiff and stiff solvers depending on the specific problem. Our numerical tests show that, compared to other methods such as the explicit RK45 and implicit BDF methods available in \texttt{scipy.integrate}, LSODA generally stands out in terms of stability and efficiency.
%We also note that different solving algorithms, maximal step values and error tolerances give consistent outputs when convergent. 

We start from a tiny positive initial value $x_0$ rather than exactly at $x=0$ to avoid singularity. Taking the limit $x\to0$ on both sides of Equation~(\ref{eq steady state numerical singularity}), we obtain
\begin{gather}
    \left(\frac{d^2f_{g0}}{d x^2}\right)_{x=0} = -\frac{1}{2t_sD_0}f_{g{\rm min}}^2 - \nonumber\\ \frac{\Omega_0}{D_0^2}(1-f_{g{\rm min}})f_{g{\rm min}}^2\left(\frac{dv'_{0y}}{dx}\right)_{x=0}\left(1\bigg/\frac{d^2f_{g0}}{dx^2}\right)_{x=0}\ .
\end{gather}
%As $u_2$ appears in the denominator in the equations, we encounter a singularity
%Even so, the problem will not allow us to solve directly from $x=0$. To illustrate, %This is circumvented by a small manual step to evolve the system. 
%Here $f_{g{\rm min}}$ is a free parameter and $dv'_{0y}/dx$ is known analytically by Equation~(\ref{eq steady state vy}).
The equation is quadratic in $(d^2f_{g0}/dx^2)_{x=0}$, which always gives one positive and one negative real root. Since the gas fraction reaches minimum at $x=0$, only the positive root is physically meaningful. We arbitrarily use $x_0=10^{-16}H$ and $(f_{g0})_{x=x_0}= f_{g{\rm min}}$ and $(df_g/dx)_{x=x_0} = x_0(d^2f_{g0}/dx^2)_{x=0}$ in the initial conditions. The solution of the IVP, when convergent, is insensitive to different choices of $x_0$.

%, but if we start to solve from $x=0$, round-off errors might lead the system towards the negative direction. Moreover, $df_{g0}/dx$ appears in the denominator and vanishes at the gas fraction minimum, which paralyzes the algorithm at the very first step. 

% Our approach is to start from a tiny positive initial value $x_0$, solve for $(d^2f_{g0}/dx^2)_{x=0}$ from the quadratic equation above, and then take $(f_{g0})_{x=x_0}= f_{g{\rm min}}$ and $(df_g/dx)_{x=x_0} = x_0(d^2f_{g0}/dx^2)_{x=0}$. This successfully leads to a growth of $f_{g0}(x)$ as $x$ increases. 

\section{Numerical techniques for the perturbation equations}
\label{append perturb}
The discretisation of Equation~(\ref{eq perturb matrix}), or equivalently Equations (\ref{eq perturb1})--(\ref{eq perturb4}), is performed as follows. We represent the four unknown perturbation function variables $\mathfrak{p}_1(x), \mathfrak{f}_{g1}(x), v'_{1x}(x),$ and $v'_{1y}(x)$ as four vectors on a grid with uniform spacing $h$ on the $x$-axis, defined by $x_0=-x_B/H, x_1=x_0+h, x_2=x_0+2h,...,x_{(N-1)/2}=0,...,x_{N-2}, x_{N-1}=x_B/H$. For example, the $n$th component of the pressure perturbation vector $\Vec{\mathfrak{p}}_1$ would be the function value at the $n$th grid point, namely, $\Vec{\mathfrak{p}}_{1n}=\Vec{\mathfrak{p}}_1(x_n),\ n=0,1,...,N-1$. For the coefficients, or linear operators, we construct a matrix of size $4N\times4N$, which is divided into 16 blocks of size $N\times N$. %Each of the four rows of the block matrix represents Equation~(\ref{eq perturb1})(\ref{eq perturb2})(\ref{eq perturb3})(\ref{eq perturb4}) respectively, and each of the four rows represents the coefficients of the four function variables $\mathfrak{p}_1, \mathfrak{f}_{g1}, v'_{1x},$ and $v'_{1y}$. For example, the third block in the second row denotes the coefficient of $v'_{1x}$ in Equation~(\ref{eq perturb2}), namely $f_{d0}/D_0$. 
Put in a formula, if we denote the block matrix as $\bm A=[\bm M_{ij}],\ i,j=0,1,2,3$, and the perturbation variables as a column vector $\Vec{u}_1=(\Vec{\mathfrak{p}}_1,\Vec{\mathfrak{f}}_{g1}, \Vec{v}'_{1x}, \Vec{v}'_{1y})^\top$, the perturbation equations can now be written as
\begin{equation}
    \bm A \Vec{u}_1 =
    \begin{bmatrix}
        \bm M_{00} & \bm M_{01} & \bm M_{02} & \bm M_{03}\\[0.5pt]
        \bm M_{10} & \bm M_{11} & \bm M_{12} & \bm M_{13}\\[0.5pt]
        \bm M_{20} & \bm M_{21} & \bm M_{22} & \bm M_{23}\\[0.5pt]
        \bm M_{30} & \bm M_{31} & \bm M_{32} & \bm M_{33}
    \end{bmatrix}
    \begin{bmatrix}
        \Vec{\mathfrak{p}}_1\\
        \Vec{\mathfrak{f}}_{g1}\\
        \Vec{v}'_{1x}\\
        \Vec{v}'_{1y}
    \end{bmatrix} = 0\ ,
\end{equation}
in parallel with Equation~(\ref{eq perturb matrix}).

Within each block $\bm M_{ij}$, each row represents the equation at a specific grid point. We start from a zero matrix. Except for the boundaries, all first and second derivative operators are approximated by the central differencing scheme. Specifically, if an operator of the form $g(x)\frac{d}{dx}$ shows up, then each row of the corresponding block $\bm M_{ij}$, whose ($p$th, $q$th) element is denoted by $\bm M_{ij,p,q}$, will be modified by
\begin{align}
    M_{ij,n,n-1} \mathrel{-}=&\ g(x_n)/2h \ ,\\
    M_{ij,n,n+1} \mathrel{+}=&\ g(x_n)/2h \ .
\end{align}
%Here $\mathrel{+}=$ denotes adding the right-hand side (a value) to the left-hand side (a variable stored in the computer). 
If a second derivative $g(x)\frac{d^2}{dx^2}$ appears, then the modification will become
\begin{align}
    M_{ij,n,n-1} \mathrel{+}=&\ g(x_n)/h^2\ ,\\
    M_{ij,n,n} \mathrel{-}=&\ 2g(x_n)/h^2\ ,\\
    M_{ij,n,n+1} \mathrel{+}=&\ g(x_n)/h^2\ .
\end{align}
For other operators without a derivative, $g(x)$, we simply add them into the diagonal:
\begin{equation}
    M_{ij,n,n} \mathrel{+}= g(x_n)\ .
\end{equation}
For all operations above, $n=1,2,...,N-2$.

The first and last rows of each block represents the boundary. Our outgoing boundary conditions, written as operators, are in the form $(d/dx-ik_x)$ (see Section~\ref{subsec boundary conditions}). Here we use the forward or backward differencing scheme to tackle the derivative, i.e., modifying the block by
\begin{align}
    M_{ij,0,0} \mathrel{+}=&\ -1/h\ ,\\
    M_{ij,0,1} \mathrel{+}=&\ 1/h-ik_x(\omega,k,-x_B) \ ,\\
    M_{ij,N-1,N-2} \mathrel{+}=&\ -1/h-ik_x(\omega,k,x_B) \ ,\\
    M_{ij,N-1,N-1} \mathrel{+}=&\ 1/h \ .
\end{align}

\begin{figure}
    \centering
    \includegraphics[width=0.5\textwidth]{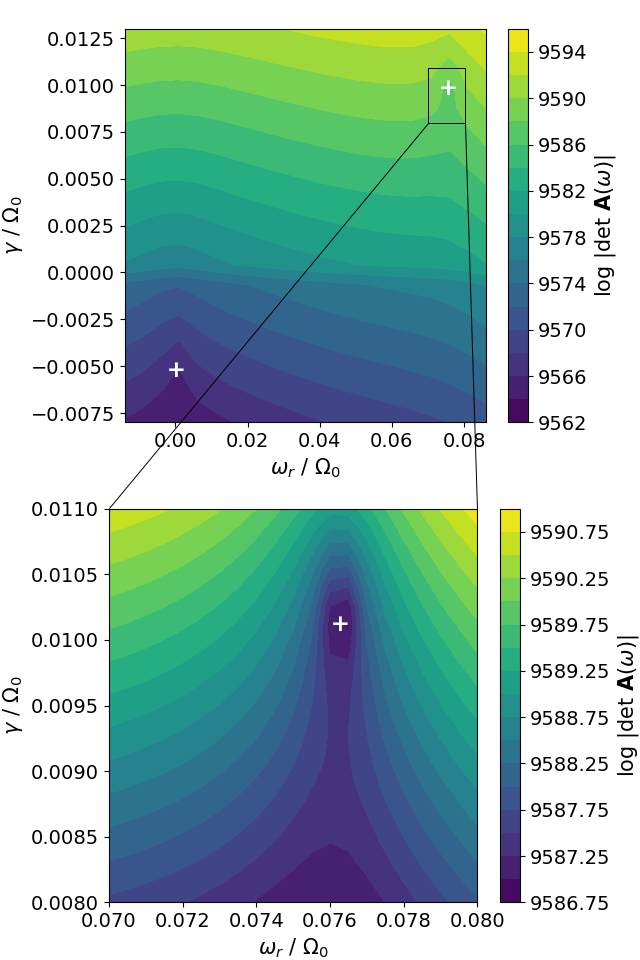}
    \caption{Local minima of $\log|\det\bm A(\omega_r+i\gamma)|$ for $A=0.8$, $\Delta w/H=1.5$, $St=0.03$, $\alpha=3\times10^{-4}$, $f_{g\rm min}=0.7$, and $k=0.1$. The first panel shows two local minima (white plus signs) at $(0.000-0.005i)\Omega_0$ and $(0.076+0.010i)\Omega_0$. The second panel is an enlargement of the second local minimum, which corresponds to a Type \RNum{2} DRWI. We do not follow up the first minimum which is a decaying mode.}
    \label{fig det minimum}
\end{figure}

All blocks are therefore sparse matrices, with nonzero terms all lying on or immediately above or below the main diagonal. In practice, the package \texttt{findiff} automatically generates such matrix representation given the operators and boundary conditions. This completes our construction of the finite difference matrix $\bm A$, which only has $\omega$ as the unknown variable if $k$ is given. The eigenvalue $\omega_m$ and eigenfunction $\Vec{u}_{1m}(x,\omega_m)=(\mathfrak{p}_{1m}(x), \mathfrak{f}_{g1m}(x), v'_{1xm}(x), v'_{1ym}(x))^\top$ satisfy
\begin{gather}
    {\rm det}\bm A(\omega_m)=0\ , \label{eq det}\\
    \bm A(\omega_m)\Vec{u}_{1m}(x,\omega_m)=0\ . \label{eq eigenfunction}
\end{gather}
Equation~(\ref{eq det}) as a nonlinear scalar equation in $\omega_m$ gives a dispersion relation of $\omega_m$ and $k$. Each $k$ corresponds to an eigenvalue problem. We follow the method described in \citet{Ono16} to solve the desired $\omega_m$ from this equation. Preceding this, the parameters $A, \Delta w, St, \alpha,$ and $f_{g{\rm min}}$ are given, and the equilibrium solutions are already calculated. In the first step, we broadly explore the magnitude of the determinant as a function of $\omega_r$ and $\gamma$. Plotted in a two-dimensional contour, the root $\omega_{rm}+i\gamma_m$ resides in a local minimum of $|\det\bm A(\omega_r+i\gamma)|$ which can be identified by seeing an abrupt reduction of this value by
%suddenly decreases 
by several orders of magnitude. Such minima are shown in Figure~\ref{fig det minimum} for example. Among the numerous existent minima, we focus on the few with positive $\gamma_m$. After roughly determining the location of $\omega_m$, we apply Muller's method to find the accurate eigenvalue. Muller's method uses iteration to solve for one complex root of a scalar equation. It requires three initial guesses, which we provide as three different points close to the local minimum. The iteration usually converges for a tolerance of $10^{-13}$ within about five steps in our problem.

The method above allows us to visualise the distribution of possible eigenvalues $\omega$ and then to identify which one is desired, but is inefficient and non-automated. In particular, Muller's method will not easily converge to the desired solution if the initial guesses are not in the small local minimum region, so one cannot start from an arbitrary guess. Therefore, after grasping the general distribution patterns of the eigenvalues, we use the following alternative method to evaluate $\omega_m$, which is fast, automatic, and only requires a rough initial estimate of $\omega_m$. Note that we have arranged Equations~(\ref{eq perturb1})--(\ref{eq perturb4}) such that the angular frequency appears only once in each equation and only takes the form $-i\omega$. In our construction of the matrix $\bm A(\omega)$, then, all $\omega$ lie on the diagonal as $-i\omega$ with the exception of eight boundary rows that involve nonlinear relations of $\omega$. This implies that $\bm A(\omega)\approx \bm A(0)-i\omega\bm I$, where $\bm I$ is the $4N\times4N$ identity matrix. Therefore, the desired eigenvalue $\omega_m$ might be very close to another complex number $\lambda_m$, the latter satisfying
\begin{equation}
    \det(\bm A(0)-i\lambda_m\bm I) = 0\ .
\end{equation}
In other words, $i\omega_m$ is approximately equal to one of the eigenvalues, $i\lambda_m$, of the matrix $\bm A(0)$ (not to be confused with the eigenvalue of the perturbation problem). We verify by calculating all matrix eigenvalues that the approximation is valid for small $|\omega_m|$, which includes almost all cases of interest in this paper.\footnote{In the rare circumstance that the matrix eigenvalue deviates from $i\omega_m$ (in particular, near the bifurcation point in Appendix~\ref{append bifurcation}), we made experiential corrections to $\lambda$ to assist convergence.} In particular, %initial guesses set near $\lambda_m$ 
choosing $\lambda_m$ and two arbitrary points nearby are generally found to be sufficient
%close to $\omega_m$ 
to ensure fast convergence toward $\omega_m$ using Muller's method. One would in fact expect an exact equation of $\omega_m=\lambda_m$ if the boundary conditions were linear in $\omega$.

To obtain $\lambda_m$, we %We can thus 
set a rough initial guess $\omega_{\rm guess}$ and find among the matrix eigenvalues the one closest to $i\omega_{\rm guess}$ on the complex plane. This is done using the shifted inverse power method (in \texttt{ARPACK} software, available via \texttt{scipy.sparse.linalg.eigs}). The resulting $\lambda_m$ and two arbitrary points nearby are then %Then, $\lambda_m$ and two arbitrary points nearby can be 
provided as initial guesses to Muller's method to give the accurate $\omega_m$. %in magnitude, denoted by $i\lambda_0$. 
The method will converge as long as $i\omega_{\rm guess}$ is closer to $i\lambda_m$ than any other irrelevant matrix eigenvalues, granting much greater tolerance compated to directly using Muller's method. %immense tolerance for the choice of $\omega_{\rm guess}$: 
For example, in Figure~\ref{fig det minimum}, $\omega_{\rm guess}=(0.03+0.05i)\Omega_0, (0.05+0i)\Omega_0$ or $(0.2+1i)\Omega_0$ all yield $\omega_m=(0.076+0.010i)\Omega_0$. In contrast, %, is well-separated from other irrelevant matrix eigenvalues, so the method robustly returns $\lambda_0$ that matches our desired $\lambda_m$. 
if we attempted to use Muller's method directly, a deviation of $0.002\Omega_0$ of the initial guesses from the true root would almost certainly fail the algorithm. %; on the other hand, the initial estimate $\omega_0$ likely correctly converges even if put at a distance of $0.04$ from $\omega_m$.

We first noticed the Type~\RNum{2} DRWI by brute-force scan in the $\omega_r$--$\gamma$ plane before developing the method above, but the discovery would have been very straightforward by noting two distinct types of $\lambda_m$ with positive imaginary parts among all matrix eigenvalues. Also, as one progressively increases $k$, no new unstable candidates showed up from the matrix eigenvalues. This, along with a few confirmatory brute-force scans at high $k$ that show no evidence of new roots,
%detected nothing interesting
indicates that short-range instabilities unlikely exist in our setting.

After obtaining $\omega_m$, we proceed to solve the eigenfunction $\Vec{u}_{1m}(x,\omega_m)$, following \citet{Ono16}, by introducing a random vector with a %of 
norm of order the machine precision on the right-hand side of Equation~(\ref{eq eigenfunction}) and then solving the linear equation with LAPACK (via \texttt{scipy.linalg.solve}). An alternative method is to calculate the eigenvector of $\bm{A}(\omega_m)$ of matrix eigenvalue zero, which yields the same result.

\section{Bifurcation of the two types of instabilities}
\label{append bifurcation}
\begin{figure*}
    \centering
    \includegraphics[width=\textwidth]{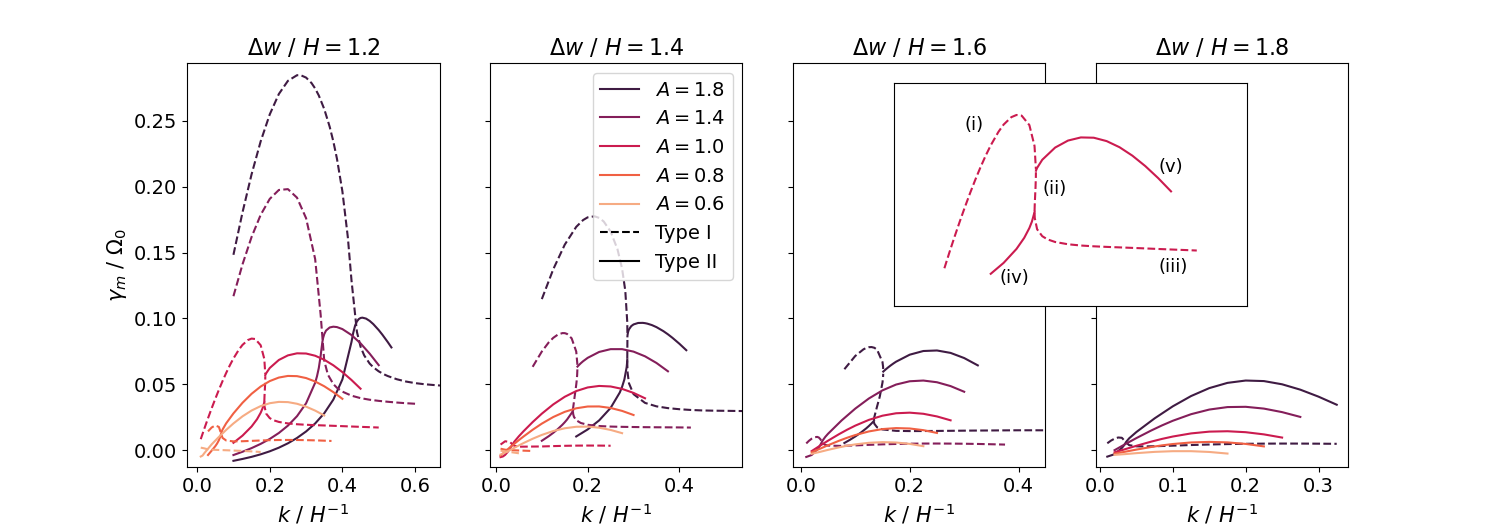}
    \includegraphics[width=\textwidth]{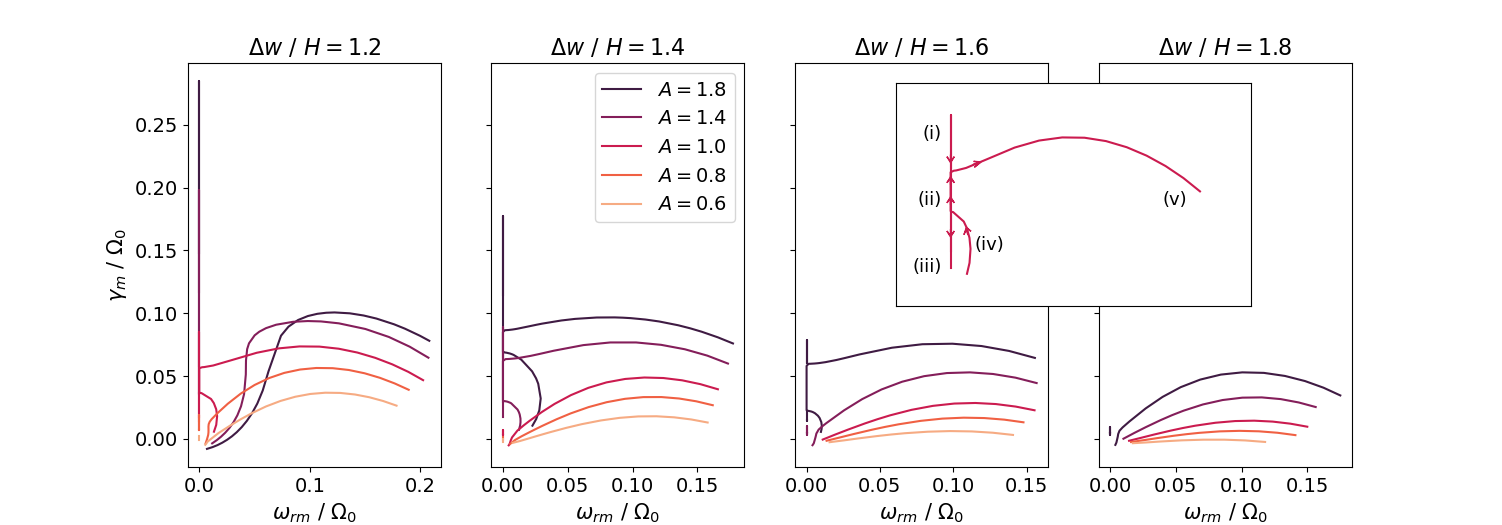}
    \caption{Dispersion relation of different pressure bumps with $f_{g\rm min}=0.7, St=0.03$, and $\alpha=1\times10^{-4}$. Inset boxes on the upper and lower right specifically display $A=1.0$ and $\Delta w/H=1.2$ as an example. The Type \RNum{1} modes for $A=1.6, \Delta w/H\leq1.0$ and for $A=1.8, \Delta w/H\leq1.4$ are stable and hence not shown. \textit{Top:} growth rate $\gamma_m$ versus azimuthal wavenumber $k$. \textit{Bottom:} $\gamma_m$ versus phase angular velocity $\omega_{rm}$ of the instability. Type \RNum{2} modes with negative $\omega_{rm}$ are not drawn. Arrows in the inset box mark the direction of increasing $k$. See text for the explanation of segments (i) to (v).}
    \label{fig bifurcation}
\end{figure*}

Despite the neat distinction in Figure~\ref{fig dispersion sharp bump}, the Type~\RNum{1} and \RNum{2} DRWIs are subtly intertwined in a broader parameter space, especially for moderately large $A$ and small $\Delta w$ combined with large $St$ and small $\alpha$. Shown in Figure~\ref{fig bifurcation} is a collection of dispersion relations of various pressure bump heights and widths. The dispersion relation may be thought of residing in the 3D space of $(k, \omega_{rm}, \gamma_m)$, of which we display the 2D projections. Within the parameter space explored in this work, we categorise these conditions as follows:

a) Moderately smooth %Modestly sharp 
bumps, e.g., $A=0.6, \Delta w/H=1.2$, or $A=1.8, \Delta w/H=1.8$. The setting in Figure~\ref{fig dispersion sharp bump} also belongs to this group. The Type \RNum{1} DRWI has slow growth rates while the Type \RNum{2} shows one single curve in the $(k, \omega_{rm}, \gamma_m)$ space with a maximum $\gamma_m$. The curves of these two modes do not intersect in the 3D space. %These two modes do not coincide in the sense that, for a given set of ($A,\Delta w$), the two types of curves in the $\gamma_m$--$\omega_{rm}$ plot do not intersect. 
In other words, it is impossible to switch from one type to the other continuously: as $k$ increases, the real part of the Type \RNum{1} eigenvalue remains zero but that of the Type \RNum{2} monotonically deviates from zero. However, a kink emerges from, e.g., the Type \RNum{2} $\gamma_m$--$\omega_{rm}$ curve of $A=0.8,\Delta w/H=1.2$ at $(\omega_{rm}/\Omega_0, \gamma_m/\Omega_0)=(0.01,0.01)$, which heralds the reversal of the Type \RNum{2} curve towards $\omega_{rm}=0$ for sharper bumps.

b) Moderately sharp bumps, e.g., $A=1.0, \Delta w/H=1.2$, or $A=1.8, \Delta w/H=1.6$. In the $\gamma_m$--$\omega_{rm}$ plot, the kink described above develops until touching the $y$-axis, forming two bifurcation points with the Type \RNum{1} curve. The trend is best understood viewing the bottom inset panel in Figure~\ref{fig bifurcation}. Initially ($k\ll1$), the Type \RNum{2} mode corresponds to the lower end of the branch (iv). With the continuous increase of $k$ (along the arrow), the phase angular velocity reaches a local maximum before reducing to zero. Now, upon arrival at the lower bifurcation point, an additional increment of $k$ may lead the curve to two directions: upwards, i.e., branch (ii), or downwards, i.e., branch (iii). Importantly, both have $\omega_{rm}=0$ and hence are Type \RNum{1}, indicating a type transition. Along the branch (ii), further increase of $k$ encounters the descending Type \RNum{1} branch (i) at the upper bifurcation point. Another type transition occurs here with the emergence of the Type \RNum{2} branch (v). In short, the previously monolithic Type \RNum{2} curve is now inserted with a Type \RNum{1} segment (ii). 

The description above traces the increase of $k$ along branches (iv)--(ii)--(v). Alternatively, from the perspective of the $\gamma_m$--$k$ plot, one may imagine that the branches (i)--(ii)--(iii) and the branches (iv)--(v) previously rested in different planes in the 3D space and thus did not interfere with each other. Enhancement of the bump sharpness,
%from modest to moderate,
however, brings the two curves to the same plane, and the Type \RNum{1} curve cuts the Type \RNum{2} curve into two pieces (iv) and (v) while itself distorted. Indeed, at the bifurcation points, the Type \RNum{1} branches (i) and (ii) or (ii) and (iii) are connected without abrupt turning, but branch (iv) or (v) shoots off obliquely (not tangential to other branches). The Type \RNum{1} trend in the $\gamma_m$--$k$ plot can therefore be described as an upward convex (i) followed by a temporary reversal (ii) and then a flat tail (iii).

We only show the Type \RNum{2} modes with positive $\omega_{rm}$ in Figure~\ref{fig bifurcation}. The negative counterpart would overlap with the shown Type \RNum{2} curves in the $\gamma_m$--$k$ plots and would appear as mirrored images of the positive-$\omega_{rm}$ Type \RNum{2} curves about the $y$-axis in the $\gamma$--$\omega_{rm}$ plots. This reconciles with the mathematical notion that a bifurcation point should have two inbound and two outbound branches. 
%The branch (iii) as viewed in the $\gamma_m$--$k$ plot has a positive slope. Then, there are two viewpoints of the relation between the five segments. Tracing the increase of $k$ suggests considering branch (iii) as a result of type transition, i.e., viewing (iv)--(iii)

c) Very
%Substantially 
sharp bumps, e.g., $A=1.4, \Delta w/H=1.2$. The two types of modes again have independent curves. In the $\gamma_m$--$k$ plot, the Type \RNum{1} curve still has a maximum followed by a long tail, whereas the low-$k$ flank of the Type \RNum{2} curve becomes concave up (showing a positive second derivative). %becomes convex down. %The maximum $\gamma_m$ of the Type \RNum{2} DRWI even appears saturated: the peak of the $A=1.8, \Delta w/H=1.2$ curve is only slightly above that of the $A=1.4, \Delta w/H=1.2$ curve, and such saturation indeed occurs for $f_{g\rm min}=0.5$.

The smooth shift between the two types of instabilities implies a smooth transition of the instability mechanism. In a transition from Type~\RNum{2} to Type~\RNum{1}, the eigenmodal phase velocity $\omega_{rm}/k$ gradually reduces to zero, and the co-rotation radius moves from outside to the center of the dust bump. Also, cross-correlation calculations similar to those in Section~\ref{subsubsec vor analysis} show that the phase difference between $q_1$ and $S_{\rm bar1}$ rapidly increases from a small angle to well over $90^\circ$; %breaks \xb{from} $90^\circ$; %upon bifurcation from Type~\RNum{2} to Type~\RNum{1}, 
meanwhile, the phase difference between the vortensity perturbation and advection diminishes to zero. This indicates that, during the transition, advection takes over the role from the baroclinic source in driving the instability in the dust bump, as might be expected. Further studies of the mechanism behind the bifurcation would be more complex and lie %complicated and lie % While presumably the dust vortical wave more readily couples with the Rossby waves in some particular parameteric settings and wavelengths, a detailed study of the mechanism behind the bifurcation is 
beyond the scope of this paper. % The bifurcation is presumably governed by the effect Here we speculate, based on the idea of resonance, that the sharpness of the pressure bump modifies the resonated characteristic velocity of the dust bump so that 
% vortical wave gradually becomes dominant as against the sound wave in the dust wave region, which couples with the Rossby waves that have $v_{\rm phase}$ approaching $v_{\rm shear}$.

\section{Tentative propagation mechanism of the Type~\RNum{2} DRWI}
\label{append propagation mechanism}
\begin{figure}
    \centering
    \includegraphics[width=0.46\textwidth]{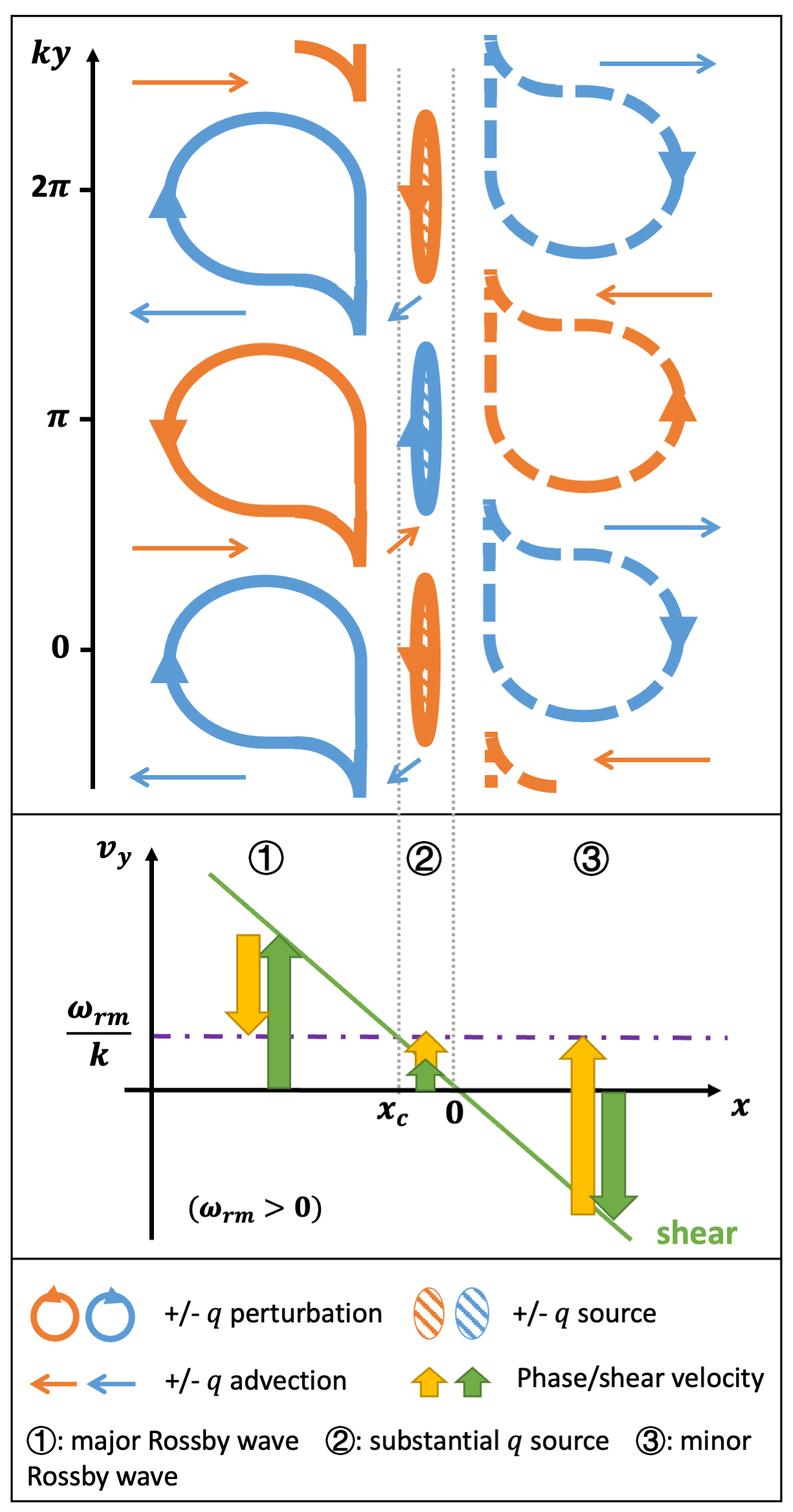}
    \caption{Schematic illustration of the tentative propagation mechanism of the Type \RNum{2} DRWI in the rotating frame with $\omega_{rm}>0$. \textit{Top:} vortensity perturbation, advection and source patterns on the $x$-$y$ plane, extracted from Figure~\ref{fig Type2 eigenfunction 0.70}. The patterns are separated by grey dotted lines into three regions, namely, \protect\textcircled{{1}}, \protect\textcircled{{2}}, and \protect\textcircled{{3}} from left to right. \textit{Bottom:} velocity decomposition in each region. The phase and shear velocities combine to give the modal azimuthal velocity $\omega_{rm}/k$. The two panels share the $x$-axis.}
    \label{fig mechanism}
\end{figure}

One main feature of the Type~\RNum{2} DRWI is a non-zero phase velocity. We have explained the propagation of perturbation patterns in the dust bump in Section~\ref{subsec vor sources}, but the Rossby waves are also important components of this instability. Here, we tentatively explain the propagation of the Type \RNum{2} DRWI with the entire mode in mind, as illustrated in Figure~\ref{fig mechanism}. With a simplified physical picture, we focus on how the different components share one modal phase velocity despite (or perhaps as a result of) the asymmetric eigenfunction patterns.
%In short, we observe that two travelling Rossby waves of different magnitudes couple with a ``dust wave" (vortensity wave strongly modulated by the baroclinic source), the three sharing one global phase velocity despite (or perhaps as a result of) differential background shear.

We first apply the interpretation in \cite{Ono16} to explain the propagation of the gas perturbation patterns. In the following, we will use the words ``upper'' and ``lower'' to describe location with higher or lower $y$-coordinate. The leftmost region (labelled as \textcircled{{1}}) in the top panel in Figure~\ref{fig mechanism}, where $f_d\ll1$ and the background vortensity satisfies $dq_0/dx<0$, illustrates a Rossby wave with a negative phase velocity $v_{\rm phase}$ in $\hat{y}$. For sake of explanation, we temporarily ignore any interaction between regions \textcircled{{1}} and \textcircled{{2}}. 
Positive vortensity perturbations imply clockwise loops of gas flow, while negative counterclockwise. Such flows drive $\hat{x}$-direction advection
of vortensity along the $q_0$ gradient, which induces $y$-direction migration of the loops themselves. For example, the flows at $ky\sim\pi/2$ and $ky\sim3\pi/2$ in region \textcircled{1} increase and decrease local vortensity respectively, the former extending the lower side of the $+q_1$ loop and the latter paring the upper side. The result is a downward shift of the perturbation patterns. However, here the Rossby wave travels on a background with shear velocity $v_{\rm shear}=v'_{0y}-3\Omega_0x/2$, and thus the patterns overall move upwards in the frame co-rotating with $x=0$. %co-rotating frame at $x=0$.

The decomposition of the motion of the Rossby wave into $v_{\rm phase}$ and $v_{\rm shear}$ is essentially built in the vortensity equation (\ref{eq one fluid vor perturb}). Formally, the phase velocity of the Rossby wave can be defined as
\begin{equation}
    v_{\rm phase} = {\rm Re}\left[i\frac{-(dq_0/dx)v'_{1x}}{kq_1}\right]\ , \label{eq v phase}
\end{equation}
where the numerator appeared in Equation~(\ref{eq one fluid vor perturb}) as the radial advection term. Replacing it with the full vortensity advection term  would recover $\omega_{rm}/k$ assuming conservation of vortensity, while replacing it with the azimuthal advection term would give $v_{\rm shear}$. Equation~(\ref{eq v phase}) implies that, when vortensity is conserved, $v_{\rm phase}$ will depend on the strength of the radial vortensity advection relative to that of the vortensity perturbation. 

Application of the argument above on both regions \textcircled{{1}} and \textcircled{{3}} gives two Rossby waves with $v_{\rm phase}$ in the opposite direction. In region \textcircled{{1}} resides what we term the ``major Rossby wave'' (solid) with much more conspicuous perturbation compared to the ``minor Rossby wave'' (dashed) in region \textcircled{{3}}. However, the $x$-direction vortensity advection has comparable strengths in the two regions, hence more influential relative to the minor Rossby wave in terms of shifting the perturbation patterns. This
leads to a faster phase velocity in region \textcircled{3}, as illustrated in the bottom panel in Figure~\ref{fig mechanism}. Then, taking the similar but opposite shearing velocity into account, one observes that $|v_{\rm phase,\textcircled{{3}}}|> |v_{\rm shear,\textcircled{{3}}}| \simeq |v_{\rm shear,\textcircled{{1}}}|> |v_{\rm phase,\textcircled{{1}}}|$.This explains why the two regions share an overall modal phase velocity $\omega_{rm}/k = v_{\rm phase,\textcircled{{3}}} - v_{\rm shear,\textcircled{{3}}} = v_{\rm shear,\textcircled{{1}}} - v_{\rm phase,\textcircled{{1}}}$. Our explanation suggests a close link between the non-zero real part of the eigenvalue and the asymmetry of the eigenfunction.

The propagation of $q_1$ patterns in region \textcircled{2} also fits in this picture. As shown in Section~\ref{subsec vor sources}, here the vortensity advection governs the propagation with a $\sim90^\circ$ spatial phase lead over $q_1$, thus inducing the $\pm q_1$ loops in region \textcircled{2} to travel upwards. This process is similar to that of the Rossby wave, except that the background $v_{\rm shear}$ is now in the same direction as $v_{\rm phase}$. Here, the material flow advects vortensity across different regions (e.g., the inclined arrows between region \textcircled{1} and \textcircled{2} in Figure~\ref{fig mechanism}), indicating a close interaction between the Rossby wave and the dust bump.

\bsp	% typesetting comment
\label{lastpage}
\end{document}